\documentclass[a4paper,11pt]{article}

\usepackage{amsmath,amssymb,amscd}
\usepackage{graphicx}
\usepackage{palatino}

\makeatletter

 \@addtoreset{equation}{section}
\makeatother

\newcounter{Enumerate}

\DeclareFontFamily{U}{rsf}{}
\DeclareFontShape{U}{rsf}{m}{n}{
  <5> <6> rsfs5 <7> <8> <9> rsfs7 <10-> rsfs10}{}
\DeclareMathAlphabet\Scr{U}{rsf}{m}{n}

\newcommand{\bra}[1]{\langle \, {#1} \, |}
\newcommand{\ket}[1]{| \, {#1} \, \rangle}
\newcommand{\Bra}[1]{\big{\langle} \, {#1} \, \big|}
\newcommand{\Ket}[1]{\big| \, {#1} \, \big{\rangle}}

\newcommand{\bracket}[2]{\langle \, {#1} \, | \, {#2}\, \rangle}

\newcommand{\vev}[1]{\langle {#1} \rangle}
\newcommand{\Vev}[1]{\big{\langle} {#1} \big{\rangle}}
\newcommand{\VEV}[1]{\Big{\langle} {#1} \Big{\rangle}}
\newcommand{\cvev}[1]{\langle\!\langle {#1} \rangle\!\rangle}

\newcommand{\cVEV}[1]{\Big{\langle}\!\!\Big{\langle} 
{#1} \Big{\rangle}\!\!\Big{\rangle}}

\newcommand{\Slash}[1]{\ooalign{\hfil/\hfil\crcr$#1$}}

\newcommand{\del}{\partial}

\newcommand{\half}{\frac{1}{2}}

\newcommand{\LS}{\ \ \ \ \ \ \ \ \ \ }
\newcommand{\ls}{\ \ \ \ \ }
\newcommand{\wt}{\widetilde}
\newcommand{\wh}{\widehat}
\newcommand{\ve}{\varepsilon}
\newcommand{\ol}{\overline}

\newcommand{\bsubeq}{\begin{subequations}}
\newcommand{\esubeq}{\end{subequations}}

\newcommand{\tr}{{\rm tr}}

\newcommand{\slb}{\scalebox}

\newcommand{\w}{\wedge}
\renewcommand{\d}{{\rm d}}

\newcommand{\gp}{g^{\frac{1}{4}}}
\newcommand{\gm}{g^{- \frac{1}{4}}} 

\newcommand{\nn}{\nonumber}
\newcommand{\Tr}{{\rm Tr}}
\newcommand{\e}{{\rm e}}

\begin{document}

\allowdisplaybreaks{

\thispagestyle{empty}


\begin{flushright}
YITP-07-20 \\
arXiv:0704.2111 [hep-th] \\
April 2007 
\end{flushright}

\vspace{20mm}

\begin{center}
\slb{2}{Index Theorems on Torsional Geometries}

\vspace{15mm}

\slb{1.3}{Tetsuji {\sc Kimura}}

\vspace{4mm}

\slb{1}{\sl
Yukawa Institute for Theoretical Physics, Kyoto University}

\slb{1}{\sl Sakyo-ku, Kyoto 606-8502, Japan}

\vspace{3mm}

{\tt tetsuji@yukawa.kyoto-u.ac.jp}

\end{center}

\vspace{15mm}


\begin{abstract}
We study various topological invariants on a torsional geometry in the presence
of a totally anti-symmetric torsion $H$ under the closed condition 
$\d H = 0$, which appears in string theory compactification scenarios.
By using the identification between the Clifford algebra on the geometry 
and the canonical quantization condition of fermions in quantum mechanics,
we construct ${\cal N}=1$ quantum mechanical sigma
model in the Hamiltonian formalism. 
We extend this model to ${\cal N}=2$ system,
equipped with the totally anti-symmetric tensor associated
with the torsion on the target space geometry.
Next we construct transition elements in the Lagrangian path
integral formalism and apply them to the analyses of the Witten
indices in supersymmetric systems.
We explicitly show the formulation of the Dirac index on the torsional manifold
which has already been studied. 
We also formulate the Euler characteristic and 
the Hirzebruch signature on the torsional manifold.
\end{abstract}

%

\newpage

\section{Introduction}

Flux compactification scenarios have become one of the most significant
issues in the study of low energy effective theories from string
theories (for instance, see \cite{G0509, DK0610, BKLS0610} and references therein).
Non-trivial fluxes induce a superpotential, which stabilizes 
moduli of a compactified geometry 
and decreases the number of ``redundant'' massless modes in the low energy
effective theory in four dimensional spacetime. 
This mechanism, called the moduli stabilization, also gives a new
insight into cosmology as well as string phenomenology
(\cite{KKLT0301} and a huge number of related works).

Flux compactification provides another interesting issue to 
the compactified geometry itself:
In a specific situation, for instance, 
the NS-NS three-form flux $H_{mnp}$ behaves as a torsion on the
compactified geometry and gives rise to a significant modification 
\cite{Str86}, i.e., 
the K\"{a}hler form is no longer closed.
This phenomenon indicates that the fluxes modify the background
geometry in supergravity in a crucial way.
Of course, the Calabi-Yau condition \cite{CHSW85}
should be influenced by the back reactions from the fluxes
onto the geometry.

If a certain $n$-dimensional manifold has a non-trivial structure group $G$ on its
tangent bundle, this manifold, called the $G$-structure manifold, 
admits the existence of nowhere
vanishing tensors; for example, the metric ($G \subseteq O(n)$), the Levi-Civita
anti-symmetric tensor ($G \subseteq SO(n)$), the almost complex
structure ($G \subseteq U(m)$ where $n = 2m$), 
and the holomorphic $m$-form ($G \subseteq SU(m)$). 
This classification does not exclude the existence of torsion.
(In this sense, a Calabi-Yau $n$-fold is one of the $SU(n)$-structure manifolds.)
This classification is also studied in terms of Killing spinors on the
manifold. 
In particular, the six-dimensional $SU(3)$-structure manifold has been investigated 
in terms of intrinsic torsion \cite{Sala89} and has been applied to
the string theory compactification scenarios \cite{GMW0302}.
Since we mainly study supergravity theories as 
low energy effective theories of string theories, we always assume the
existence of the metric $g_{mn}$  
and dilaton field $\Phi$ on the compactified manifold.
In a generic case of the string compactification, we can also introduce
non-trivial NS-NS three-form flux $H_{mnp}$ 
with its Bianchi identity.
In type II theories appropriate R-R fluxes are also incorporated.
All of these are strongly related via the preserved condition of
supersymmetry. In the heterotic case, supersymmetry variations of
the gravitino $\psi_m$, the dilatino $\lambda$ and the gaugino $\chi$
give rise to the Killing spinor equations 
\bsubeq
\begin{align}
0 \ &= \ 
\delta \psi_m
\ = \ 
\Big( \del_m + \frac{1}{4} \omega_{-m}{}^{ab} \, \Gamma_{ab} \Big) \eta_+
\ \equiv \ 
D_m (\omega_-) \eta_+
\; , \label{const-1-1} \\
0 \ &= \ 
\delta \lambda 
\ = \ 
- \frac{1}{4} \Big( \Gamma^m \nabla_m \Phi - \frac{1}{6} H_{mnp} \Gamma^{mnp}
\Big) \eta_+
\; , \label{const-1-2} \\
0 \ &= \ 
\delta \chi 
\ = \ 
- \frac{1}{4} F_{mn} \Gamma^{mn} \eta_+
\; , \label{const-1-3}
\end{align}
\esubeq
where $\eta_+$ is the Weyl spinor on the six-dimensional manifold
whose normalization is given as 
$\eta_+^{\dagger} \eta_+^{\phantom{\dagger}} = 1$, and
$\omega_{-mab} = \omega_{mab} - H_{mab}$ \cite{Str86}.
Then the NS-NS three-form flux $H_{mnp}$ is interpreted as a totally
anti-symmetric contorsion (or equivalently, a totally anti-symmetric
torsion) on the manifold with negative sign: 
$H^m{}_{np} = - T^m{}_{np} = - \Gamma^m{}_{[np]}$.
The analysis of the manifold 
becomes much clear when we introduce a set of mathematical
definitions such as
\bsubeq
\begin{align}
\text{Almost complex structure} \ &: &
J_m{}^n \ &\equiv \ 
i \eta_+^{\dagger} \Gamma_m{}^n \eta_+^{\phantom{\dagger}}
\; , \ls
J_m{}^p J_p{}^n 
\ = \ 
- \delta_m^n
\; , \label{const-0-0} \\
\text{Lee-form} \ &: &
\theta \ &\equiv \ 
J \, \lrcorner \, \d J \ = \ 
\frac{3}{2} J^{mn} \nabla_{[m} J_{np]} \, \d x^p 
\; , \label{const-0-1} \\
\text{Nijenhuis tensor} \ &: &
N_{mn}{}^p \ &\equiv \ 
J_m{}^q \nabla_{[q} J_{n]}{}^p - J_n{}^q \nabla_{[q} J_{m]}{}^p 
\; , \label{const-0-2} \\
\text{Bismut torsion} \ &: &
T^{\rm (B)}_{mnp} \ &\equiv \ 
\frac{3}{2} J_m{}^q J_n{}^r J_p{}^s \nabla_{[s} J_{qr]}
\ = \ - \frac{3}{2} J_{[m}{}^q \nabla_{|q|} J_{np]}
\; . && \label{const-0-3}
\end{align}
\esubeq
If there are no fermion condensations and $H$-flux
condensation in heterotic string compactified on the manifold with $SU(3)$-structure
satisfying $D_m (\omega_-) J_{np} = 0$,
the compactified manifold is complex and non-K\"{a}hler. 
Actually this is so-called a
conformally balanced manifold, on which the Nijenhuis tensor vanishes
${\cal N}_{mn}{}^p = 0$,
the dilaton field is related to
the Lee-form $\theta = 2 \d \Phi$ and $\d (\e^{-2 \Phi} J \w J) = 0$.
Furthermore, the NS-NS three-form flux $H_{mnp}$ is given by the 
Bismut torsion $T^{\rm (B)}_{mnp}$ \cite{Bismut89}.
We can classify compactified manifolds
under specific conditions in the following way 
(see also the discussions in \cite{IP0010, Pap0201, BT0509}):
\bsubeq
\begin{align}
&\theta \ = \ 2 \d \Phi \; , \ls
\d \big(\e^{-2 \Phi} J \w J \big) \ = \ 0 
&&\to \ \text{conformally balanced}
\label{const-5-1} \\
&\LS \text{if} \ \ 
\theta \ = \ 0 
&&\to \ \text{balanced}
\label{const-5-2} \\
&\LS \text{if} \ \
\d \big( \e^{- \Phi} J \big) \ = \ 0 
&&\to \ \text{conformally K\"{a}hler}
\label{const-5-3} \\
&\LS \text{if} \ \
\d H \ = \ \d T^{\rm (B)} = 0
&&\to \ \text{strong K\"{a}hler with torsion}
\label{const-5-4} 
\end{align}
\esubeq

On the contrary, however, one has not understood a lot of mathematical properties of
the $G$-structure manifold such as moduli and moduli spaces.
This is quite different from the case of Calabi-Yau manifold \cite{BTY0612}.
Because of the lack of knowledge, 
one has not been able to discuss the massless modes on the ground
state in the effective theory derived from string theory compactified
on the $G$-structure manifold.

Similarly, various kinds of topological invariants on torsional
geometries have not been analyzed, although many topological invariants
on Riemannian manifolds have been well investigated. 
Here let us briefly introduce some invariants:
Suppose there exist Dirac fermions in an even dimensional geometry.
We define chirality on the Dirac fermions and find the difference
between the number of fermions with positive chirality and the number
of fermions with negative chirality at the massless level. 
This difference is a topological invariant, which is
called the index of the Dirac operator, or the Dirac index
\cite{Alvarez-gaume83a, ASZ84, FOY86}.
We also introduce the Euler characteristic as the difference between
the number of harmonic even-forms and the number of odd-forms on the
manifold, and the Hirzebruch signature as the difference between the
number of self-dual forms and the number of anti-self-dual forms.
These invariants are described in terms of polynomials of Riemann
curvature two-form (see, for example, \cite{EGH, Nakahara, GSW}).
So far the index of the Dirac operator in the presence of torsion has been
studied \cite{Mav88, Yajima89, PW99, SS0010}.
Unfortunately, however, the other indices on a torsional manifold have not been
analyzed so much.
In particular, it is quite worth studying 
the Euler characteristic on a complex manifold in the
presence of torsion, which will give a new insight on the number of
generation in the flux compactification scenarios.

The main discussion of this paper is to analyze such kinds of
topological invariants derived from the Dirac operator, which appears
in the following equations of motion for fermionic fields in the
supergravity \cite{KY0605}:
\bsubeq \label{eom-fer-SUGRA}
\begin{align}
0 \ &= \ 
\Slash{D} (\omega) \lambda 
- \frac{1}{12} H_{mnp} \Gamma^{mnp} \lambda
\ = \ 
\Slash{D} (\omega - \tfrac{1}{3} H) \lambda 
\; , \\
0 \ &= \ 
\Slash{D} (\omega, A) \chi
- \frac{1}{12} H_{mnp} \Gamma^{mnp} \chi
\ = \ 
\Slash{D} (\omega - \tfrac{1}{3} H, A) \chi 
\; .
\end{align}
\esubeq
First, we define the index of the Dirac operator on the
torsional manifold in the infinity limit of $\beta$:
\begin{align}
{\rm index} \Slash{D} 
\ \equiv \ 
\lim_{\beta \to \infty} \Tr \big\{ \Gamma_{(5)} \e^{- \beta \Scr{R}} \big\}
\ = \ 
\lim_{\beta \to 0} \Tr \big\{ \Gamma_{(5)} \e^{- \beta \Scr{R}} \big\}
\; , \label{Dirac-def}
\end{align}
where $\Scr{R}$ is an appropriate regulator, given by the square of the
Dirac operator (or, equivalently, the Laplacian) in a usual case.
Notice that since a topological value is definitely independent of the continuous
parameter $\beta$, we can take the zero limit $\beta \to 0$. 
This topological invariant can be represented as an appropriate
quantum number in supersymmetric quantum mechanics
\cite{Alvarez-gaume83a} via the identification of the cohomology on
the manifold with the supersymmetric states in the quantum mechanics.
To investigate this,
we define the Witten index in the quantum mechanics 
\begin{align}
\lim_{\beta \to 0} \Tr \big\{ (-1)^F \e^{- \frac{\beta}{\hbar} \Scr{H}} \big\}
\ = \ 
\lim_{\beta \to 0} \int \d X \, 
\bra{X} (-1)^F \e^{- \frac{\beta}{\hbar} \Scr{H}} \ket{X}
\; . \label{Dirac-intro}
\end{align}
We identify (\ref{Dirac-def}) with (\ref{Dirac-intro})
via the identification of 
the the regulator $\Scr{R}$ and the chirality operator $\Gamma_{(5)}$  
on the manifold with the Hamiltonian $\Scr{H}$ and the fermion number operator
$(-1)^F$ in the quantum mechanics, respectively.
The trace $\Tr$ denotes the sum of all transition elements whose final
states $\bra{X}$ correspond to the initial states $\ket{X}$.
Second, we rewrite the Witten index from the Hamiltonian formalism, as
described above, to the Lagrangian path integral formalism.
During this process, we introduce  discretized transition elements
and adopt the Weyl-ordered form in order to avoid any ambiguous
ordering of quantum operators.
Then we integrate out momentum variables and obtain the transition
elements described in the configuration space path integral.
Third, we discuss the Feynman rule which defines 
free propagators and interaction terms 
in the supersymmetric systems.
Finally, we evaluate the Witten indices in the quantum mechanical
nonlinear sigma models in appropriate ways.
This procedure is summarized in a clear way by de Boer, Peeters,
Skenderis and van Nieuwenhuizen \cite{dBPSN9509}, and Bastianelli and
van Nieuwenhuizen \cite{BvN06}.
We will apply this technique to the analysis of index theorems on the
torsional manifold.
To simplify the discussion, we impose the closed condition $\d H = 0$
on the NS-NS three-form in the same way as \cite{Yajima89, Mav88}.
This indicates that we only focus on the index theorems on the strong
K\"{a}hler with torsion (\ref{const-5-4}).
Although this condition is too strong to find the suitable solution in the
heterotic string compactification with non-trivial fluxes
\cite{IP0008, KY0605}, it is still of importance to analyze the
manifold with such condition, which also appears in type II string
theory compactifications.

This paper is organized as follows:
In section \ref{Dirac-index-2} we construct ${\cal N}=1$ and ${\cal N}=2$ quantum
supersymmetric Hamiltonians equipped with a non-vanishing totally
anti-symmetric field $H_{mnp}$, which can be regarded as the torsion
on the manifold considered.
In section \ref{SUSY-Lagrangian} we describe the transition elements in the
Hamiltonian formalism and rewrite them to functional path integrals in the
Lagrangian formalism. We also prepare bosonic and fermionic
propagators in the quantum mechanics. This transition elements play
significant roles in the evaluation of the Witten indices in next
sections.
In section \ref{QM1} and \ref{QM-gauge} the Witten index in 
${\cal N}=1$ supersymmetric quantum mechanical nonlinear sigma model
is analyzed. First we review the Witten index associated with the
Dirac index on a usual Riemannian manifold without boundary.
Next we generalize the index on the manifold in the presence of non-trivial torsion
$H$.
We obtain an explicit expression of the Pontrjagin class 
and of the Chern character on the torsional manifold. 
The Euler characteristic corresponding to the Witten index in 
${\cal N}=2$ supersymmetric system is discussed in section \ref{QM-Euler}.
This topological invariant is also discussed on the
torsional manifold.
In section \ref{QM-Hir} we also analyze the derivation of the Hirzebruch
signature on the manifold with and without torsion
from the ${\cal N}=2$ supersymmetric quantum mechanics.
We summarize this paper and discuss open problems and future works in
section \ref{summary}.
We attach some appendices in the last few pages. 
In appendix \ref{app-convention} we list the convention of 
differential geometry which we adopt in this paper.
In appendix \ref{app-formula} a number of useful formulae, which play 
important roles in the computation of Feynman graphs, are listed.


\section{Supersymmetric quantum Hamiltonians} \label{Dirac-index-2}

First of all, we prepare a bosonic operator $x^m$ and its canonical conjugate
momentum $p_m$ in quantum mechanics, whose canonical quantization
condition is defined as a commutation relation between them
in such a way as $[ x^m , p_n ] = i \hbar \, \delta^m_n$.
Since we consider a quantum mechanical nonlinear sigma model, 
we regard $x^m$ as a coordinate on the target space of the sigma
model, where its index runs $m = 1, \dots, D$.
Since the target space is curved, the differential representation of
the canonical momentum operator is given as 
$\gp p_m \gm = - i \hbar \, \del_m$ equipped
with the determinant of the target space metric $g = \det g_{mn}$.
We also introduce a real fermionic operator $\psi^a$ in the
quantum mechanics, equipped with the local Lorentz index 
$a = 1, \dots, D$.
In the quantum mechanics of real fermions, we define the canonical
quantization condition as an anti-commutation relation 
$\{ \psi^a , \psi^b \} = \hbar \, \delta^{ab}$.
Since, under the identification $\psi^a \equiv \sqrt{\frac{\hbar}{2}} \Gamma^a$,
 the structure of this quantization condition can be interpreted as the
$SO(D)$ Clifford algebra given by the anti-commutation relation
 between the Dirac gamma matrices
$\{ \Gamma^a , \Gamma^b \} = 2 \delta^{ab}$ on the target geometry,
we will investigate the Dirac index on this curved geometry 
in terms of the Witten index in the quantum mechanics.
First let us discuss ${\cal N}=1$ supersymmetry, and extend this to 
${\cal N}=2$ supersymmetry under a certain condition.
We should choose
${\cal N}=1$ or ${\cal N}=2$ in the case when we want to study the index density for
the Pontrjagin classes, or for the Euler characteristics, respectively
\cite{Alvarez-gaume83a}\footnote{Alvarez-Gaum\'{e}
  \cite{Alvarez-gaume83a} and Mavromatos \cite{Mav88} refer the 
${\cal N}=2$ (${\cal N}=1$) model to ${\cal N}=1$ (${\cal N}=1/2$) supersymmetric
  quantum mechanics.}.

\subsection{${\cal N}=1$ real supersymmetry}

Now let us introduce the ${\cal N}=1$ supersymmetry algebra 
with respect to a real fermionic charge $Q^1$:
\begin{gather}
\{ Q^1 , Q^1 \} \ = \ 2 \hbar \, \Scr{H}^1
\; . \label{SUSY-algebra-N1}
\end{gather}
Note that $\Scr{H}^1$ is the quantum Hamiltonian in ${\cal N}=1$
system, where the superscript ``$1$''  indicates ${\cal N}=1$.
We will realize this algebra in terms of quantum operators $x^m$,
$p_m$ and $\psi^a$.
It is useful to introduce a covariant momentum operator associated with a
covariant derivative $D_m (\omega - \tfrac{1}{3}H)$ which appears in
the equation of motion in the supergravity (\ref{eom-fer-SUGRA}). 
The covariant momentum operator is 
\begin{align}
\pi^{(-1/3)}_m
\ &\equiv \ 
p_m - \frac{\hbar}{2} \Big( \omega_{mab} - \frac{1}{3} H_{mab} \Big) \Sigma^{ab}
\; .
\end{align}
Later we sometimes use the description $\hat{\omega}_{mab} \equiv 
\omega_{mab} - \frac{1}{3} H_{mab}$.
Since the Dirac operator acts on spinors on the geometry,
the Lorentz generator $\Sigma^{ab}$ is given in the spinor representation, 
which can be described in terms of 
the real fermions via the identification 
$\Gamma^a = \sqrt{\frac{2}{\hbar}} \psi^a$ such as
\begin{align}
\Sigma^{ab}
\ = \ 
\frac{i}{4} \big( \Gamma^a \Gamma^b - \Gamma^b \Gamma^a \big)
\ = \ 
\frac{i}{2 \hbar} \big( \psi^a \psi^b - \psi^b \psi^a \big)
\ \equiv \
\frac{i}{\hbar} \psi^{ab}
\; .
\end{align}
We should also define the action of the covariant momentum on the
fermionic operator:
\begin{align}
\gp [ \pi^{(-1/3)}_m , \psi^a ] \gm
\ &= \ 0
\; , \ls
\gp [ \pi^{(-1/3)}_m , \psi^n ] \gm
\ = \ 
i \hbar \, \Gamma^n_{(-1/3)pm} \, \psi^p
\ = \ 
i \hbar \Big( \Gamma^n_{0pm} - \frac{1}{3} H^n{}_{pm} \Big) \psi^p
\; , 
\end{align}
where $\Gamma^n_{0pm}$ is the Levi-Civita connection defined in
appendix \ref{app-convention}. 
Actually, the above commutator is associated with the covariant
derivative of the Dirac gamma matrix on the target geometry.

By using the covariant momentum $\pi^{(-1/3)}_m$,
let us represent the supercharge $Q_H^1$ and the Hamiltonian
$\Scr{H}_H^1$ (where the subscript $H$ denotes that the operator
contains the torsion $H$) as follows:
\bsubeq \label{SUSY-real}
\begin{align}
Q_H^1 \ &\equiv \ 
\psi^m \gp \pi_m^{(-1/3)} \gm
\ = \ 
\psi^m \gp \Big(
p_m - \frac{i}{2} \big( \omega_{mab} - \tfrac{1}{3} H_{mab} \big)
  \psi^{ab}
\Big) \gm
\; , \\
\Scr{H}_H^1 
\ &= \ 
\half \gm \pi^{(-1)}_m g^{mn} \sqrt{g} \, \pi^{(-1)}_n
\, \gm
+ \frac{\hbar^2}{8} R (\omega)
- \frac{\hbar^2}{24} H_{mnp} H^{mnp}
\; . \label{SQM-H1-3}
\end{align}
\esubeq
Note that we used the closed condition $\d H = 0$.
Since we used the complete square in $\Scr{H}_H^1$, the
magnitude of the torsion in the covariant momentum is changed to
$\pi^{(-1)}_m$. This is consistent with the analysis of the Killing spinor
equation in the heterotic theory \cite{KY0605}. 
We can also formulate the ${\cal N}=1$ supersymmetric charges 
with introducing a (non-abelian) gauge fields on the target space:
\bsubeq \label{SUSY-real-A}
\begin{align}
Q_H^1 \ &= \ 
\psi^m \gp \wt{\pi}_m^{(-1/3)} \gm
\; , \ls
\{ Q_H^1 , Q_H^1 \} \ = \ 2 \hbar \, \Scr{H}_H^1
\; , \\
\Scr{H}_H^1
\ &= \ 
\half \gm \wt{\pi}^{(-1)}_m g^{mn} \sqrt{g} \,
\wt{\pi}^{(-1)}_n \gm
+ \frac{\hbar^2}{8} \Big[ R (\omega) - \frac{1}{3} H_{mnp} H^{mnp} \Big]
\nn \\
\ & \ \ \ \ 
- \half F^{\alpha}_{mn} \psi^{mn} (\hat{c}^{\dagger} T_{\alpha} \hat{c})
\; , \\
\wt{\pi}^{(\alpha)}_m
\ &= \ 
p_m - \frac{i}{2} \Big( \omega_{mab}
+ \alpha H_{mab} \Big) \psi^{ab}
- i A^{\alpha}_m (\hat{c}^{\dagger} T_{\alpha} \hat{c}) 
\; ,
\end{align}
\esubeq
where we used the anti-hermitian matrix $T_{\alpha}$ as a generator of
the gauge symmetry group. We also introduced a complex ghost field
$\hat{c}^i$ living in the quantum mechanics.


\subsection{${\cal N}=2$ complex supersymmetry}

Now we introduce two sets of real fermionic operators $\psi_{\alpha}^a$
($\alpha = 1,2$)
and perform the complexification of fermionic operators 
via linear combination
\bsubeq
\begin{align}
\varphi^a \ &\equiv \ 
\frac{1}{\sqrt{2}} (\psi_1^a + i \psi_2^a) 
\; , \ls
\ol{\varphi}{}^a \ \equiv \ 
\frac{1}{\sqrt{2}} (\psi_1^a - i \psi_2^a) 
\; .
\end{align}
Note that we used the convention $\ol{\varphi}{}^a = (\varphi^a)^{\dagger}$.
Then the canonical quantization condition is extended in such a way as
\begin{align}
\{ \varphi^a , \varphi^b \} \ = \ 0 \; , \ls
\{ \ol{\varphi}{}^a , \ol{\varphi}{}^b \} \ = \ 0 \; , \ls
\{ \varphi^a , \ol{\varphi}{}^b \} \ = \ \hbar \, \delta^{ab}
\; . \label{CQC2}
\end{align}
\esubeq
This is nothing but the $SO(D,D)$ Clifford algebra.
This complex fermion $\varphi^a$ plays a central role in ${\cal N}=2$ supersymmetry,
while $\psi^a$ consists of ${\cal N}=1$ supersymmetry.
Now let us construct the ${\cal N}=2$ supersymmetric model.
Let us define the commutation relations between the covariant momentum operator
$\pi_m$ and the complex fermions, 
which are given in terms of the affine connection 
$\Gamma^p_{0mn}$ in the same analogy as in the ${\cal N}=1$
system:
\begin{align}
\gp [ \pi_m , \varphi^n ] \gm
\ &= \ i \hbar \, \Gamma^n_{0pm} \varphi^p
\; , \ls
\gp [ \pi_m , \varphi_n ] \gm
\ = \ -i \hbar \, \Gamma^p_{0nm} \varphi_p
\; . \label{cov-pi-comm}
\end{align}
The Lorentz generator coupled to the spin connection
and the curvature tensor are expressed as
\bsubeq
\begin{gather}
\Sigma^{ab}
\ = \ 
\frac{i}{\hbar} 
\Big( \varphi^a \ol{\varphi}{}^b - \varphi^b \ol{\varphi}{}^a \Big)
\; , \\
\gp [ \pi_m , \pi_n ] \, \gm 
\ = \ 
\frac{i\hbar^2}{2} R_{abmn} (\omega) \Sigma^{ab}
\ = \ 
- \hbar \, R_{abmn} (\omega) \varphi^a \ol{\varphi}{}^b
\; .
\end{gather}
\esubeq

Next, let us express ${\cal N}=2$ supercharge $Q$
and extend it as the supercharge equipped 
with the torsion given by three-form flux $H$.
In the same way as the ${\cal N}=1$ supercharge,
we will identify the 
de Rham cohomology on the manifold with the ${\cal N}=2$ supersymmetry algebra.
In the case on the Riemannian manifold, we identify the exterior
derivative $\d$ on the geometry with the ${\cal N}=2$ supercharge 
$Q \equiv \varphi^m \gp \pi_m \gm$,
where $\pi_m$ is the covariant momentum in the ${\cal N}=2$ quantum
mechanics defined as
\begin{align}
\pi_m \ &= \ 
p_m - \frac{\hbar}{2} \omega_{mab} \, \Sigma^{ab}
\ = \ 
p_m - i \omega_{mab} \, \varphi^a \ol{\varphi}{}^b
\; .
\end{align}
Let us introduce the torsion on the geometry.
Following the discussions \cite{Braden85, RW86, Mav88, Gua0401, KL0407}, 
we extend the exterior derivative $\d$ to $\d_H$ in such a way as
\begin{align}
\d_H \ &\equiv \ \d + H \w 
\; , \ls
(\d_H)^2 \ = \ (\d H ) \w 
\; . \label{ex-dH}
\end{align}
This means that $\d_H$ is nilpotent up to the
derivative $\d H$, i.e., this yields the equivariant cohomology.
In this paper we always impose the vanishing condition $\d H = 0$.
In addition, by using the Darboux theorem, 
we can identify the one-form with the
holomorphic variable, while the adjoint of the one-form can be
identified with the anti-holomorphic variable.
Thus, we identify the exterior derivative $\d_H$ and its adjoint $\d_H^{\dagger}$
with appropriate operators 
in terms of complex fermions $\varphi^m$ and $\ol{\varphi}{}^m$ in
the quantum mechanics:
\bsubeq \label{def-extend-2}
\begin{align}
\d_H 
\ls \leftrightarrow \ls
Q_H
\ &\equiv \ 
\varphi^m \gp \pi_m \gm 
+ \alpha i H_{mnp} \, \varphi^m \varphi^n \varphi^p
\; , \\
\d_H^{\dagger} 
\ls \leftrightarrow \ls
\ol{Q}_H \ &\equiv \ 
\ol{\varphi}{}^m \gp \pi_m \gm 
+ \ol{\alpha} i H_{mnp} \, \ol{\varphi}{}^m \ol{\varphi}{}^n
\ol{\varphi}{}^p 
\; .
\end{align}
\esubeq
We wish to interpret $Q_H$ as the ``supercharge'', associated with
the exterior derivative $\d_H \equiv \d + H \w$, while $\ol{Q}{}_H$ 
associated with $\d^{\dagger}_{H}$, i.e., the adjoint of the derivative $\d_H$.
Here we also introduced the scale factor $\alpha$, which should be
fixed compared with the ${\cal N}=1$ supercharge.
In order to fix the coefficient $\alpha$, 
let us truncate the supercharge $Q_H$ 
to the supercharge $Q_H^1$ in the ${\cal N}=1$ supersymmetry (\ref{SUSY-real})
via the restriction $\psi_2^a = 0$ and $\psi_1^a \equiv \psi^a$:
\begin{align}
Q_H 
\ &\to \ 
\frac{1}{\sqrt{2}} \psi^m 
\Big\{ \gp p_m \, \gm
- \frac{i}{2} \big( \omega_{mab} - \alpha H_{mab} \big) \psi^a \psi^b \Big\}
\ = \ 
\frac{1}{\sqrt{2}} Q_H^1
\; .
\end{align}
Since we have already known the ${\cal N}=1$ supercharge $Q_H^1$, 
we can fix the coefficient 
\begin{align}
\alpha \ &= \ \frac{1}{3} \ = \ \ol{\alpha}
\; .
\end{align}

Due to the first Bianchi identity $R_{[mnp]q} (\omega) = 0$ 
and $D_{[d} (\omega) H_{cab]} = \frac{1}{4} (\d H)_{dcab} = 0$, 
we find that the supersymmetry algebra is given by
\bsubeq \label{SUSY-algebra-N2}
\begin{gather}
\{ Q_H , Q_H \} \ = \ \frac{\hbar}{6} \, (\d H)_{abcd} \, \varphi^{abcd}
\ = \ 0
\; , \ls
\{ \ol{Q}{}_H , \ol{Q}{}_H \} \ = \ 
\frac{\hbar}{6} \, (\d H)_{abcd} \, \ol{\varphi}{}^{abcd}
\ = \ 0
\; , \\
\{ Q_H , \ol{Q}{}_H \} \ = \ 2 \hbar \, \Scr{H}_H 
\; , \ls
[ Q_H , \Scr{H}_H ]
\ = \ 
- \frac{1}{2 \hbar} [ \ol{Q}{}_H , (Q_H)^2 ]
\ = \ 
0 \; .
\end{gather}
\esubeq
The vanishing condition of the last commutator guarantees the
supersymmetric system, in which the energy levels of the bosonic and
fermionic states are degenerated.
Now we explicitly express the Hamiltonian $\Scr{H}_H$ in
terms of the complex fermions:
\bsubeq \label{SUSY-complex-H}
\begin{align}
Q_H \ &= \ 
\varphi^m \Big( \gp p_m \, \gm
- i \omega_{mab} \, \varphi^a \ol{\varphi}{}^b
+ \frac{i}{3} H_{mab} \, \varphi^{ab} \Big) 
\ = \ 
\varphi^m \Big( \gp \pi_m \, \gm
+ \frac{i}{3} H_{mab} \, \varphi^{ab} \Big) 
\; , \\
\Scr{H}_H
\ &= \ 
\half \gm \Big\{ \pi_m + \frac{i}{2} H_{mab} 
\big( \varphi^{ab} + \ol{\varphi}{}^{ab} \big) \Big\}
g^{mn} \sqrt{g} 
\Big\{ \pi_n + \frac{i}{2} H_{ncd} 
\big( \varphi^{cd} + \ol{\varphi}{}^{cd} \big) \Big\} \gm
\nn \\
\ & \ls
- \half R_{abmn} (\omega) \, 
\varphi^m \ol{\varphi}{}^n \varphi^a \ol{\varphi}{}^b
+ \frac{1}{6} \del_m (H_{npq}) \Bigg(
\ol{\varphi}{}^m \varphi^{npq}
+ \varphi^m \ol{\varphi}{}^{npq}
- \frac{3 \hbar}{2} g^{mn} \varphi^{pq}
- \frac{3 \hbar}{2} g^{mn} \ol{\varphi}{}^{pq}
\Bigg)
\nn \\
\ & \ls
+ \frac{1}{8} H_{mnr} H_{pq}{}^r 
\Big(
\varphi^{mnpq} + \ol{\varphi}{}^{mnpq}
- 2 \varphi^{mn} \ol{\varphi}{}^{pq}
\Big)
- \frac{\hbar}{2} H_{mpq} H_n{}^{pq} \, \varphi^m \ol{\varphi}{}^n
+ \frac{\hbar^2}{12} H_{mnp} H^{mnp}
\; . 
\end{align}
\esubeq

There exists a comment on the Hamiltonians in the ${\cal N}=1$ and
in the ${\cal N}=2$ systems.
The ${\cal N}=1$ Hamiltonian cannot be obtained by truncation of the ${\cal N}=2$
Hamiltonian, because the truncation $\psi_2^a = 0$ is no longer
consistent at the quantum level since 
the anti-commutation relation
$\{ \psi_1^a + i \psi_2^a , \psi_1^b + i \psi_2^b \}$ becomes non-zero
via the truncation.  
On the other hand, we need not use such anti-commutation relation 
when we reduce the ${\cal N}=2$ supercharge to the charge in 
the ${\cal N}=1$ system.


\section{Path integral formalism from Hamiltonian formalism}
\label{SUSY-Lagrangian}

In this section we will discuss a generic strategy to obtain 
the transition element $\bra{x} \e^{- \frac{\beta}{\hbar} \Scr{H}} \ket{y}$ 
which appears in (\ref{Dirac-intro}).
We will introduce a number of useful tools to investigate the quantum
mechanical path integral, i.e., the complete sets of eigenstates, and
the Weyl-ordered form. Next we will move to the concrete constructions of 
the transition elements in the ${\cal N}=1$ and in the ${\cal N}=2$ systems. 
In this paper we omit many technical details which can be seen in the
works \cite{dBPSN9509, BvN06}. We mainly follow the convention defined
in \cite{BvN06}.
Before going to the main discussion,
for later convenience,
let us take a rescaling on the fermionic operators which we introduced
in the previous section: 
\bsubeq \label{rescale-fer}
\begin{align}
\text{${\cal N}=2$ system:} \ls
\varphi^a \ &\to \ \sqrt{\hbar} \, \varphi^a
\; , \\
\text{${\cal N}=1$ system:} \ls
\psi^a \ &\to \ \sqrt{\hbar} \, \psi^a
\; , \LS
\text{ghost fields:} \ls
\varphi^{\rm gh} \ \to \ \sqrt{\hbar} \, \varphi^{\rm gh}
\; .
\end{align}
\esubeq

\subsection{General discussion}

In order to formulate the transition elements we should prepare 
a number of tools.
Let $\wh{x}^m$ and $\wh{p}_m$ be the operators of the
coordinate and the momentum, respectively, while $x^m$ and $p_m$
denote their eigenvalues\footnote{The symbol ``$\wh{\phantom{\cal O}}$'' on
an operator is omitted if there are no confusions.}.
According to \cite{dBPSN9509, BvN06}, 
let us introduce the complete set of the
$\wh{x}$-eigenfunctions and the complete set of the
$\wh{p}$-eigenfunctions 
\begin{align}
\int \d^D x \, \ket{x} \sqrt{g(x)} \bra{x}
\ &\equiv \ 
1
\ \equiv \ 
\int \d^D p \, \ket{p} \bra{p}
\; , \label{complete-sets-boson}
\end{align}
where $g(x) = \det g_{mn} (x)$.
We also define the inner products and the plane wave such as
\bsubeq \label{inn-prod}
\begin{gather}
\bracket{x}{y} \ \equiv \ 
\frac{1}{\sqrt{g(x)}} \delta^D (x - y)
\; , \ls
\bracket{p}{p'} \ \equiv \ 
\delta^D (p - p')
\; , \\
\bracket{x}{p} \ \equiv \ 
\frac{1}{(2 \pi \hbar)^{D/2}} 
\exp \Big( \frac{i}{\hbar} p \cdot x \Big) \gm
\; , 
\end{gather}
where the plane wave is normalized to 
\begin{align}
\int \d^D p \, \exp \Big( \frac{i}{\hbar} p \cdot (x-y) \Big)
\ &= \ 
(2 \pi \hbar)^{D/2} \delta^D (x-y)
\; ,
\end{align}
\esubeq
which appears when we evaluate the transition elements with
infinitesimal short period.
In order to discuss the path integrals for Dirac fermion operators,
let us also introduce a set of coherent states for fermionic
operators in terms of the operator $\wh{\varphi}{}^a$ satisfying 
$\{ \wh{\varphi}{}^a , \wh{\ol{\varphi}}{}^b \} = \delta^{ab}$, and 
a complex Grassmann odd variable $\eta$:
\bsubeq
\begin{align}
\ket{\eta} \ &\equiv \ \e^{\wh{\ol{\varphi}}{}^a \eta^a} \ket{0}
\; , \ls 
\wh{\varphi}{}^a \ket{0} \ = \ 0 
\; , \ls
\wh{\varphi}{}^a \ket{\eta} \ = \ \eta^a \ket{\eta}
\; , \\
\bra{\ol{\eta}} \ &\equiv \ \bra{0} \e^{\ol{\eta}{}^a \wh{\varphi}{}^a} 
\; , \ls
\bra{0} \wh{\ol{\varphi}}{}^a \ = \ 0
\; , \ls
\bra{\ol{\eta}} \wh{\ol{\varphi}}{}^a \ = \ \bra{\ol{\eta}} \ol{\eta}{}^a
\; .
\end{align}
\esubeq
The inner product of these coherent state is given by
$\bracket{\ol{\eta}}{\zeta} = \e^{\ol{\eta}{}^a \zeta^a}$.
In the same analogy as (\ref{complete-sets-boson}), we introduce a
complete set of the Dirac fermion coherent states:
\bsubeq \label{complete-sets-fermion}
\begin{gather}
1 \ = \ 
\int \prod_{a=1}^D \d \ol{\eta}{}^a \d \eta^a \, \ket{\eta} 
\, \e^{ - \ol{\eta}{}^a \eta^a} \bra{\eta}
\; , \\
\prod_{a=1}^D 
\d \ol{\eta}{}^a \ \equiv \ \d \ol{\eta}{}^D \d \ol{\eta}{}^{D-1} \cdots 
\d \ol{\eta}{}^1
\; , \ls
\prod_{a=1}^D \d \eta^a \ \equiv \ \d \eta^1 \d \eta^2 \cdots \d \eta^D
\; .
\end{gather}
\esubeq

Generically we define the following 
matrix element $M(z,y)$ in the quantum mechanics:
\begin{align}
M (z,y) \ &= \ \bra{z} \wh{\cal O} (\wh{x}, \wh{p}) \ket{y}
\; ,
\end{align}
where $\ket{y}$ and $\bra{z}$ are the initial and final state,
respectively. 
Now we are quite interested in the transition element with respect to
the quantum Hamiltonian $\wh{\Scr{H}}$ and a parameter $\beta$:
\begin{align}
T (z, \ol{\eta};y, \zeta; \beta) \ &\equiv \ 
\bra{z, \ol{\eta}} \exp \Big( - \frac{\beta}{\hbar} \wh{\Scr{H}} \Big)
\ket{y, \zeta}
\; .
\end{align}
Next we introduce $N-1$ complete sets of position
eigenstates $x_k$ and of the fermion coherent states $\lambda_k$
into the above transition elements.
At the same time let us 
also insert $N$ complete sets of momentum eigenstates $p_k$ and of another
fermion coherent states $\xi_k$ to yield
\begin{align}
&\bra{z, \ol{\eta}} \exp \Big( - \frac{\beta}{\hbar} \wh{\Scr{H}} \Big)
\ket{y, \zeta}
\nn \\
\ &\ls= \ 
\int \prod_{i=1}^{N-1} \d^D x_i 
\prod_{i'=1}^{N-1} \d \ol{\lambda}{}_{i'} \lambda_{i'} \, 
\e^{- \ol{\lambda}{}_{i'} \lambda_{i'}} 
\prod_{j=1}^{N} \d^D p_j 
\prod_{j'=0}^{N-1} \d \ol{\xi}{}_{j'} \d \xi_{j'} \, 
\e^{- \ol{\xi}{}_{j'} \xi_{j'}}
\nn \\
\ &\LS \ls \times \prod_{k=0}^{N-1}
\bracket{x_{k+1}, \ol{\lambda}{}_{k+1}}{p_k, \xi_k}
\exp \Big( - \frac{\epsilon}{\hbar} \Scr{H}_{\rm W} 
(x_{k+\half}, p_{k+1}; \ol{\xi}{}_k, \tfrac{1}{2} (\xi_k + \lambda_k))
\Big) \bracket{p_k, \ol{\xi}{}_k}{x_k , \lambda_k}
\nn \\
\ &\ls= \ 
\big[ g(z) g(y) \big]^{- \frac{1}{4}} 
\int \prod_{j=1}^N \frac{\d^D p_j}{(2 \pi \hbar)^D}
\prod_{i=1}^{N-1} \d^D x_i
\prod_{j'=0}^{N-1} \d \ol{\xi}{}_{j'} \d \xi_{j'} 
\nn \\
\ &\LS \times
\exp \Bigg(
\ol{\eta} \cdot \xi_{N-1}
+ \frac{\epsilon}{\hbar} \sum_{k=0}^{N-1} 
\Big[
i p_{k+1} \cdot \frac{x_{k+1} - x_k}{\epsilon} 
- \hbar \, \ol{\xi}{}_k \cdot \frac{\xi_k - \xi_{k-1}}{\epsilon} 
- \Scr{H}_W (x_{k+\half}, p_{k+1}; \ol{\xi}{}_k, \xi_{k-\half}) 
\Big]
\Bigg)
\; .
\end{align}
Notice that the subscript $k$ denotes the $k$-th complete set of the
bosonic eigenstates, or the $k$-th complete set of the fermionic coherent states.
We also note that $y = x_0$, $z = x_N$, $\ol{\eta} = \ol{\lambda}_N$,
$\zeta = \lambda_0 = \xi_{-1}$.
We adopt the midpoint rule $x_{k+\half} = \half (x_{k+1} + x_k)$ 
and $\xi_{k-\half} = \half (\xi_k + \xi_{k-1})$.
The factors $\sqrt{g(x_k)}$ compensate exactly the $\gp$
factors from the plane waves in the inner products.
Furthermore, we integrated the arguments $\lambda_k$ and
$\ol{\lambda}_k$ to yield a useful equation
\begin{gather}
\int \d \ol{\lambda}{}_k \d \lambda_k \, 
\e^{- \ol{\lambda}_k \cdot (\lambda_k - \xi_{k-1})} f (\lambda_k)
\ = \ 
f (\xi_{k-1})
\; ,
\end{gather}
where $f (\lambda)$ is an arbitrary function of the fermionic variable $\lambda$.
Notice that $\wh{\Scr{H}}$ is the quantum Hamiltonian in terms 
of quantum operators, while $\Scr{H}_{\rm W}$ is
its Weyl-ordered form.
The translation from the operator to the Weyl-ordered form is given 
in terms of the symmetrized form $\wh{\Scr{H}}_{\rm S}$ by
\begin{align}
\wh{\Scr{H}} \ &= \ \wh{\Scr{H}}_{\rm S} + \text{further terms} 
\ = \ 
\Scr{H}_{\rm W}
\; . \label{def-Weyl-order}
\end{align}
Integrating out the (discretized) momenta and  
taking the continuum limit $N \to \infty$, ${\epsilon}/{\beta} \to
\d \tau$ with $\sum_{k=0}^{N-1} {\epsilon}/{\beta} \to
\int_{-1}^0 \d \tau$, 
we obtain the continuum path integral description in a following form:
\begin{align}
T (z, \ol{\eta}; y, \zeta; \beta)
\ &= \ 
\Bigg( \frac{g(z)}{g(y)} \Bigg)^{\frac{1}{4}}
\frac{1}{(2 \pi \beta \hbar)^{D/2}} \, \e^{\ol{\eta}{}_a \zeta^a}
\VEV{\exp \Big( - \frac{1}{\hbar} S^{({\rm int})} - \frac{1}{\hbar}
  S^{({\rm source})} \Big)}_0
\; .
\end{align}
Note the followings: 
The action $S^{(\rm int)}$ is given in terms of the interaction terms
in the Lagrangian derived from the Legendre transformation of
Weyl-ordered Hamiltonian, which we will explicitly show later.
We introduced the external source of fields contained in the action
$S^{({\rm source})}$ to define their propagators.
The additional factor $\sqrt{g(z)}$ appears due to the expanding the
metric in $S^{({\rm int})}$ at the point $z$ and due to the
integrating out the free kinetic terms of fields (see, for detail,
section {\it 2.1} in \cite{BvN06}).
The symbol $\vev{\cdots}_0$ denotes the contraction of interaction
terms in terms of propagators and setting the external source to zero.
From now on we simply abbreviate 
$\Vev{\e^{- \frac{1}{\hbar} S^{({\rm int})} 
- \frac{1}{\hbar} S^{({\rm source})}}}_0$ as 
$\Vev{\exp (- \frac{1}{\hbar} S^{({\rm int})})}$.

\subsection{Weyl-ordered form of quantum Hamiltonians}

The next task is to study the Weyl-ordered form of the
Hamiltonians $\Scr{H}_H^{\rm W}$ and obtain the actions $S^{({\rm int})}$ 
in the ${\cal N}=2$ and the ${\cal N}=1$ systems, respectively. 
The symmetrized form of the bosonic operators is defined by
\bsubeq \label{symm-boson} 
\begin{align}
\prod_{m,n} N! \big\{ (\wh{p}_m)^{k_m} (\wh{x}^n)^{\ell_n} \big\}_{\rm S}
\ &\equiv \
\prod_{m,n}
\Big( \frac{\del}{\del \alpha^m} \Big)^{k_m}
\Big( \frac{\del}{\del \beta_n} \Big)^{\ell_n}
\big( \alpha^m \wh{p}_m + \beta_n \wh{x}^n \big)^N
\; , \\
N \ &\equiv \ \sum_m k_m + \sum_n \ell_n 
\; .
\end{align}
\esubeq
In the ${\cal N}=2$ complex fermions' case 
we define the following anti-symmetrized
form:
\bsubeq \label{symm-fermion} 
\begin{align}
\prod_{a,b} N! 
\big\{ (\wh{\varphi}{}^a)^{m_a} (\wh{\ol{\varphi}}{}_b)^{n_b} \big\}_{\rm S}
\ &\equiv \
\prod_{a,b}
\Big( \frac{\del}{\del \alpha_a} \Big)^{m_a}
\Big( \frac{\del}{\del \beta^b} \Big)^{n_b}
\big( \alpha_a \wh{\varphi}{}^a + \beta^b \wh{\ol{\varphi}}{}_b \big)^N
\; , \\
N \ &\equiv \ \sum_a m_a + \sum_b n_b 
\; ,
\end{align}
\esubeq
where we perform the left derivative with respect to the Grassmann odd
variables $\alpha_a$ and $\beta^b$.
In the ${\cal N}=1$ real fermions' case, the anti-symmetrized form is defined by
\begin{align}
\big( \psi^{a_1} \cdots \psi^{a_N} \big)_{\rm S}
\ &\equiv \ 
\frac{1}{N!} \prod_i \Big( \frac{\del}{\del \alpha_{a_i}} \Big)
\big( \alpha_{a} \psi^a \big)^N
\; . \label{symm-M-fer}
\end{align}
By using the above rules, we obtain the Weyl-ordered form of
the ${\cal N}=2$ Hamiltonian
\bsubeq \label{Weyl-H2}
\begin{align}
\Scr{H}_H^{\rm W} 
\ &= \ 
\half \Big( g^{mn} \pi^{(-1)}_m \pi^{(-1)}_n \Big)_{\rm S}
+ \frac{\hbar^2}{8} \Big[
g^{mn} \Gamma^p_{0mq} \Gamma^q_{0np} 
+ g^{mn} \, \omega_{mab} \, \omega_n{}^{ab}
\Big]
\nn \\
\ & \ls
- \frac{\hbar^2}{2} R_{pqmn} (\Gamma_0) \, 
(\varphi^m \ol{\varphi}{}^{n} \varphi^p \ol{\varphi}{}^q)_{\rm S}
+ \frac{\hbar^2}{12} H_{mnp} H^{mnp}
+ \frac{\hbar^2}{6} \del_m (H_{npq}) \, 
\Big[ (\ol{\varphi}{}^m \varphi^{npq})_{\rm S}
+ (\varphi^m \ol{\varphi}{}^{npq})_{\rm S} \Big]
\nn \\
\ & \ls
+ \frac{\hbar^2}{8} H_{mnr} H_{pq}{}^r \Big[ 
( \varphi^{mnpq} )_{\rm S} + (\ol{\varphi}{}^{mnpq})_{\rm S}
- 2 (\varphi^{mn} \ol{\varphi}{}^{pq})_{\rm S}
\Big]
\; , \\
\pi_m^{(-1)}
\ &\equiv \ 
\pi_m + \frac{i \hbar}{2} H_{mab} \big( \varphi^{ab} + \ol{\varphi}{}^{ab} \big)
\ = \ 
p_m - i \hbar \, \omega_{mab} \, \varphi^a \ol{\varphi}{}^b
+ \frac{i \hbar}{2} H_{mab} \big( \varphi^{ab} + \ol{\varphi}{}^{ab} \big)
\; ,
\end{align}
\esubeq
and of the ${\cal N}=1$ Hamiltonian
\bsubeq  \label{Weyl-H1}
\begin{align}
\Scr{H}_H^{\rm 1;W}
\ &= \
\half \Big( g^{mn} \wt{\pi}^{(-1)}_m \wt{\pi}^{(-1)}_n \Big)_{\rm S}
+ \frac{\hbar^2}{8} \Big\{ 
g^{mn} \Gamma^p_{0mq} \Gamma^q_{0np} 
+ \half g^{mn} \omega_{-mab} \, \omega_{-n}{}^{ab} \Big\}
\nn \\
\ & \ls
- \frac{\hbar^2}{24} H_{mnp} H^{mnp}
- \frac{\hbar^2}{2} F^{\alpha}_{mn} (\psi^{mn})_{\rm S} 
(\hat{c}^{\dagger} T_{\alpha} \hat{c})
\; , \\
\wt{\pi}_m^{(-1)}
\ &\equiv \ 
p_m - \frac{i \hbar}{2} \omega_{-mab} \, \psi^{ab}
- i \hbar \, A^{\alpha}_m (\hat{c}^{\dagger} T_{\alpha} \hat{c})
\; .
\end{align}
\esubeq

To proceed computations in path integral formalism in the ${\cal N}=1$ system,
we would like to add a second set of ``free''
Majorana fermions in order to simplify the path integral 
in the ${\cal N}=1$ system in the same way as the one in the ${\cal N}=2$ system. 
Denoting the original Majorana fermions $\psi^a$ by $\psi_1^a$, and
the new ones by $\psi_2^a$, and combining them, 
we again construct Dirac fermions $\chi^a$ and
$\ol{\chi}{}^a$ as
\begin{gather}
\chi^a \ = \ 
\frac{1}{\sqrt{2}} \big( \psi_1^a + i \psi_2^a \big)
\; , \ls
\ol{\chi}{}^a \ = \ 
\frac{1}{\sqrt{2}} \big( \psi_1^a - i \psi_2^a \big)
\; . \label{M2D}
\end{gather}
Notice that, in this context, 
$\psi_2^a$ differs from the second component of
the previously defined Dirac fermions $\varphi^a$ because now
$\psi_2^a$ is introduced as a ``free'' fermion in the ${\cal N}=1$
Hamiltonian.

\subsection{Explicit form of the transition element in ${\cal N}=2$ system} 

We are ready to discuss the explicit form of the transition element 
in the ${\cal N}=2$ system in the framework of the Lagrangian formalism.
Let us first decompose the bosonic and fermionic variables
into two parts, i.e., the background fields and quantum
fluctuations in such a way as
$x^m (\tau) = x_{\rm bg}^m (\tau) + q^m (\tau)$ and 
$\xi^a (\tau) = \xi_{\rm bg}^a (\tau) + \xi_{\rm qu}^a (\tau)$, respectively.
These background fields follow the free equations of motion whose 
solutions are
\begin{align}
x_{\rm bg}^m (\tau) 
\ &= \ 
z^m + \tau ( z^m - y^m ) 
\; , \ls
\xi_{\rm bg}^a (\tau) \ = \ \zeta^a 
\; , \ls 
\ol{\xi}{}_{\rm bg}^a (\tau) \ = \ \ol{\eta}{}^a 
\; , \label{decomp-N2}
\end{align}
with constraints (via the mean-value theorem)
\bsubeq
\begin{gather}
q^m (-1) \ = \ q^m (0) \ = \ 0
\; , \ls
\int_{-1}^0 \d \tau \, q^m (\tau) \ = \ 0
\; , \label{mvt} \\
\xi_{\rm qu}^a (-1) \ = \
\ol{\xi}{}_{\rm qu}^a (0) \ = \ 0
\; . \label{bound-D-fer}
\end{gather}
\esubeq
Then the description of the transition element in the
configuration space path integral is given in the following form
(see eq.({\it 2.81}) in \cite{BvN06}):
\bsubeq \label{transition-N2}
\begin{align}
&\bra{z,\ol{\eta}} \exp \Big( - \frac{\beta}{\hbar} \wh{\Scr{H}}_H
\Big) \ket{y, \zeta}
\ = \ 
\Bigg( \frac{g (z)}{g (y)} \Bigg)^{\frac{1}{4}} 
\frac{1}{(2 \pi \beta \hbar)^{D/2}} \, \e^{\ol{\eta}{}_a \zeta^a} 
\VEV{\exp \Big( - \frac{1}{\hbar} S_H^{(\rm int)} \Big) }
\; , \\
&\ol{\eta}{}_a \zeta^a - \frac{1}{\hbar} S_H^{(\rm int)} 
\ = \
- \frac{1}{\hbar} \big( S_H - S^{(0)} \big) 
\; , \\
- \frac{1}{\hbar} S_H
\ &= \
- \frac{1}{\beta \hbar} \int_{-1}^0 \d \tau
\, \half g_{mn} (x) \Bigg(
\frac{\d x^m}{\d \tau} \frac{\d x^n}{\d \tau} 
+ b^m c^n
+ a^m a^n
\Bigg)
+ \ol{\eta}{}_a \zeta^a
- \int_{-1}^0 \d \tau \, \delta_{ab} \, \ol{\xi}{}_{\rm qu}^a
\frac{\d}{\d \tau} \xi_{\rm qu}^b
\nn \\
\ & \ls 
- \int_{-1}^0 
\d \tau \, \frac{\d x^m}{\d \tau} \Bigg( \omega_{mab} (x)
\ol{\xi}{}^a \xi^b
- \half H_{mab} (x) (\xi^{ab} + \ol{\xi}{}^{ab}) \Bigg)
\nn \\
\ & \ls 
+ \frac{\beta \hbar}{2} \int_{-1}^0 \d \tau \, R_{cdab} (\omega (x)) \, 
\xi^a \ol{\xi}{}^b \xi^c \ol{\xi}{}^d
- \frac{\beta \hbar}{8} \int_{-1}^0 \d \tau \, H_{abe} (x) H_{cd}{}^e (x)
\Big( \xi^{abcd} + \ol{\xi}{}^{abcd} - 2 \xi^{ab} \ol{\xi}{}^{cd} \Big)
\nn \\
\ & \ls 
- \frac{\beta \hbar}{6} \int_{-1}^0 \d \tau \, \del_m (H_{npq} (x)) 
\Big( \ol{\xi}{}^m \xi^{npq} + \xi^m \ol{\xi}{}^{npq} \Big)
\nn \\
\ & \ls 
- \frac{\beta \hbar}{8} \int_{-1}^0 \d \tau \, 
{\cal G}_2 (x)
\; , \label{S-origin-N2} \\
{\cal G}_2 (x)
\ &\equiv \ 
g^{mn} (x) \Big\{ \Gamma^p_{0mq} (x) \Gamma^q_{0np} (x) 
+ \omega_{mab} (x) \omega_n{}^{ab} (x) \Big\}
+ \frac{2}{3} H_{mnp} (x) H^{mnp} (x)
\; , \label{F-N2} \\
- \frac{1}{\hbar} S^{(0)}  
\ &= \
- \frac{1}{\beta \hbar} \int_{-1}^0 \d \tau
\, \half g_{mn} (z) \Bigg(
\frac{\d q^m}{\d \tau} \frac{\d q^n}{\d \tau} 
+ b^m c^n
+ a^m a^n
\Bigg)
- \int_{-1}^0 \d \tau \, \delta_{ab} \, \ol{\xi}{}_{\rm qu}^a
\frac{\d}{\d \tau} \xi_{\rm qu}^b
\; .
\end{align}
\esubeq
Note that we introduced anti-commuting ghost fields $b^m$, $c^m$ and a
commuting ghost field $a^m$ associated with the integrating out of
momentum variables. They also appear in the ${\cal N}=1$ system.
We should notice that the metric in $S^{(0)}$ is given at the
point $z$, not at the
intermediate point $x$, while the metric, spin connection, and the
fluxes in $S_H$ are given at the intermediate point $x$.
We can also define the propagators in this system:
\bsubeq \label{propagators-boson}
\begin{align}
\Vev{q^m (\sigma) q^n (\tau)}
\ &= \ 
- \beta \hbar \, g^{mn} (z) \, \Delta (\sigma , \tau)
\; , \\
\Vev{a^m (\sigma) a^n (\tau)}
\ &= \ 
\beta \hbar \, g^{mn} (z) \, \delta (\sigma - \tau)
\; , \\
\Vev{b^m (\sigma) c^n (\tau)}
\ &= \ 
-2 \beta \hbar \, g^{mn} (z) \, \delta (\sigma - \tau)
\; , \\
\Vev{\xi_{\rm qu}^a (\sigma) \ol{\xi}{}_{\rm qu}^b (\tau)}
\ &= \ 
\delta^{ab} \, \theta (\sigma - \tau)
\; , \\
\Vev{\xi_{\rm qu}^a (\sigma) \xi_{\rm qu}^b (\tau)}
\ &= \ 0 \ = \ 
\Vev{\ol{\xi}{}_{\rm qu}^a (\sigma) \ol{\xi}{}_{\rm qu}^b (\tau)}
\; ,
\end{align}
\esubeq
where the $\delta (\sigma - \tau)$ is the ``Kronecker delta'',
and $-1 \leq \tau , \sigma \leq 0$. 
The definitions of various functions are defined as
$\Delta (\sigma , \tau)
= \sigma (\tau + 1) \theta (\sigma - \tau)
+ \tau (\sigma + 1 ) \theta (\tau - \sigma)
= \Delta (\tau , \sigma)$, 
$\theta (\tau - \tau) = \half$,
$\theta (\tau - \sigma) = - \theta (\sigma - \tau) + 1$, and so forth,
which we list in (\ref{formula-delta}) (see also \cite{BvN06}).


\subsection{Explicit form of the transition element in ${\cal N}=1$ system} 

We can also describe the transition element in the ${\cal N}=1$
supersymmetric quantum system in terms of the dynamical bosonic and
fermionic fields and free Majorana fields
 (see eq.({\it 2.81}) in \cite{BvN06}):
\bsubeq \label{transition-N1}
\begin{align}
&\hspace{-10mm}
\bra{z,\ol{\eta}, \ol{\eta}{}_{\rm gh}} 
\exp \Big( - \frac{\beta}{\hbar} \wh{\Scr{H}}_H^1 \Big) 
\ket{y, \zeta, \zeta_{\rm gh}}
\ = \ 
\Bigg( \frac{g (z)}{g (y)} \Bigg)^{\frac{1}{4}} 
\frac{1}{(2 \pi \beta \hbar)^{D/2}} 
\, \e^{\ol{\eta}{}_a \zeta^a}
\, \e^{\ol{\eta}{}_{\rm gh} \cdot \zeta_{\rm gh}} 
\Vev{ \e^{- \frac{1}{\hbar} S_{1, H}^{(\rm int)}}}
\; , \\ 
&\ol{\eta}{}_a \zeta^a + \ol{\eta}{}_{\rm gh} \cdot \zeta_{\rm gh}
- \frac{1}{\hbar} S_{1, H}^{(\rm int)} 
\ = \
- \frac{1}{\hbar} \big( S_{1, H} - S_1^{(0)} \big) 
\; , \\
- \frac{1}{\hbar} S_{1, H}  
\ &= \
\ol{\eta}{}_a \zeta^a
+ \ol{\eta}{}_{\rm gh} \cdot \zeta_{\rm gh}
\nn \\
\ &\ls
- \frac{1}{\beta \hbar} \int_{-1}^0 \d \tau
\, \half g_{mn} (x) \Bigg(
\frac{\d x^m}{\d \tau} \frac{\d x^n}{\d \tau} 
+ b^m c^n 
+ a^m a^n 
\Bigg)
- \int_{-1}^0 \d \tau \, \Bigg( \delta_{ab} \, \ol{\xi}{}_{\rm qu}^a
\frac{\d}{\d \tau} \xi{}_{\rm qu}^b
+ \hat{c}^{\dagger}_{i, {\rm qu}} \frac{\d}{\d \tau} \hat{c}_{\rm qu}^i
\Bigg)
\nn \\
\ &\ls
- \half \int_{-1}^0 
\d \tau \, \frac{\d x^m}{\d \tau} 
\omega_{-mab} (x) \,
\psi_1^a \psi_1^b
- \int_{-1}^0 \d \tau \, \frac{\d x^m}{\d \tau} 
A^{\alpha}_m (x) \, 
\big( \ol{\xi}{}_{\rm gh} \, T_{\alpha} \, \xi_{\rm gh} \big)
\nn \\
\ &\ls
+ \frac{\beta \hbar}{2} \int_{-1}^0 \d \tau \, F^{\alpha}_{mn} (x) \,
\psi_1^m \psi_1^n \, 
\big( \ol{\xi}{}_{\rm gh} \, T_{\alpha} \, \xi_{\rm gh} \big)
\nn \\
\ &\ls
- \frac{\beta \hbar}{8} \int_{-1}^0 \d \tau \,
{\cal G}_1 (x)
\; , \\
{\cal G}_1 (x)
\ &\equiv \ 
g^{mn} (x) \Big\{ \Gamma^p_{0mq} (x) \Gamma^q_{0np} (x)
+ \half \omega_{-mab} (x) \omega_{-n}{}^{ab} (x) \Big\}
- \frac{1}{3} H_{mnp} (x) H^{mnp} (x) 
\; , \label{F-N1} \\
- \frac{1}{\hbar} S_1^{(0)}  
\ &= \
- \int_{-1}^0 \d \tau \Bigg(
\frac{1}{2 \beta \hbar} g_{mn} (z) \Big\{
\frac{\d q^m}{\d \tau} \frac{\d q^n}{\d \tau} 
+ b^m c^n
+ a^m a^n
\Big\}
+ \Big\{
\delta_{ab} \, \ol{\xi}{}_{\rm qu}^a
\frac{\d}{\d \tau} \xi{}_{\rm qu}^b
+ \hat{c}^{\dagger}_{i, {\rm qu}} \frac{\d}{\d \tau} \hat{c}_{\rm qu}^i
\Big\} \Bigg)
\; .
\end{align}
\esubeq
In the same way as (\ref{decomp-N2}), 
the dynamical fields are decomposed into the background fields
and the quantum fields
\bsubeq
\begin{gather}
\psi_1^a (\tau) \ = \ 
\psi_{1,{\rm bg}}^a (\tau) + \psi_{1,{\rm qu}}^a (\tau)
\; , \ls
\psi_{1,{\rm bg}}^a (\tau)
\ = \ 
\frac{1}{\sqrt{2}} \big( \zeta^a + \ol{\eta}{}^a \big)
\; , \label{def-qu-Ma} \\
\xi_{\rm gh}^i (\tau)
\ = \ 
\zeta_{\rm gh}^i
+ \hat{c}_{\rm qu}^i (\tau)
\; , \ls
\ol{\xi}{}_{i, {\rm gh}} (\tau)
\ = \ 
\ol{\eta}{}_{i, {\rm gh}} 
+ \hat{c}^{\dagger}_{i, {\rm qu}} (\tau)
\; . 
\end{gather}
\esubeq
Notice that the metric in $S_1^{(0)}$ is given at the point $z$, not
at the intermediate point $x$, while the metric, spin connection, and the
fluxes in $S_{1, H}$ are given at the intermediate point $x$.
In the same analogy as the ${\cal N}=2$ system, we introduce the bosonic and
fermionic propagators.
The propagators with respect to the bosonic quantum fields $q^m$ and the
ghost fields $b^m$, $c^m$ and $a^m$ are same as the ones
(\ref{propagators-boson}) in the ${\cal N}=2$ system.
Here we newly introduce the propagators with respect to
the real fermion $\psi_{1,{\rm qu}}^a$ given by the combination with two Dirac
fermions (\ref{def-qu-Ma}). 
Since we have already introduced the propagators with respect to the
Dirac (complex) fermions $\xi_{\rm qu}^a$, we can derive the propagators of
$\psi_{1,{\rm qu}}^a$ in such a way as
\begin{align}
\Vev{\psi_{1,{\rm qu}}^a (\sigma) \psi_{1,{\rm qu}}^b (\tau)}
\ &= \ 
\half \delta^{ab} \Big( \theta (\sigma - \tau) 
- \theta (\tau - \sigma) \Big)
\; . \label{propagators-M-fer} 
\end{align}
The propagator of ghost field $\hat{c}_{\rm gh}^i$ is also given as
\begin{align}
\Vev{\hat{c}^i_{\rm qu} (\sigma) \hat{c}^{\dagger}_{j, {\rm qu}} (\tau)}
\ &= \ 
\delta^i_j \, \theta (\sigma - \tau)
\; .
\end{align}


\section{Witten index in ${\cal N}=1$ quantum mechanics} \label{QM1}

In this section we will discuss the Witten index in the ${\cal N}=1$
quantum mechanical system derived from the path integral formalism.
To obtain this, we will analyze Feynman path integral in terms of
Feynman (dis)connected graphs. Since the form of the Witten index (or
equivalently, the Dirac index)
is same as the one of the chiral anomaly, we refer to the derivation
of the chiral anomaly given in section {\it 6.1} and {\it 6.2} of \cite{BvN06}.

\subsection{Formulation}

As mentioned before, 
by using the identification between the Clifford algebra on the target
geometry and the
anti-commutation relations of fermions in the quantum mechanics,
we can describe the Dirac index equipped with the regulator $\Scr{R}$
in terms of the transition element
of ${\cal N}=1$ quantum mechanics
\begin{align}
{\rm index} \Slash{D} (\hat{\omega})
\ &\equiv \
\lim_{\beta \to 0} \Tr \big\{ \Gamma_{(5)} \e^{- \beta \Scr{R}} \big\}
\ = \
\lim_{\beta \to 0}
\Tr \big\{ (-1)^F \e^{- \frac{\beta}{\hbar} \wh{\Scr{H}}_H^1 } \big\}
\nn \\
\ &= \
\lim_{\beta \to 0} \frac{(-i)^{D/2}}{2^{D/2}} \,
\Tr \prod_{a=1}^D \big( \wh{\varphi}{}^a + \wh{\ol{\varphi}}{}^a
\big) \, \e^{- \frac{\beta}{\hbar} \wh{\Scr{H}}_H^1}
\; .
\end{align}
Note that the chirality operator $\Gamma_{(5)}$ on the target geometry
can be identified with the fermion number operator $(-1)^F$ in the 
${\cal N}=1$ quantum mechanics, i.e.,
the chirality operator is defined as $\Gamma_{(5)} = (-i)^{D/2} \Gamma^1
\Gamma^2 \cdots \Gamma^D$,  
the number operator $(-1)^F$ is replaced in terms of the fermion
operators 
\begin{gather}
\Gamma^a \ \equiv \ \sqrt{2} \psi_{1}^a
\ = \ 
\big( \wh{\varphi}{}^a + \wh{\ol{\varphi}}{}^a \big)
\; , \ls
\Gamma_{(5)} \ \equiv \ 
(-i)^{D/2} \prod_{a=1}^D 
\big( \wh{\varphi}{}^a + \wh{\ol{\varphi}}{}^a \big)
\; .
\end{gather}
Notice that the fermion $\psi_{2}^a$, which is now 
included in the path integral measure while does not appear in the
Hamiltonian, has dimension $2^{D/2}$. 
Then we should divide by $2^{D/2}$ from the formulation $(-i)^{D/2} \prod_{a=1}^D 
( \wh{\varphi}{}^a + \wh{\ol{\varphi}}{}^a)$ by hand.
(See the explanation in section {\it 6.1} in \cite{BvN06} and we will find
that this factor is canceled out via the fermionic measure computation.)
The symbol $\Tr$ in the above expression of the index is defined as
\begin{align}
\Tr \, {\cal O} \ &\equiv \ 
\int \d^D x_0 \sqrt{g (x_0)} \prod_{a=1}^D \big( \d \zeta^a \d
\ol{\zeta}{}_a \big) \, \e^{\ol{\zeta} \zeta} \,
\bra{x_0, \ol{\zeta}} {\cal O} \ket{x_0, \zeta}
\; . 
\end{align}
Then, inserting the complete set of the fermion coherent states
(\ref{complete-sets-fermion}),
we obtain the explicit form of the Dirac index, i.e., the Witten
index with respect
to the ${\cal N}=1$ quantum mechanical path integral: 
\bsubeq \label{chiral-anomaly-PI1}
\begin{align}
{\rm index} \Slash{D} (\hat{\omega})
\ &= \ 
\lim_{\beta \to 0} \frac{(-i)^{D/2}}{2^{D/2}}
\int \d^D x_0 \sqrt{g (x_0)} \prod_{a=1}^D 
\Big( \d \ol{\eta}{}_a \d \eta^a \, \d \zeta^a \d \ol{\zeta}{}_a \Big) 
\, \e^{\ol{\zeta} \zeta} \,
\bra{\ol{\zeta}} \prod_{b=1}^D 
\big( \wh{\varphi}{}^b + \wh{\ol{\varphi}}{}^b \big) \ket{\eta} \, 
\e^{- \ol{\eta} \eta} 
\nn \\
\ &\LS \LS \times 
\bra{x_0 , \ol{\eta}} 
\exp \Big( - \frac{\beta}{\hbar} \wh{\Scr{H}}_H^1 \Big) \ket{x_0, \zeta}
\; .
\end{align}
Here the appearing transition element has already described in the
previous section such as 
\begin{align} 
&\bra{x_0,\ol{\eta}} \exp \Big( - \frac{\beta}{\hbar} \wh{\Scr{H}}_H^1
\Big) \ket{x_0, \zeta}
\ = \ 
\frac{1}{(2 \pi \beta \hbar)^{D/2}} \, \e^{\ol{\eta}{}_a \zeta^a} 
\VEV{\exp \Big(- \frac{1}{\hbar} S_{1, H}^{(\rm int)} \Big) }
\; , \\
- \frac{1}{\hbar} S_{1,H}^{({\rm int})}
\ &= \ 
- \frac{1}{2 \beta \hbar} \int_{-1}^0 \d \tau \,
\Big\{ g_{mn} (x) - g_{mn} (x_0) \Big\}
\Big( \dot{q}^m \dot{q}^n + b^m c^n + a^m a^n \Big)
\nn \\
\ & \ls
- \frac{1}{2} \int_{-1}^0 
\d \tau \, \dot{q}^m \omega_{-mab} (x) \,
\psi_1^{ab}
- \frac{\beta \hbar}{8} \int_{-1}^0 \d \tau \,
{\cal G}_1 (x)
\; ,
\end{align}
\esubeq
where $x = x_0 + q$, $\omega_{-mab} (x) = \omega_{mab} (x) - H_{mab} (x)$
and $\psi_1^a = \psi_{1,{\rm bg}}^a + \psi_{1,{\rm qu}}^a (\tau)$.
The functional ${\cal G}_1 (x)$ is defined in (\ref{F-N1}).
The fermionic terms are summarized as
\bsubeq \label{N1-fer-measure}
\begin{align}
&\int \prod_{a=1}^D 
\Big( 
\d \ol{\eta}{}_a \d \eta^a \d \zeta^a \d \ol{\zeta}{}_a 
\Big)
\, \e^{\ol{\zeta} \zeta - \ol{\eta} \eta} 
\bra{\ol{\zeta}} \prod_{b=1}^D 
\big( \wh{\varphi}{}^b + \wh{\ol{\varphi}}{}^b \big) \ket{\eta}
\ = \ 
\int \prod_{a=1}^D 
\Big( 
\d \ol{\eta}{}_a \d \eta^a \d \zeta^a \d \ol{\zeta}{}_a 
\Big)
\, \e^{\ol{\zeta} \zeta - \ol{\eta} \eta + \ol{\zeta} \eta} 
\prod_{b=1}^D \big( \eta^b + \ol{\zeta}{}^b \big) 
\nn \\
\ &\LS\ls= \ 
\int \prod_{a=1}^D 
\Big( 
\d \ol{\eta}{}_a \d \eta^a \d \zeta^a \d \ol{\zeta}{}_a 
\Big)
\, \e^{\ol{\zeta} \zeta - \ol{\eta} \eta}
\prod_{b=1}^D \big( \eta^b + \ol{\zeta}{}^b \big) 
\; . 
\end{align}
The last factor becomes a fermionic delta function $\delta (\eta + \ol{\zeta})$,
hence $\bracket{\ol{\zeta}}{\eta} = \e^{\ol{\zeta} \eta}$ can be replaced by unity.
For the same reason, we rewrite other exponential factor in such a way
as $\ol{\zeta} \zeta - \ol{\eta} \eta = - \half (\eta - \ol{\zeta})
(\zeta - \ol{\eta})$.
Let us see the measure:
\begin{align}
\prod_{a=1}^D \d \ol{\eta}{}_a \d \eta^a \d \zeta^a \d \ol{\zeta}{}_a 
\ &= \ 
\prod_a \d \ol{\eta}{}_a \d \zeta^a 
\cdot 2^D \d (\ol{\zeta} + \eta)^D \cdots \d (\ol{\zeta} + \eta)^1
\d (\eta - \ol{\zeta})^1 \cdots \d (\eta - \ol{\zeta})^D
\; .
\end{align}
Thus, combining the above two equations, we show
\begin{align}
&\int \prod_a \d \ol{\eta}{}_a \d \zeta^a
\Big[ 2^D \d (\ol{\zeta} + \eta)^D \cdots \d (\ol{\zeta} + \eta)^1
\d (\eta - \ol{\zeta})^1 \cdots \d (\eta - \ol{\zeta})^D \Big]
\, \e^{- \half (\eta - \ol{\zeta}) (\zeta - \ol{\eta})}
\prod_b \big( \eta^b + \ol{\zeta}{}^b \big)
\nn \\
\ &\LS= \ 
\int \prod_a \d \ol{\eta}{}_a \d \zeta^a
\prod_b \big( \zeta^b - \ol{\eta}{}^b \big)
\; .
\end{align}
This is again the fermionic delta function, which annihilates the
exponential factor $\e^{\ol{\eta} \zeta}$ from the Weyl-ordered Hamiltonian.
We perform this fermionic delta function to the transition element.
Generically we consider the following equation in the ${\cal N}=1$
system:
\begin{align}
\int \prod_a \d \ol{\eta}{}_a \d \zeta^a
\prod_b \big( \zeta^b - \ol{\eta}{}^b \big) \, \e^{\ol{\eta} \zeta}
F \Big( \frac{\zeta + \ol{\eta}}{\sqrt{2}} \Big)
\ &= \ 
2^{D/2} \int \prod_a \d \psi_{1,{\rm bg}}^a \, F(\psi_{1,{\rm bg}}^a) 
\; . 
\end{align}
\esubeq
The factor $2^{D/2}$ cancels the factor $2^{-D/2}$ in 
(\ref{chiral-anomaly-PI1}), which we
introduced caused by the free fermion $\psi_2^a$.
Next, rescaling the fermions $\psi_1^a$ by a factor $(\beta
\hbar)^{-\half}$ as $\psi_1^a \to (\beta \hbar)^{- \half} \psi_1^a$, 
we remove the $\beta \hbar$ dependence in the path integral measure. 
Here we show the Witten index in the path integral formalism:
\bsubeq \label{chiral-anomaly-PI1'}
\begin{align}
{\rm index} \Slash{D} (\hat{\omega})
\ &= \ 
\lim_{\beta \to 0} \frac{(-i)^{D/2}}{(2 \pi)^{D/2}}
\int \d^D x_0 \sqrt{g(x_0)} 
\prod_{a=1}^D \d \psi_{1,{\rm bg}}^a 
\VEV{\exp \Big( - \frac{1}{\hbar} S_{1,H}^{({\rm int})} \Big)}
\; , \label{chiral-anomaly-PI1'-1} \\
- \frac{1}{\hbar} S_{1, H}^{(\rm int)}  
\ &= \
- \frac{1}{\beta \hbar} \int_{-1}^0 \d \tau
\, \half \Big\{ g_{mn} (x) - g_{mn} (x_0) \Big\}
\Big(
\dot{q}^m \dot{q}^n
+ b^m c^n
+ a^m a^n
\Big)
\nn \\
\ & \ls
- \frac{1}{2 \beta \hbar} 
\int_{-1}^0 \d \tau \, \dot{q}^m
\omega_{-mab} (x) 
\big( \psi_{1,{\rm bg}} + \psi_{1,{\rm qu}} \big)^{ab} 
- \frac{\beta \hbar}{8} \int_{-1}^0 \d \tau \,
{\cal G}_1 (x)
\; , \label{chiral-anomaly-PI1'-2}
\end{align}
\esubeq
where $x = x_0 + q$.
In addition, 
all the bosonic and fermionic propagators are proportional to $\beta \hbar$:
\bsubeq \label{prop-N1}
\begin{align}
\Vev{q^m (\sigma) q^n (\tau)}
\ &= \ 
- \beta \hbar \, g^{mn} (x_0) \, \Delta (\sigma , \tau)
\; , \\
\Vev{q^m (\sigma) \dot{q}{}^n (\tau)}
\ &= \ 
- \beta \hbar \, g^{mn} (x_0) \, \Big( \sigma + \theta (\tau - \sigma) \Big)
\; , \\
\Vev{\dot{q}{}^m (\sigma) \dot{q}{}^n (\tau)}
\ &= \ 
- \beta \hbar \, g^{mn} (x_0) \, \Big( 1 - \delta (\tau - \sigma) \Big)
\; , \\
\Vev{a^m (\sigma) a^n (\tau)}
\ &= \ 
\beta \hbar \, g^{mn} (x_0) \, \delta (\sigma - \tau)
\; , \\
\Vev{b^m (\sigma) c^n (\tau)}
\ &= \ 
-2 \beta \hbar \, g^{mn} (x_0) \, \delta (\sigma - \tau)
\; , \\
\Vev{\psi_{1,{\rm qu}}^a (\sigma) \psi_{1,{\rm qu}}^b (\tau)}
\ &= \ 
\half \beta \hbar \, \delta^{ab} \Big( \theta (\sigma - \tau) 
- \theta (\tau - \sigma) \Big)
\; .
\end{align}
\esubeq
The properties of these functions are seen in (\ref{formula-delta}).
In the end of the evaluation of the path integral, 
we should take a limit $\beta \to 0$.
There are a number of comments to verify the path integral:
\begin{itemize}
\item {Disconnected} graphs should
    contribute to the functional integrals, called the Feynman
    amplitudes \cite{PW99, BvN06}. 

\item {Graphs} of higher order
    in $\beta \hbar$ do not contribute to Feynman amplitudes in the
    vanishing limit $\beta \to 0$.

\item {Terms} linear in the quantum fields
    $\dot{q}^m$ do not contribute because of the periodic boundary condition
    $q^m (-1) = q^m (0) =0$. 

\item {Terms} linear in the quantum fields $q^m$ do not
  contribute because of the periodic boundary condition and the
  mean-value theorem (\ref{mvt}),
while the terms linear in $\psi_{1,{\rm qu}}^a$
    contribute because there are no restrictions on the quantum
    fermion fields except for 
$\xi_{\rm qu}^a (-1) = \ol{\xi}{}_{\rm qu}^a (0) = 0$. 

\item {We} could, for convenience, choose a
    frame with $\del_m g_{pq} (x_0) = 0$, called the Riemann normal
    coordinate frame. 
Due to this we find $\del_m
    e_n{}^a = \del_m E_a{}^n = 0$, $\Gamma^p_{0nq} (x_0) = 0$ and
    $\omega_{mab} (x_0) = 0$. 
Notice, however that
    $\del_p \del_q e_m{}^a (x_0) \neq 0$,
$\del_m \omega_{nab} (x_0) \neq 0$ and so forth.

\item {The} torsion given by the NS-NS
    flux $H_{mnp}$ (or, in mathematically equivalent form, the Bismut
    torsion $T^{({\rm B})}$) is also expanded in the Riemann normal
    coordinate frame around $x_0$.

\item {The} Feynman amplitudes
    should be independent of the target space metric, at least
    invariant under the rescale of the metric.
\end{itemize}
The torsion is given by the NS-NS three-form flux $H_{mnp}$, which is
represented in terms of the Bismut torsion $T^{({\rm B})}$ in the
supergravity \cite{KY0605}:
\begin{align}
H_{mnp} (x) \ &= \ 
\frac{3}{2} J_m{}^q J_n{}^r J_p{}^s \nabla_{[q} J_{rs]} (x)
\ = \ 
\frac{3}{2} J_m{}^q J_n{}^r J_p{}^s \del_{[q} J_{rs]} (x)
\; . \label{H-TB}
\end{align}
As mentioned in the above comment, we will take the
Riemann normal coordinate frame at the point $x_0$.
At this point we can set the flat metric at the lowest order
approximation in the following way:
\begin{align}
g_{mn} (x_0)
\ &= \ 
\delta_{mn} \; , \ls
\del_p g_{mn} (x_0) \ = \ 0
\; , \ls
\del_p \del_q g_{mn} (x_0) \ \neq \ 0
\; . \label{g-x0}
\end{align}
Due to (\ref{H-TB}), and since the complex structure is proportional
to the metric, the flux (or the torsion) should be also expanded
around the point $x_0$ with the values
\begin{align}
H_{mnp} (x_0) 
\ &= \ 
\frac{3}{2} J_m{}^q J_n{}^r J_p{}^s \del_{[q} J_{rs]} (x_0)
\ = \ 0
\; , \ls
\del_q H_{mnp} (x_0) 
\ \neq \ 0
\; . \label{H-x0}
\end{align}
By using this, the evaluation of the path integral becomes much simpler.

Note that we rewrite the derivative of the spin connection in such a way as
\begin{align}
\del_n \omega_{-mab} (x_0) 
\int_{-1}^0 \d \tau \, \dot{q}^m q^n 
\ &= \ 
- \half \Big( \del_m \omega_{-nab} (x_0) - \del_n \omega_{-mab} (x_0) \Big) 
\int_{-1}^0 \d \tau \, \dot{q}^m q^n
\nn \\
\ &= \ 
\half R_{abmn} (\omega_- (x_0) )
\int_{-1}^0 \d \tau \, q^m \dot{q}^n
\nn \\
\ &= \ 
\half R_{mnab} (\omega_+ (x_0) )
\int_{-1}^0 \d \tau \, q^m \dot{q}^n
\; , \label{omega-R+}
\end{align}
where we used the symmetricity on a Riemann tensor with torsion 
$R_{pqmn} (\omega_-) = R_{mnpq} (\omega_+) - (\d H)_{pqmn}$ and
the periodicity of the bosonic quantum fields $q^m (0) = q^m (-1)$.
Furthermore we also generalized the derivative to the covariant derivative 
because now we analyze on a point $x_0$ 
on which the torsion free connections vanish: 
$\Gamma^p_{0mn} (x_0) = \omega_{mab} (x_0) = H_{mab} (x_0) = 0$.

Let us evaluate the functional integral in terms of the bosonic
propagators (\ref{prop-N1}) at the point $x_0$.
The exponent $\vev{\exp (- \frac{1}{\hbar} S_{1, H}^{({\rm int})})}$ contains both
connected and disconnected Feynman graphs.
First we analyze connected graphs, then we summarize them to obtain
the products of connected graphs.
Let us introduce the effective action $W_H$ by 
$\e^{- \frac{1}{\hbar} W_H} = 
\vev{ \exp( - \frac{1}{\hbar} S_{1, H}^{({\rm int})})}$, which is
expanded as
\begin{align}
- \frac{1}{\hbar} W_H 
\ &= \ 
\sum_{k=1}^{\infty} \frac{1}{k!} 
\cVEV{ \Big( - \frac{1}{\hbar} S_{1, H}^{({\rm int})} \Big)^k }
\; , \label{connected-sum}
\end{align}
where $\cvev{\cdots}$ indicates the value given only by the
connected Feynman graphs. 

For later discussions, it is also 
worth mentioning that the volume form and the Riemann curvature two-form are given
in terms of the vielbein one-form $e^a = e_m{}^a \d x^m$ in the
following way:
\begin{align}
\d^{2n} x_0 \sqrt{g(x_0)} \, {\cal E}^{b_1 \cdots b_{2n}}
\ &= \ 
e^{b_1} \w \cdots \w e^{b_{2n}}
\; . \label{vol-R-form}
\end{align}
Furthermore, we also find the following formula: 
\begin{align}
\int \prod_{a=1}^D \d \psi_{1,{\rm bg}}^a \, 
\psi_{1,{\rm bg}}^{a_1 \cdots a_D}
\ &= \ 
(-)^{D/2} \, {\cal E}^{a_1 a_2 \cdots a_D}
\; . \label{int-M-fer}
\end{align}
The trace of the odd number of the curvature two-form vanishes
because the permutation of the two-form is symmetric but the flip of
the indices is anti-symmetric
$\tr ( R^{2k-1} ) = 0$.


\subsection{Pontrjagin classes}

\subsubsection{Riemannian manifold}

In this case $S_1^{({\rm int})}$ becomes much simpler than
(\ref{chiral-anomaly-PI1'-1})
because there are no terms from $H$-flux.
The spin connection $\omega_-$ is also reduced to $\omega$.
We also easily find that the terms equipped with higher derivatives carrying more
than three bosonic quantum fields $q^m$ always generate higher-loops Feynman
graphs because of the absence of the tadpole graphs.
Furthermore, the terms of order in $\beta \hbar$ do not
contribute to the final result.
Then we truncate $S_1^{({\rm int})}$ in the following way:
\begin{align}
- \frac{1}{\hbar} S_1^{({\rm int})}
\ &= \ 
- \frac{1}{2 \beta \hbar} 
R_{mn} (\omega (x_0)) 
\int_{-1}^0 \d \tau \, q^m \dot{q}^n
\; , \ls
R_{mn} 
\ \equiv \ 
\half R_{mnab} (\omega (x_0)) 
\, \psi_{1,{\rm bg}}^a \psi_{1,{\rm bg}}^b
\; , \label{S1-int-riemann} 
\end{align}
where we used (\ref{omega-R+}) with $H = \d H = 0$. 
Then, the path integral form of the Witten index without
$H$-flux is reduced to
\begin{align}
{\rm index} \Slash{D} (\omega)
\ &= \ 
\lim_{\beta \to 0} \frac{(-i)^{D/2}}{(2 \pi)^{D/2}}
\int \d^D x_0 \sqrt{g(x_0)} 
\prod_{a=1}^D \d \psi_{1,{\rm bg}}^a 
\VEV{ \exp \Big( 
- \frac{1}{2 \beta \hbar} 
R_{mn} 
\int_{-1}^0 \d \tau \, q^m \dot{q}^n
\Big) }
\; . \label{chiral-R}
\end{align}
Let us first evaluate the sum of connected graphs:
\begin{align}
- \frac{1}{\hbar} W 
\ &= \ 
\log \VEV{ \exp \Big( 
- \frac{1}{2 \beta \hbar} 
R_{mn} \int_{-1}^0 \d \tau \, q^m \dot{q}^n
\Big) }
\nn \\
\ &= \ 
\sum_{k=1}^{\infty} \frac{1}{k!} \Big( - \frac{1}{2 \beta \hbar} \Big)^k
R_{m_1 n_1} \cdots R_{m_k n_k}
\int_{-1}^0 \d \tau_1 \cdots \d \tau_k 
\cVEV{\big( q^{m_1} \dot{q}^{n_1} \big) (\tau_1) \cdots
\big( q^{m_k} \dot{q}^{n_k} \big) (\tau_k) }
\; . \label{R-W1}
\end{align}
Since the two indices in the Riemann tensors are anti-symmetric whereas
the propagators are symmetric with respect to the exchanging of
bosonic quantum fields, we easily find that the contraction at the
same ``time'' $\tau_i$ yields a vanishing amplitude.
We also know that 
the partial integration is allowed since $q^m (\tau_i) = 0$ at the end points.
Then, there are $(k-1)!$ ways to contract $k$ vertices and the
symmetry of each vertex in both $q$ yields a factor $2^{k-1}$.
Then we find that the effective action (\ref{R-W1}) is
described as
\bsubeq \label{eff-Pont-Riemann}
\begin{align}
- \frac{1}{\hbar} W 
\ &= \ 
\sum_{k=1}^{\infty} \frac{1}{k!} \Big( - \frac{1}{2 \beta \hbar} \Big)^k
(k-1)! \, 2^{k-1} \, \big( - \beta \hbar \big)^k \cdot
\, R_{m_1 n_1} R_{m_2 n_2} \cdots R_{m_k n_k} 
\, g^{n_1 m_2} g^{n_2 m_3} \cdots g^{n_k m_1}
\nn \\
\ &\LS \times
\int_{-1}^0 \d \tau_1 \cdots \d \tau_k \,
\del_{\tau_1} \Delta (\tau_1, \tau_2)
\del_{\tau_2} \Delta (\tau_2, \tau_3)
\cdots 
\del_{\tau_{k-1}} \Delta (\tau_{k-1}, \tau_{k})
\del_{\tau_k} \Delta (\tau_k, \tau_1)
\nn \\
\ &\equiv \ 
\half \sum_{k=2}^{\infty} \frac{1}{k} 
\tr ( R^k ) I_k
\; , \\
I_k \ &\equiv \ 
\int_{-1}^0 \d \tau_1 \cdots \d \tau_k
\big[ \tau_2 + \theta (\tau_1 - \tau_2) \big]
\big[ \tau_3 + \theta (\tau_2 - \tau_3) \big]
\cdots 
\big[ \tau_1 + \theta (\tau_k - \tau_1) \big]
\; ,
\end{align}
\esubeq
where we used $\tr R^1 = 0$.
By using the formula (see appendix {\it A.4} in \cite{dBPSN9509})
\begin{align}
\sum_{k=2}^{\infty} \frac{y^k}{k} I_k
\ &= \ 
\log \frac{y/2}{\sinh (y/2)}
\ = \ 
- \frac{1}{3!} \Big( \frac{y}{2} \Big)^2 + \cdots
\; ,
\end{align}
we summarize the form of the effective action
\begin{align}
- \frac{1}{\hbar} W 
\ &= \ 
\half \tr \log \Big( \frac{R/2}{\sinh (R/2)} \Big)
\; .
\end{align}
Furthermore, in order to remove the overall factor in front of the
path integral (\ref{chiral-anomaly-PI1}), 
we rescale the background fermions 
$\psi_{1,{\rm bg}}^a \to
\sqrt{\frac{-i}{2 \pi}} \psi_{1,{\rm bg}}^a$.
Then we obtain 
the path integral form of the Witten index in such a way as
\bsubeq \label{chiral-anomaly-R-final}
\begin{gather}
{\rm index} \Slash{D} (\omega) 
\ = \ 
\int \d^D x_0 \sqrt{g(x_0)}
\prod_{a=1}^D \d \psi_{1,{\rm bg}}^a \, 
\exp \Bigg[ \half \tr \log \Bigg( \frac{-i R/4 \pi}{\sinh (-iR/4\pi)} \Bigg) \Bigg]
\; , \\
\tr (R^k) 
\ = \ 
R_{m_1 n_1} R_{m_2 n_2} \cdots R_{m_k n_k}
\, g^{n_1 m_2} g^{n_2 m_3} \cdots g^{n_k m_1} 
\; .
\end{gather}
\esubeq
Due to the property of $\tr (R^k)$, this value becomes zero when $D = 4k+2$. 
Let us simplify the formula (\ref{chiral-anomaly-R-final}) 
by integrating the background fermion 
$\psi_{1,{\rm bg}}^a$ of (\ref{chiral-anomaly-R-final}) 
with noticing the formulae (\ref{vol-R-form}) (in particular, eq.(\ref{int-M-fer})):
\begin{gather}
{\rm index} \Slash{D} (\omega)
\ = \ 
\int_{\cal M} \exp \Bigg[ \half \tr \log 
\Bigg( \frac{i R/4 \pi}{\sinh (iR/4\pi)} \Bigg) \Bigg]
\; , \ls
R_{mn} \ = \ 
\half R_{mnab} (\omega) \, e^a \w e^b
\; . \label{D-index-R}
\end{gather}
This is the well-known form of Dirac index on the Riemannian manifold
${\cal M}$.
The integrand is called the (Dirac) $\hat{\cal A}$-genus.


\subsubsection{Torsional manifold}

This case is still simple. 
Since there does not exist an interaction term with single quantum fermion,
all the Feynman amplitudes are of order in $(\beta \hbar)^k$, 
where $k$ is a non-negative integer.
Thus, since we are interested only in the amplitudes of order in
$(\beta \hbar)^0$ which remain in the vanishing limit $\beta \to 0$,
we can neglect the last term in (\ref{chiral-anomaly-PI1'-2})
which yields graphs of higher order in $\beta \hbar$.
We can also neglect the interaction terms including more than three
quantum fields which yield more than two-loops graphs.
Thus we truncate $S_{1,H}^{({\rm int})}$ carrying only two
bosonic and fermionic quantum fields to
\bsubeq
\begin{align}
- \frac{1}{\hbar} S_{1,H}^{(\rm int)}
\ &= \
- \frac{1}{2 \beta \hbar} R^{(+)}_{mn} \int_{-1}^0 \d \tau \,
q^m \dot{q}^n
\; , \label{S1-int-dH0}
\end{align} 
where we used (\ref{omega-R+}) with $\d H = 0$ and
\begin{gather}
R^{(+)}_{mn}
\ \equiv \ 
\half 
R_{mnab} (\omega_+ (x_0)) 
\psi_{1,{\rm bg}}^a \psi_{1,{\rm bg}}^b
\; . \label{R+-2HH} 
\end{gather}
\esubeq
The effective action, or the functional integral of the connected
graphs are given in terms of (\ref{connected-sum}):
\begin{align}
- \frac{1}{\hbar} W_H
\ &= \ 
\sum_{N=1}^{\infty} \frac{1}{N!} 
\cVEV{
\Big( - \frac{1}{\hbar} S_1^{({\rm int})} \Big)^N
}
\ = \
\sum_{N=1}^{\infty} \frac{1}{N!} 
\Big( - \frac{1}{2 \beta \hbar} \Big)^N
\cVEV{
\Big( R^{(+)}_{mn} \int_{-1}^0 \d \tau \, q^m \dot{q}^n
\Big)^N
}
\nn \\
\ &= \ 
\sum_{N=1}^{\infty} \frac{1}{N!} \Big( - \frac{1}{2 \beta \hbar} \Big)^N
R^{(+)}_{m_1 n_1} \cdots R^{(+)}_{m_N n_N}
\int_{-1}^0 \d \tau_1 \cdots \d \tau_N
\cVEV{
(q^{m_1} \dot{q}^{n_1}) (\tau_1) 
\cdots 
(q^{m_N} \dot{q}^{n_N}) (\tau_N) 
} 
\; ,
\end{align}
where we abbreviated $\psi_{1,{\rm qu}}^a \equiv \psi^a$.
This is exactly same equation as (\ref{R-W1}) except for the Riemann
curvature tensors.
Then, after the rescaling of the background fermion fields, 
the result is given by (\ref{chiral-anomaly-R-final}) in the following way:
\bsubeq \label{chiral-anomaly-RH-final}
\begin{gather}
{\rm index} \Slash{D} (\hat{\omega}) 
\ = \ 
\int \d^D x_0 \sqrt{g(x_0)}
\prod_{a=1}^D \d \psi_{1,{\rm bg}}^a \, \exp
\Bigg[ \half \tr \log \Big( 
\frac{-i R^{(+)}/4 \pi}{\sinh (-i R^{(+)}/4 \pi)}
\Big) \Bigg]
\; , \\
\tr (R_{(+)}^k)
\ = \ R^{(+)}_{m_1 n_1} R^{(+)}_{m_2 n_2} 
\cdots R^{(+)}_{m_k n_k}
\, g^{n_1 m_2} g^{n_2 m_3} \cdots g^{n_k m_1}
\; , \\
R^{(+)}_{mn}
\ \equiv \ 
\half R_{mnab} (\omega_+ (x_0))
\psi_{1,{\rm bg}}^a \psi_{1,{\rm bg}}^b
\; .
\end{gather}
\esubeq
{\bf The most significant point} is that we obtained the same result
which appears in the Mavromatos' work \cite{Mav88, Yajima89, PW99}.
This is exactly same equation as (\ref{R-W1}) except for the Riemann
curvature tensors.
Then, after the rescaling of the background fermion fields, 
the result is given in the following way:
Finally, let us integrate the background fermion $\psi_{1,{\rm bg}}^a$ 
of (\ref{chiral-anomaly-RH-final}) in the same analogy as (\ref{D-index-R}):
\bsubeq \label{D-index-RH}
\begin{gather}
{\rm index} \Slash{D} (\hat{\omega}) 
\ = \ 
\int \exp \Bigg[ \half \tr \log \Big( 
\frac{i R^{(+)}/4 \pi}{\sinh (i R^{(+)}/4 \pi)}
\Big) \Bigg]
\; , \\
\tr (R_{(+)}^k)
\ = \ R^{(+)}_{m_1 n_1} R^{(+)}_{m_2 n_2} 
\cdots R^{(+)}_{m_k n_k}
\, g^{n_1 m_2} g^{n_2 m_3} \cdots g^{n_k m_1}
\; , \\
R^{(+)}_{mn}
\ \equiv \ 
\half R_{mnab} (\omega_+)
e^a \w e^b
\; .
\end{gather}
\esubeq


\section{${\cal N}=1$ quantum mechanics for internal gauge symmetry}
\label{QM-gauge}

In this section we will focus on the gauge field and the invariant
polynomial derived from the path integral. 
The transition element is described in terms of the quantum Hamiltonian in 
(\ref{SUSY-real-A}).
Since the $\hat{c}$-ghost field in (\ref{SUSY-real-A}) are
independent of the other fields, the path integral of this
$\hat{c}$-ghost can be evaluated on a flat geometry and can be 
applied to an arbitrary curved manifold. 
Thus let us first formulate the path integral of this ghost field on a
flat geometry, and we apply this result on the computation on a
generic curved geometry. Here we again follow the convention in \cite{BvN06}.

\subsection{Formulation}

The Dirac index is given by the Witten index in a
following way:
\bsubeq \label{chiral-a-HA}
\begin{gather}
{\rm index} \Slash{D} (\hat{\omega}, A)
\ \equiv \ 
\lim_{\beta \to 0} \Tr' 
\big\{ (-1)^F \e^{- \frac{\beta}{\hbar} \wh{\Scr{H}}_H^1} \big\}
\ = \
\lim_{\beta \to 0} \frac{(-i)^{D/2}}{2^{D/2}} \Tr \prod_{a=1}^D
\big( \wh{\varphi}{}^a + \wh{\ol{\varphi}}{}^a \big) 
P_{\rm gh} \, \e^{- \frac{\beta}{\hbar} \wh{\Scr{H}}_H^1} 
\; , \\
P_{\rm gh}
\ \equiv \ 
: x \, \e^{- x} : 
\; , \ls
x \ \equiv \ \hat{c}^{\dagger}_i \hat{c}^i
\; ,
\end{gather}
\esubeq
where we expressed the trace with prime in order 
to evaluate the trace only over the one-particle ghost sector.
We also introduce the one-particle ghost ``projection operator'' $P_{\rm gh}$ 
instead of the trace with prime. We should also define the
completeness relation of the fermionic states as
\begin{align}
I_{\rm gh} 
\ &\equiv \ 
\int \prod_{i=1}^{\dim R} \d \ol{\eta}{}_{i , {\rm gh}} \d \eta_{\rm gh}^i
 \, \ket{\eta_{\rm gh}} \, \e^{- \ol{\eta}{}_{\rm gh} \cdot \eta_{\rm gh}}
\bra{\ol{\eta}{}_{\rm gh}}
\; , \ls
I_{\rm f} 
\ \equiv \ 
\int \prod_{a=1}^{D} \d \ol{\eta}{}_{a , {\rm f}} \d \eta_{\rm f}^a
 \, \ket{\eta_{\rm f}} \, \e^{- \ol{\eta}{}_{\rm f} \cdot \eta_{\rm f}}
\bra{\ol{\eta}{}_{\rm f}}
\; .
\end{align}
The trace formulae for the ghost and physical fermionic states are
also independently defined by
\begin{align}
\tr_{\rm gh} {\cal O}
\ &\equiv \ 
\int \prod_{i=1}^{\dim R} \d \chi_{\rm gh}^i \d \ol{\chi}{}_{i,{\rm gh}}
\, \e^{\ol{\chi}{}_{\rm gh} \cdot \chi_{\rm gh}}
\bra{\ol{\chi}{}_{\rm gh}} {\cal O} \ket{\chi_{\rm gh}}
\; , \ls
\tr_{\rm f} {\cal O}
\ \equiv \ 
\int \prod_{a=1}^{D} \d \chi_{\rm f}^a \d \ol{\chi}{}_{a,{\rm f}}
\, \e^{\ol{\chi}{}_{\rm f} \cdot \chi_{\rm f}}
\bra{\ol{\chi}{}_{\rm f}} {\cal O} \ket{\chi_{\rm f}}
\; .
\end{align}
In a usual case this trace formula gives the anti-periodic boundary
condition on the fermion. The fermion number operator $(-1)^F$, which
acts on the physical fermion states, flips the condition 
to the periodic boundary condition (see section {\it 2.4} in \cite{BvN06}). 
By using these formulae, we rewrite the Dirac index given by (\ref{chiral-a-HA}):
\begin{align}
{\rm index} \Slash{D} (\hat{\omega}, A)
\ &= \ 
\lim_{\beta \to 0} \frac{(-i)^{D/2}}{2^{D/2}} 
\int \d^D x_0 \sqrt{g(x_0)} 
\nn \\
\ &\LS \times \tr_{\rm f} \, \tr_{\rm gh}
\Bra{x_0, \ol{\chi}{}_{\rm gh}, \ol{\chi}{}_{\rm f}} 
\prod_{a=1}^D \big( \wh{\varphi}{}^a + \wh{\ol{\varphi}}{}^a \big) 
P_{\rm gh} I_{\rm gh} I_{\rm f} \, \e^{- \frac{\beta}{\hbar} \wh{\Scr{H}}_H^1} 
\Ket{x_0, \chi_{\rm f} , \chi_{\rm gh}}
\; .
\end{align}
Of course the ghost Hilbert space and the physical fermion Hilbert
space are independent of each other. Then these completeness relation
act on the individual spaces without any interruption.
Now let us evaluate the trace in the ghost sector:
\begin{align}
&\tr_{\rm gh} \bra{\ol{\chi}{}_{\rm gh}} P_{\rm gh} I_{\rm gh} 
\e^{- \frac{\beta}{\hbar} \wh{\Scr{H}}_H^1} \ket{\chi_{\rm gh}}
\nn \\
\ &\LS= \ 
\int \prod_i \d \chi_{\rm gh}^i \d \ol{\chi}{}_{i,{\rm gh}} \, 
\e^{\ol{\chi}{}_{\rm gh} \cdot \chi_{\rm gh}}
\prod_j \d \ol{\eta}{}_{j,{\rm gh}} \d \eta_{\rm gh}^j \,
\e^{- \ol{\eta}{}_{\rm gh} \cdot \eta_{\rm gh}}
\, \bra{\ol{\chi}{}_{\rm gh}} P_{\rm gh} \ket{\eta_{\rm gh}}
\bra{\ol{\eta}{}_{\rm gh}} 
\e^{- \frac{\beta}{\hbar} \wh{\Scr{H}}_H^1} \ket{\chi_{\rm gh}}
\; .
\end{align}
Since $P_{\rm gh} = :x \e^{-x}:$ projects the ghost coherent state
$\ket{\eta_{\rm gh}}$ onto its one-particle part 
$P_{\rm gh} \ket{\eta_{\rm gh}} = c^{\dagger}_i \eta^i_{\rm gh} \ket{0}$, 
the matrix element of the ghost projection operator $P_{\rm gh}$ is 
easily computed and yields
\begin{align}
\bra{\ol{\chi}_{\rm gh}} P_{\rm gh} \ket{\eta_{\rm gh}}
\ &= \ 
\sum_{i=1}^{\dim R} \ol{\chi}{}_{i,{\rm gh}} \, \eta^i_{\rm gh}
\ = \ 
\ol{\chi}{}_{\rm gh} \cdot \eta_{\rm gh}
\; .
\end{align}
Then we can integrate out the ghost variables $\eta_{\rm gh}^i$ and 
$\ol{\chi}{}_{i,{\rm gh}}$ and define a new kind of projection
operator in the following way:
\begin{align}
\int \prod_i \d \eta^i_{\rm gh} \d \ol{\chi}{}_{i,{\rm gh}} \,
\e^{\ol{\chi}{}_{\rm gh} \cdot \chi_{\rm gh} 
- \ol{\eta}{}_{\rm gh} \cdot \eta_{\rm gh}} \, 
\bra{\ol{\chi}_{\rm gh}} P_{\rm gh} \ket{\eta_{\rm gh}}
\ &= \
\sum_{i=1}^{\dim R} \prod_{\ell \neq i}
\Big( \ol{\eta}{}_{\ell,{\rm gh}} \chi_{\rm gh}^{\ell} \Big)
\ \equiv \ 
P_{\ol{\eta}, \chi}^{\rm gh}
\; . \label{project-eta-chi}
\end{align}
This operator annihilates all terms containing more than two ghost
fields $\ol{\eta}{}_{\rm gh}$ and $\chi_{\rm gh}$.
Because of this we interpret this operator as a kind of 
``projection operator'' onto terms which are linear
in $\ol{\eta}{}_{\rm gh}$ and $\chi_{\rm gh}$, and onto terms
independent of any ghost fields.
  
By using (\ref{N1-fer-measure}), (\ref{project-eta-chi}) and
(\ref{transition-N1}), and rescaling physical fermions as 
$\psi_1^a \to (\beta \hbar)^{- \half} \psi_1^a$, while keeping the
scale of the ghost fields unchanged,
we can evaluate the Dirac index (\ref{chiral-a-HA}):
\bsubeq \label{chiral-a-HA-2}
\begin{align}
{\rm index} \Slash{D} (\hat{\omega}, A)
\ &= \ 
\lim_{\beta \to 0} \frac{(-i)^{D/2}}{(2 \pi)^{D/2}}
\int \d^D x_0 \sqrt{g(x_0)} 
\prod_{i=1}^{\dim R} 
\d \chi_{\rm gh}^i \d \ol{\eta}{}_{i,{\rm gh}} \, P_{\ol{\eta}, \chi}^{\rm gh}
\, \e^{\ol{\eta}{}_{\rm gh} \cdot \chi_{\rm gh}} 
\prod_{a=1}^D \d \psi_{1,{\rm bg}}^a 
\Vev{\e^{ - \frac{1}{\hbar} S_{1,H}^{({\rm int})} }}
\; , \\
- \frac{1}{\hbar} S_{1,H}^{({\rm int})}
\ &= \ 
- \frac{1}{2 \beta \hbar} \int_{-1}^0 \d \tau \,
\Big\{ g_{mn} (x) - g_{mn} (x_0) \Big\}
\Big( \dot{q}^m \dot{q}^n + b^m c^n + a^m a^n \Big)
\nn \\
\ & \ls
- \frac{1}{2 \beta \hbar} \int_{-1}^0 
\d \tau \, \dot{q}^m \omega_{-mab} (x) \,
\psi_1^{ab}
- \frac{\beta \hbar}{8} \int_{-1}^0 \d \tau \,
{\cal G}_1 (x)
\nn \\
\ & \ls
- \int_{-1}^0 \d \tau \, \dot{q}^m 
A^{\alpha}_m (x) \, 
\big( \ol{\xi}{}_{\rm gh} \, T_{\alpha} \, \xi_{\rm gh} \big)
+ \frac{1}{2} \int_{-1}^0 \d \tau \, F^{\alpha}_{ab} (x) \,
\, \psi_1^a \psi_1^b \,  
\big( \ol{\xi}{}_{\rm gh} \, T_{\alpha} \, \xi_{\rm gh} \big)
\; , \label{S1-int-HA} 
\end{align}
\esubeq
with $x = x_0 + q$
and the boundary conditions
$q^m (-1) = q^m (0) = 0$,
$\int_{-1}^0 \d \tau \, q^m (\tau) = 0$, and 
\begin{gather}
\psi_{1,{\rm qu}}^a (-1) \ = \ 
\frac{1}{\sqrt{2}} \ol{\xi}{}_{\rm qu}^a (-1)
\; , \ls
\psi_{1, {\rm qu}}^a (0) \ = \ 
\frac{1}{\sqrt{2}} \xi_{\rm qu}^a (0)
\; , \ls
\hat{c}^i_{\rm qu} (-1) \ = \ 
\hat{c}^{\dagger}_{i,{\rm qu}} (0) \ = \ 0
\; .
\end{gather}
In addition we can rewrite the expansion of gauge field in such a way as
\bsubeq
\begin{align}
\del_n A^{\alpha}_m (x_0) \int_{-1}^0 \d \tau \, \dot{q}^m q^n
\ &= \ 
- \half \Big( \del_m A^{\alpha}_n (x_0) 
- \del_n A^{\alpha}_m (x_0) \Big) \int_{-1}^0 \d \tau \, \dot{q}^m q^n
\nn \\
\ &= \ 
\half \Big[ F^{\alpha}_{mn} (x_0)
- f^{\alpha}{}_{\beta \gamma} \, A^{\beta}_m (x_0) A^{\gamma}_n (x_0) \Big]
\int_{-1}^0 \d \tau \, q^m \dot{q}^n
\nn \\
\ &\equiv \ 
\half {\cal F}^{\alpha}_{mn} (x_0) \int_{-1}^0 \d \tau \, q^m \dot{q}^n
\; , \\
F^{\alpha}_{mn} (x_0)
\ &= \ 
\del_m A^{\alpha}_n (x_0) - \del_n A^{\alpha}_m (x_0)
+ f^{\alpha}{}_{\beta \gamma} \, A^{\beta}_m (x_0) A^{\gamma}_n (x_0)
\; ,
\end{align}
\esubeq
where $F^{\alpha}_{mn}$ is the field strength of the gauge field and
$f^{\alpha}{}_{\beta \gamma}$ is the structure constant of the gauge group.
Notice that the ghost fermions
$\xi_{\rm gh}$ and $\ol{\xi}{}_{\rm gh}$ obey the anti-periodic
boundary condition, while the physical fermions $\xi_{\rm f}$ and
$\ol{\xi}{}_{\rm f}$ follow the periodic boundary condition because of
the insertion of $(-1)^F$. 
This indicates that any closed-loop graphs of the ghost fields
yield zero amplitudes and that only tree graphs contribute to
non-vanishing amplitudes.
Because of this, disconnected graphs with respect to the
$\hat{c}$-ghost amplitudes does not appear in this path integral
transition element. This statement is quite strong.


\subsection{Chern character}


\subsubsection{Chern character on flat geometry without $H$-flux}

Let us first consider the simplest system on a flat geometry with
vanishing flux $H = \d H = 0$. 
In this case there are no (background) interaction terms which carries negative
powers of $\beta \hbar$, contractions of any physical fields $q^m$
and $\psi_{1,{\rm qu}}^a$ 
 become irrelevant under the vanishing limit $\beta \to 0$.
Then we can neglect the term linear in $A_m^{\alpha} (x_0 + q)$ and
the path integral (\ref{chiral-a-HA-2}) is reduced to
\bsubeq \label{chiral-a-HA-2-simplest}
\begin{align}
{\rm index} \Slash{D} (A)
\ &= \ 
\lim_{\beta \to 0} \frac{(-i)^{D/2}}{(2 \pi)^{D/2}}
\int \d^D x_0 
\prod_{i=1}^{\dim R} 
\d \chi_{\rm gh}^i \d \ol{\eta}{}_{i,{\rm gh}} \, P_{\ol{\eta}, \chi}^{\rm gh}
\, \e^{\ol{\eta}{}_{\rm gh} \cdot \chi_{\rm gh}} 
\prod_{a=1}^D \d \psi_{1,{\rm bg}}^a 
\Vev{\e^{ - \frac{1}{\hbar} S_{1,H}^{({\rm int})} }}
\; , \\
- \frac{1}{\hbar} S_{1,H}^{({\rm int})}
\ &= \ 
(F (x_0))^i{}_j 
\int_{-1}^0 \d \tau \, 
\big( \ol{\xi}{}_{\rm gh} + \hat{c}^{\dagger}_{\rm qu} (\tau) \big)_i \,
\big( \xi_{\rm gh} + \hat{c}_{\rm qu} (\tau) \big)^j
\; , \label{S1-int-HA-simplest} 
\end{align}
\esubeq
where $(F (x_0))^i{}_j = \half F_{ab}^{\alpha} (x_0) 
\psi_{1,{\rm bg}}^{ab} (T_{\alpha})^i{}_j$.
As we mentioned before, we only analyze the ghost tree graphs via the
expansion of the above form:
\begin{align}
&\VEV{\exp \Big( F^i{}_j 
\int_{-1}^0 \d \tau \, 
\big( \ol{\xi}{}_{\rm gh} + \hat{c}^{\dagger}_{\rm qu} (\tau) \big)_i \,
\big( \xi_{\rm gh} + \hat{c}_{\rm qu} (\tau) \big)^j \Big) } 
\nn \\
\ &\ls = \ 
1 + \sum_{k=1}^{\infty} \frac{1}{k!} \ol{\eta}{}_{j,{\rm gh}} \, 
\big( F^k \big)^j{}_l 
\, \chi_{\rm gh}^l
\Bigg[ k!
\int_{-1}^0 \d \sigma_1 \cdots \d \sigma_k \,
\theta (\sigma_1 - \sigma_2) \, \theta (\sigma_2 - \sigma_3) \cdots
\theta (\sigma_{k-1} - \sigma_k) 
\Bigg]
\nn \\
\ &\ls = \ 
1 + \sum_{k=1}^{\infty} \frac{1}{k!} \ol{\eta}{}_{j,{\rm gh}} \, 
\big( F^k \big)^j{}_l 
\, \chi_{\rm gh}^l
\; . \label{F-k-gh-tree}
\end{align}
Note that the factor $k!$ in the square bracket in the second line 
is due to the fact that we can order the $k$ vertices into a tree in
$k!$ ways. We also used the following integral:
\begin{align}
\int_{-1}^0 \d \sigma_1 \cdots \d \sigma_k \,
\theta (\sigma_1 - \sigma_2) \, \theta (\sigma_2 - \sigma_3) \cdots
\theta (\sigma_{k-1} - \sigma_k) 
\ &= \ 
\frac{1}{k!}
\; .
\end{align}
Integral of the ghost fields of (\ref{F-k-gh-tree}) 
gives the following simple result:
\begin{align}
&\int \prod_{i=1}^{\dim R} 
\d \chi_{\rm gh}^i \d \ol{\eta}{}_{i,{\rm gh}} \, P_{\ol{\eta}, \chi}^{\rm gh}
\, \e^{\ol{\eta}{}_{\rm gh} \cdot \chi_{\rm gh}} 
\VEV{\exp \Big( F^i{}_j 
\int_{-1}^0 \d \tau \, 
\big( \ol{\xi}{}_{\rm gh} + \hat{c}^{\dagger}_{\rm qu} (\tau) \big)_i \,
\big( \xi_{\rm gh} + \hat{c}_{\rm qu} (\tau) \big)^j \Big) } 
\nn \\
\ &\LS = \
\int \prod_{i=1}^{\dim R} 
\d \chi_{\rm gh}^i \d \ol{\eta}{}_{i,{\rm gh}} \, P_{\ol{\eta}, \chi}^{\rm gh}
\, \e^{\ol{\eta}{}_{\rm gh} \cdot \chi_{\rm gh}} 
\Bigg(
1 + \sum_{k=1}^{\infty} \frac{1}{k!} \ol{\eta}{}_{j,{\rm gh}} \, 
\big( F^k \big)^j{}_l \, \chi_{\rm gh}^l
\Bigg)
\nn \\
\ &\LS = \
\sum_{j=1}^{\dim R} \Bigg[ \delta^j{}_j 
+ \sum_{k=1}^{\infty} \frac{1}{k!} 
\big( F^k \big)^j{}_j
\Bigg]
\ = \ 
\sum_{i=1}^{\dim R} \exp \big( F \big)^i{}_i
\ \equiv \ 
\Tr_R \exp \big( F \big)
\; ,
\end{align}
where the symbol $\Tr_R$ denotes the trace in the $R$ representation of
the gauge group. 
Summarizing the integral and rescaling the background fermion in such
a way as 
$\psi_{1,{\rm bg}}^a \to \sqrt{\frac{-i}{2 \pi}} \psi_{1,{\rm bg}}^a$, 
we obtain 
\begin{align}
{\rm index} \Slash{D} (A)
\ &= \ 
\int \d^D x_0 
\prod_{a=1}^D \d \psi_{1,{\rm bg}}^a 
\Tr_R \exp \Big( - \frac{i}{2 \pi} F \Big)
\; , \ls 
F \ = \
\half F_{ab}^{\alpha} (x_0) \, \psi_{1,{\rm bg}}^a \psi_{1,{\rm bg}}^b 
\, T_{\alpha} 
\; . \label{Chern-simplest}
\end{align}
This is nothing but the Chern character of the gauge
fields $A_m^{\alpha}$.
When we explicitly calculate, we should use the formulae (\ref{vol-R-form}).
In the same way as (\ref{D-index-R}), let us integrate the background
fermions with respect to (\ref{vol-R-form}) and obtain
\begin{align}
{\rm index} \Slash{D} (A)
\ &= \ 
\int_{\cal M} 
\Tr_R \exp \Big( \frac{i}{2 \pi} F \Big)
\; , \ls 
F \ = \
\half F_{ab} \, e^a \w e^b
\ = \ 
\d A + A \w A
\; . \label{D-index-F}
\end{align}


\subsubsection{Torsional manifold}

Let us easily generalize the equation (\ref{Chern-simplest}) to the one on a
curved manifold ${\cal M}$ (in the presence of torsion $H$).
Since the Hilbert spaces of the physical states and the
$\hat{c}$-ghost states are independent of each other, 
the functional integrals of the Dirac index are also performed
independently. Then, combining the functional integral of the physical
field sector (\ref{chiral-anomaly-R-final}) and the functional integral of the
$\hat{c}$-ghost sector (\ref{Chern-simplest}), we obtain the Dirac index 
in the following representation:
\bsubeq \label{D-index-SKT}
\begin{gather}
{\rm index} \Slash{D} (\hat{\omega}, A)
\ = \
\int_{\cal M} 
\exp \Bigg[ \half \tr \log 
\Bigg( \frac{i R^{(+)}/4 \pi}{\sinh (i R^{(+)}/4\pi)} \Bigg) \Bigg]
\Tr_R \exp \Big( \frac{i}{2 \pi} F \Big)
\; , \\
R^{(+)}_{mn} \ = \ 
R_{mnab} (\omega_+) \, e^a \w e^b
\; , \ls
F \ = \ 
\half F_{ab} \, e^a \w e^b
\; .
\end{gather}
\esubeq
The index on a Riemannian manifold without torsion can be easily obtained
when we choose $H = 0$ in this form.


\section{Witten index in ${\cal N}=2$ quantum mechanics} \label{QM-Euler}

In this section let us analyze the Euler characteristics 
on the manifold with torsion $H$. 
In the case of vanishing torsion, we will find a form of the
Gauss-Bonnet theorem.

\subsection{Formulation}

The Euler characteristics $\chi$ on the target space geometry 
can also be expressed in terms of the ${\cal N}=2$ supersymmetric quantum
mechanics (see section {\it 14.3} in \cite{GSW})
\begin{align}
\chi
\ &\equiv \ 
\lim_{\beta \to 0} 
\Tr \big\{ \Gamma_{(5)} \wt{\Gamma}_{(5)} \e^{- \beta \Scr{R}} \big\}
\ = \ 
\lim_{\beta \to 0} 
\Tr \prod_{a=1}^D 
\big( \wh{\varphi}{}^a + \wh{\ol{\varphi}}{}^a \big)
\prod_{b=1}^D \big( \wh{\varphi}{}^b - \wh{\ol{\varphi}}{}^b \big)
\, \e^{- \frac{\beta}{\hbar} \wh{\Scr{H}}}
\; .
\end{align}
The chirality operators $\Gamma_{(5)}$ and $\wt{\Gamma}{}_{(5)}$ are
given in terms of $\Gamma^a = \sqrt{2} \wh{\psi}_{1}^a$
and $\wt{\Gamma}{}^a = \sqrt{2} \wh{\psi}_{2}^a$, respectively:
\bsubeq
\begin{align}
\Gamma_{(5)} \ &\equiv \ 
(-i)^{D/2} \Gamma^1 \cdots \Gamma^D
\ = \ 
(-i)^{D/2} 2^{D/2} \wh{\psi}_{1}^1 \cdots \wh{\psi}_{1}^D
\ = \ 
(-i)^{D/2} \prod_{a=1}^D 
\big( \wh{\varphi}{}^a + \wh{\ol{\varphi}}{}^a \big)
\; , \\
\wt{\Gamma}{}_{(5)} \ &\equiv \ 
(-i)^{D/2} \wt{\Gamma}{}^1 \cdots \wt{\Gamma}{}^D
\ = \ 
(-i)^{D/2} 2^{D/2} \wh{\psi}_{2}^1 \cdots \wh{\psi}_{2}^D
\ = \ 
(-i)^{D/2} (-i)^D \prod_{a=1}^D 
\big( \wh{\varphi}{}^a - \wh{\ol{\varphi}}{}^a \big)
\; .
\end{align}
\esubeq
Notice that since the non-trivial values are given when $D$ is even number, 
we find $(-i)^{2D} = 1$.
Then we formulate the Euler characteristic in terms of the transition
element and effective action (where $x = x_0 + q$):
\bsubeq \label{Euler} 
\begin{align}
\chi \ &= \ 
\lim_{\beta \to 0} 
\int \d^D x_0 \sqrt{g (x_0)} 
\prod_{a=1}^D \Big( \d \ol{\eta}{}_a \d \eta^a \d \zeta^a \d
\ol{\zeta}{}^a \d \ol{\lambda}{}_a \d \lambda^a \Big) \,
\e^{\ol{\zeta} \zeta} \, \e^{- \ol{\lambda} \lambda} \, \e^{- \ol{\eta} \eta} 
\nn \\
\ & \LS  \times
\bra{\ol{\zeta}} \prod_{b=1}^D 
\big( \wh{\varphi}{}^b + \wh{\ol{\varphi}}{}^b \big) \ket{\lambda}
\bra{\ol{\lambda}} \prod_{c=1}^D 
\big( \wh{\varphi}{}^c - \wh{\ol{\varphi}}{}^c \big) \ket{\eta}
 \bra{x_0 , \ol{\eta}} 
\exp \Big( - \frac{\beta}{\hbar} \wh{\Scr{H}}_H \Big) \ket{x_0, \zeta}
\; , \\
&\ls \bra{x_0 , \ol{\eta}} 
\exp \Big( - \frac{\beta}{\hbar} \wh{\Scr{H}}_H \Big) \ket{x_0, \zeta}
\ = \ 
\frac{1}{(2 \pi \beta \hbar)^{D/2}} \e^{\ol{\eta} \zeta} 
\VEV{\exp \Big(- \frac{1}{\hbar} S_H^{({\rm int})} \Big) }
\; , \label{tr-element-N2} \\
- \frac{1}{\hbar} S_H^{({\rm int})}
\ &= \ 
- \frac{1}{\beta \hbar} \int_{-1}^0 \d \tau \, 
\half \Big[ g_{mn} (x) - g_{mn} (x_0) \Big] 
\Big( \dot{q}^m \dot{q}^n + b^m c^n + a^m a^n \Big)
\nn \\
\ & \ls 
- \int_{-1}^0 \d \tau \, 
\dot{q}^m \Bigg( \omega_{mab} (x) \, 
\big( \ol{\eta} + \ol{\xi}{}_{\rm qu} \big)^a 
\big( \zeta + \xi_{\rm qu} \big)^b
- \half H_{mab} (x) 
\Big\{ \big( \zeta + \xi_{\rm qu} \big)^{ab} 
+ \big( \ol{\eta} + \ol{\xi}{}_{\rm qu} \big)^{ab} 
\Big\} \Bigg)
\nn \\
\ & \ls 
+ \frac{\beta \hbar}{2} \int_{-1}^0 \d \tau \, 
R_{cdab} (\omega (x)) \, 
(\zeta + \xi_{\rm qu})^a ( \ol{\eta} + \ol{\xi}{}_{\rm qu})^b
(\zeta + \xi_{\rm qu})^c ( \ol{\eta} + \ol{\xi}{}_{\rm qu})^d
\nn \\
\ & \ls 
- \frac{\beta \hbar}{8} \int_{-1}^0 \d \tau \, 
H_{abe} (x) H_{cd}{}^e (x)
\Big\{ (\zeta + \xi_{\rm qu})^{abcd} + 
(\ol{\eta} + \ol{\xi}{}_{\rm qu})^{abcd} - 2 (\zeta + \xi_{\rm qu})^{ab} 
(\ol{\eta} + \ol{\xi}{}_{\rm qu})^{cd} \Big\}
\nn \\
\ & \ls 
- \frac{\beta \hbar}{6} \int_{-1}^0 \d \tau \, 
\del_m (H_{npq}(x))
\Big\{ (\ol{\eta} + \ol{\xi}{}_{\rm qu})^m 
(\zeta + \xi_{\rm qu})^{npq}
+ (\zeta + \xi_{\rm qu})^m (\ol{\eta} + \ol{\xi}{}_{\rm qu})^{npq}
\Big\}
\nn \\
\ & \ls 
- \frac{\beta \hbar}{8} \int_{-1}^0 \d \tau \, 
{\cal G}_2 (x)
\; , \label{S-int-1}
\end{align}
\esubeq
where the functional ${\cal G}_2 (x)$ is given in (\ref{F-N2}).
Now let us analyze fermionic measure in the form (\ref{Euler}).
The effective action $S^{({\rm int})}$ contains $\xi^a$ and
$\ol{\xi}{}^a$ whose boundaries are $\zeta$ and $\ol{\eta}$,
respectively, and $\eta$, $\ol{\zeta}$ and $\lambda$, $\ol{\lambda}$
do not appear in $S^{({\rm int})}$.
Then let us rewrite the path integral measure with fermions:
\bsubeq
\begin{align}
&\prod_a \d \ol{\eta}{}_a \d \eta^a \, 
\d \zeta^a \d \ol{\zeta}{}_a \, \d \ol{\lambda}{}_a \d \lambda^a 
\ = \
\prod_a \big( \d \ol{\eta}{}_a \d \zeta^a \big)
\big( \d \ol{\lambda}{}_a \d \eta^a \big)
\big( \d \ol{\zeta}{}_a \d \lambda^a \big)
\nn \\
\ &\LS= \ 
\prod_a \big( \d \ol{\eta}{}_a \d \zeta^a \big)
\Big( 2^D \d (\ol{\lambda} + \eta){}_a \d (\eta - \ol{\lambda})^a \Big)
\Big( 2^D \d (\ol{\zeta} + \lambda){}_a \d (\lambda - \ol{\zeta})^a \Big)
\; ,
\end{align}
where we implicitly used the orderings of $\d \ol{\eta}$ and $\d
\zeta$ (\ref{complete-sets-fermion}). 
Under the integral with $\prod_b ( \lambda^b + \ol{\zeta}{}^b ) 
\prod_c ( \eta^b - \ol{\lambda}{}^b )$ which can be regarded as the
fermionic delta functions, we can see $\ol{\zeta} = - \lambda$ and
$\eta = \ol{\lambda}$. 
Then, after a tedious computation, we obtain 
\begin{align}
\int \prod_a \d \ol{\eta}{}_a \d \eta^a \,
\d \zeta^a \d \ol{\zeta}{}_a \, \d \ol{\lambda}{}_a \d \lambda^a
\, \e^{\ol{\zeta} \zeta - \ol{\lambda} \lambda - \ol{\eta} \eta 
+ \ol{\zeta} \lambda + \ol{\lambda} \eta + \ol{\eta} \zeta}
\prod_b \big( \lambda^b + \ol{\zeta}{}^b \big) 
\prod_c \big( \eta^b - \ol{\lambda}{}^b \big) 
\ &= \ 
\int \prod_{a} \d \ol{\eta}{}_a \d \zeta^a 
\; ,
\end{align}
where we used the fermionic delta functions:
\begin{gather}
\int
\prod_a \d (\ol{\zeta} + \eta){}_a \prod_b (\eta^b + \ol{\zeta}{}^b) 
\ = \ 
1
\; , \ls
(-1)^D \int \prod_{a=1}^D \d \ol{\zeta} \, \e^{- \ol{\zeta} (\eta - \zeta)}
\ = \ 
\prod_{a=1}^D \big( \eta^a - \zeta^a \big)
\; .
\end{gather}
\esubeq
Then we rescale the fermion to remove the $\beta \hbar$ dependence on the
measure
in such a way as
\begin{align}
\frac{1}{(\beta \hbar)^{D/2}} \prod_{a=1}^D \d \ol{\eta}{}_a \d \zeta^a
\ \equiv \ 
\prod_{a=1}^D \d \ol{\eta}{}'_a \d \zeta'{}^a
\; , \ls
\xi^a \ \equiv \ ( {\beta \hbar} )^{-\frac{1}{4}} \xi'{}^a
\; .
\end{align}
Then the rescaled $S^{({\rm int})}$ is given by
(where we omit the prime symbol)
\begin{align}
- \frac{1}{\hbar} S^{({\rm int})}
\ &= \ 
- \frac{1}{\beta \hbar} \int_{-1}^0 \d \tau \, 
\half \Big[ g_{mn} (x) - g_{mn} (x_0) \Big] 
\Big( \dot{q}^m \dot{q}^n + b^m c^n + a^m a^n \Big)
\nn \\
\ & \ls 
- \frac{1}{\sqrt{\beta \hbar}} \int_{-1}^0 \d \tau \,
\dot{q}^m \Bigg( \omega_{mab} (x) 
( \ol{\eta} + \ol{\xi}{}_{\rm qu} )^a 
( \zeta + \xi_{\rm qu} )^b
- \half H_{mab} (x) 
\Big\{ ( \zeta + \xi_{\rm qu} )^{ab} 
+ ( \ol{\eta} + \ol{\xi}{}_{\rm qu} )^{ab} 
\Big\} \Bigg)
\nn \\
\ & \ls 
+ \frac{1}{2} \int_{-1}^0 \d \tau \, 
R_{cdab} (\omega (x)) \, 
(\zeta + \xi_{\rm qu})^a ( \ol{\eta} + \ol{\xi}{}_{\rm qu})^b
(\zeta + \xi_{\rm qu})^c ( \ol{\eta} + \ol{\xi}{}_{\rm qu})^d
\nn \\
\ & \ls 
- \frac{1}{8} \int_{-1}^0 \d \tau \, 
H_{abe} (x) H_{cd}{}^e (x)
\Big\{ (\zeta + \xi_{\rm qu})^{abcd} + 
(\ol{\eta} + \ol{\xi}{}_{\rm qu})^{abcd} - 2 (\zeta + \xi_{\rm qu})^{ab} 
(\ol{\eta} + \ol{\xi}{}_{\rm qu})^{cd} \Big\}
\nn \\
\ & \ls 
- \frac{1}{6} \int_{-1}^0 \d \tau \, 
\del_m (H_{npq} (x)) 
\Big\{ (\ol{\eta} + \ol{\xi}{}_{\rm qu})^m 
(\zeta + \xi_{\rm qu})^{npq}
+ (\zeta + \xi_{\rm qu})^m (\ol{\eta} + \ol{\xi}{}_{\rm qu})^{npq}
\Big\}
\nn \\
\ & \ls 
- \frac{\beta \hbar}{8} \int_{-1}^0 \d \tau \, 
{\cal G}_2 (x)
\; . \label{S-int-2}
\end{align}
Notice that the bosonic and fermionic propagators are now proportional to
$\beta \hbar$ and $\sqrt{\beta \hbar}$, respectively (we have also
rescaled the fermion propagator):
\bsubeq \label{propagators-Euler}
\begin{align}
\Vev{q^m (\sigma) q^n (\tau)}
\ &= \ 
- \beta \hbar \, g^{mn} (z) \, \Delta (\sigma , \tau)
\; , \\
\Vev{\xi_{\rm qu}^a (\sigma) \ol{\xi}{}_{\rm qu}^b (\tau)}
\ &= \ 
\sqrt{\beta \hbar} \, \delta^{ab} \, \theta (\sigma - \tau)
\; .
\end{align}
\esubeq
Then we easily find that each contraction among quantum fields yields 
Feynman graphs of higher order in $\beta \hbar$, which goes to zero in
the limit $\beta \hbar \to 0$.
Only the interaction terms given by background fields $x_0^m$,
$\zeta^a$ and $\ol{\eta}{}^a$ are independent
of $\beta$ and they give rise to the relevant Feynman graphs.
Then, we can truncate $S^{({\rm int})}$ in order to obtain the Euler
characteristics on the $D$-dimensional geometry ${\cal M}$ in the
path integral formalism:
\bsubeq \label{PI-Euler}
\begin{align}
\chi ({\cal M}) 
\ &= \ 
\frac{1}{(2 \pi)^{D/2}}
\int_{\cal M} \d^D x_0 \sqrt{g (x_0)} 
\prod_{a=1}^D \d \ol{\eta}{}_a \d \zeta^a \,
\Vev{\e^{- \frac{1}{\hbar} S^{({\rm int})}}}
\; , \\
- \frac{1}{\hbar} S^{({\rm int})}
\ &= \
- \frac{1}{4} R_{abcd} (\omega (x_0)) 
\, \zeta^{ab} \ol{\eta}{}^{cd}
- \frac{1}{6} \del_a (H_{bcd}) (x_0)
\Big( \ol{\eta}{}^a \zeta^{bcd} + \zeta^a \ol{\eta}{}^{bcd} \Big)
\; , \label{S-int-Euler}
\end{align}
\esubeq
where we used $H_{abc} (x_0) = 0$, 
$R_{cdab} (\omega) = R_{abcd} (\omega)$ and the second
Bianchi identity $R_{abcd} (\omega) + R_{acdb} (\omega) + R_{adbc} (\omega) 
= 0$ without torsion:
$R_{abcd} (\omega) \, \zeta^a \ol{\eta}{}^b \zeta^c \ol{\eta}{}^d
=  - \half R_{abcd} (\omega) \, \zeta^{ab} \ol{\eta}{}^{cd}$.
Since there exist only background fields, we do not have to introduce
quantum propagators to contract interaction terms.
The Feynman amplitude of the path integral is given only by the expansion 
of $\exp (- \frac{1}{\hbar} S^{({\rm int})})$ with noticing that 
the number of $\zeta$ should be equal to the number of $\ol{\eta}$ to
saturate the fermionic path integral measure.
Since each term in (\ref{S-int-Euler}) carries even number of
background fermions $\zeta$ and 
$\ol{\eta}$, the path integral with $D = 2 n + 1$ becomes trivial.

Next let us investigate the formulation in various geometries in
diverse dimensions.
We can easily find that the second and the third terms do not
contribute to the Feynman graphs in the case of $D=2$.
This is consistent with the fact there does not exist a totally
antisymmetric torsion in two-dimensional geometry.


\subsection{Euler characteristics}

Next let us investigate the formulation in various geometries in
diverse dimensions.
We can easily find that the second and the third terms do not
contribute to the Feynman graphs in the case of $D=2$.
This is consistent with the fact there does not exist a totally
antisymmetric torsion in two-dimensional geometry.

\subsubsection{Riemannian manifold}

It is worth reviewing the case of the Riemannian manifold without torsion.
The action is given as
\begin{align}
- \frac{1}{\hbar} S^{({\rm int})}
\ &= \ 
- \frac{1}{4} R_{abcd} (\omega) \, \zeta^{ab} \ol{\eta}{}^{cd}
\; , \label{S-int-Euler1}
\end{align}
Then the path integral formulation is described in the following way:
\begin{align}
\chi ({\cal M})
\ &= \ 
\frac{1}{(2 \pi)^{D/2}} \int_{\cal M} \!\! \d^D x_0 \sqrt{g(x_0)} 
\prod_{a=1}^D \d \ol{\eta}{}^a \d \zeta^a \,
\exp \Big( - \frac{1}{4} R_{abcd} (\omega) \, \zeta^{ab} \ol{\eta}{}^{cd} \Big)
\nn \\
\ &= \
\frac{1}{(8 \pi)^{n} n!} \, 
{\cal E}_{a_1 \cdots a_{2n}} {\cal E}^{b_1 \cdots b_{2n}} 
\int_{\cal M} \!\! \d^{2n} x_0 \sqrt{g(x_0)} \,
\Big( R^{a_1 a_2}{}_{b_1 b_2} (\omega) 
\cdots R^{a_{2n-1} a_{2n}}{}_{b_{2n-1} b_{2n}} (\omega) \Big)
\nn \\
\ &= \ 
\frac{1}{(4 \pi)^n n!} \, {\cal E}_{a_1 \cdots a_{2n}} 
\int_{\cal M} \! 
R^{a_1 a_2} (\omega) \w \cdots \w R^{a_{2n-1} a_{2n}} (\omega)
\; , \label{Euler-ch-R}
\end{align}
where we used the formulae in Euclidean space:
\begin{align}
\d^{2n} x_0 \sqrt{g(x_0)} \, {\cal E}^{b_1 \cdots b_{2n}}
\ &= \ 
e^{b_1} \w \cdots \w e^{b_{2n}}
\; , \ls
R^{ab} (\omega) \ = \ 
\half R^{ab}{}_{cd} (\omega) \, e^c \w e^d
\; .
\end{align}
Non-trivial value of $\chi ({\cal M})$ is given only when $D = 2n = 2k$
and all indices of totally antisymmetric tensor ${\cal E}_{abcd\cdots}$ are
the frame (local Lorentz) indices with Euclidean signature. 
Then we do not mind the positions of the
indices\footnote{In the case of curved indices, the positions of
  indices are quite important we should really mind whether $\ve_{mnpq\cdots}$
is a tensor or a tensor density. In the case of frame coordinate
indices, the weight $\sqrt{g(x_0)}$ does not appear.}.
We also used the following formulae in the same way as (\ref{int-M-fer}):
\begin{align}
\int \d \zeta_1 \cdots \d \zeta_{2n} \, \zeta_{1\cdots 2n}
\ &= \ 
(-1)^n
\; , \ls
\int \d \ol{\eta}{}^{2n} \cdots \d \ol{\eta}{}^1 \, 
\ol{\eta}{}^{12\cdots 2n}
\ = \ 
1
\; . \label{fer-contract}
\end{align}


\subsubsection{Torsional manifold}

In this case we should analyze the full action in (\ref{S-int-Euler}):
\begin{align}
- \frac{1}{\hbar} S^{(\rm int)}
\ &= \ 
- \frac{1}{4} R_{abcd} (\omega) 
\, \zeta^{ab} \ol{\eta}{}^{cd}
- \frac{1}{6} \del_a (H_{bcd})
\Big( \ol{\eta}{}^a \zeta^{bcd} + \zeta^a \ol{\eta}{}^{bcd} \Big)
\; .
\end{align}
We omitted the argument $x_0$.
In the same as the analysis on the Riemannian manifold,
we can only investigate the case $D = 2n$, i.e.,
the case of the even-dimensional manifolds.
The expectation value of the exponent is 
\begin{align}
\Vev{\e^{ - \frac{1}{\hbar} S^{(\rm int)}}}
\ &= \ 
\exp \Big(
- \frac{1}{4} R_{abcd} (\omega) 
\, \zeta^{ab} \ol{\eta}{}^{cd}
- \frac{1}{6} \del_a (H_{bcd})
\big( \ol{\eta}{}^a \zeta^{bcd} + \zeta^a \ol{\eta}{}^{bcd} \big)
\Big)
\nn \\
\ &= \
\exp \Big(
- \frac{1}{4} R_{abcd} (\omega) \, \zeta^{ab} \ol{\eta}{}^{cd}
\Big)
\exp \Big(
- \frac{1}{6} \del_a (H_{bcd}) \, \ol{\eta}{}^a \zeta^{bcd} 
\Big)
\exp \Big(
- \frac{1}{6} \del_a (H_{bcd}) \, \zeta^a \ol{\eta}{}^{bcd}
\Big)
\; . \label{Euler-SKT-1}
\end{align}
Since the path integral measure in (\ref{PI-Euler}) requires that the
number of the background fermions $\zeta$ should be equal to the
number of $\ol{\eta}$, the third exponent in (\ref{Euler-SKT-1})
should be contracted only with the second exponent.
The second and the third exponents cannot be contracted with the first
exponent. 
Then (\ref{Euler-SKT-1}) is truncated to
\begin{align}
\Vev{\e^{ - \frac{1}{\hbar} S^{(\rm int)}}}
\ &\sim \ 
\sum_{k + 2 \ell = n}
\frac{1}{k! \ell! \ell!} 
\Big( - \frac{1}{4} R_{abcd} (\omega) \, \zeta^{ab} \ol{\eta}{}^{cd} \Big)^k
\Big( - \frac{1}{6} \del_a (H_{bcd}) \, \ol{\eta}{}^a \zeta^{bcd} \Big)^{\ell}
\Big( - \frac{1}{6} \del_a (H_{bcd}) \, \zeta^a \ol{\eta}{}^{bcd} \Big)^{\ell}
\nn \\
\ &= \ 
\sum_{k + 2 \ell = n}
\frac{2^{2\ell}}{3^{2\ell} k! \ell! \ell!} \Big( - \frac{1}{4} \Big)^n
\Big( R_{a_1 a_2 b_1 b_1} \cdots R_{a_{2k-1} a_{2k} b_{2k-1} b_{2k}} \Big)
\nn \\
\ &\LS
\times \Big(
\del_{c_1} (H_{d_1 d_2 d_3}) \cdots
\del_{c_{\ell}} (H_{d_{3 \ell -2} d_{3 \ell -1} d_{3 \ell}}) 
\Big)
\Big(
\del_{e_1} (H_{f_1 f_2 f_3}) \cdots
\del_{e_{\ell}} (H_{f_{3 \ell -2} f_{3 \ell -1} f_{3\ell}}) 
\Big)
\nn \\
\ &\LS
\times \zeta^{a_1 \cdots a_{2 k} c_1 \cdots c_{\ell} f_1 \cdots f_{3 \ell}}
\ol{\eta}{}^{b_1 \cdots b_{2 k} d_1 \cdots d_{\ell} e_1 \cdots e_{3 \ell}}
\nn \\
\ &= \ 
\sum_{k + 2 \ell = n}
\frac{2^{2\ell}}{3^{2\ell} k! \ell! \ell!} \Big( - \frac{1}{4} \Big)^n
\, {\cal E}^{a_1 \cdots a_{2k} c_1 \cdots c_{\ell} f_1 \cdots f_{3 \ell}}
{\cal E}^{b_1 \cdots b_{2k} d_1 \cdots d_{\ell} e_1 \cdots e_{3 \ell}}
\, \zeta^{1 \cdots 2n} \, \ol{\eta}{}^{1 \cdots 2n}
\nn \\
\ &\LS \times
\Big( R_{a_1 a_2 b_1 b_1} \cdots R_{a_{2k-1} a_{2k} b_{2k-1} b_{2k}} \Big)
\Big(
\del_{c_1} (H_{d_1 d_2 d_3}) \cdots
\del_{c_{\ell}} (H_{d_{3 \ell -2} d_{3 \ell -1} d_{3 \ell}}) 
\Big)
\nn \\
\ &\LS
\times 
\Big(
\del_{e_1} (H_{f_1 f_2 f_3}) \cdots
\del_{e_{\ell}} (H_{f_{3 \ell -2} f_{3 \ell -1} f_{3\ell}}) 
\Big)
\end{align}
Substituting this into (\ref{PI-Euler}), we obtain
\begin{align}
\chi ({\cal M})
\ &= \ 
\frac{1}{(8\pi)^n} \sum_{k + 2 \ell = n}
\frac{2^{2\ell}}{3^{2\ell} k! \ell! \ell!} 
\, {\cal E}^{a_1 \cdots a_{2k} c_1 \cdots c_{\ell} f_1 \cdots f_{3 \ell}}
{\cal E}^{b_1 \cdots b_{2k} d_1 \cdots d_{\ell} e_1 \cdots e_{3 \ell}}
\nn \\
\ &\LS
\times \int_{\cal M} \!\! \d^{2n} x_0 \sqrt{g(x_0)} 
\Big( R_{a_1 a_2 b_1 b_1} \cdots R_{a_{2k-1} a_{2k} b_{2k-1} b_{2k}} \Big)
\nn \\
\ &\LS \LS
\times 
\Big(
\del_{c_1} (H_{d_1 d_2 d_3}) \cdots
\del_{c_{\ell}} (H_{d_{3 \ell -2} d_{3 \ell -1} d_{3 \ell}}) 
\Big)
\Big(
\del_{e_1} (H_{f_1 f_2 f_3}) \cdots
\del_{e_{\ell}} (H_{f_{3 \ell -2} f_{3 \ell -1} f_{3\ell}}) 
\Big)
\; .
\end{align}
Fortunately, we can furthermore reduce the above representation by
using the second Bianchi identity of the Riemann tensor
(\ref{second-B}) and the closed condition $\d H = 0$.
For simplicity, let us analyze the case $k = 1$, $\ell = 2$, from which we
can read a general statement:
\begin{align}
&{\cal E}^{a_1 a_2 c_1 c_2 f_1 \cdots f_6}
{\cal E}^{b_1 b_2 d_1 d_2 e_1 \cdots e_6}
\int \d^{2n} x_0 \sqrt{g(x_0)} 
R_{a_1 a_2 b_1 b_1} 
\del_{c_1} (H_{d_1 d_2 d_3}) \del_{c_2} (H_{d_4 d_5 d_6}) 
\del_{e_1} (H_{f_1 f_2 f_3}) \del_{e_2} (H_{f_4 f_5 f_6}) 
\nn \\
\ &\ls= \ 
{\cal E}^{a_1 \cdots f_6}
{\cal E}^{b_1 \cdots e_6}
\int \d^{2n} x_0 \del_{c_1} \Big( \text{all terms} \Big)
\nn \\
\ &\LS 
- {\cal E}^{a_1 \cdots f_6}
{\cal E}^{b_1 \cdots e_6}
\int \d^{2n} x_0 \sqrt{g(x_0)} 
\del_{c_1} \Big\{ R_{a_1 a_2 b_1 b_1} \Big\} 
(H_{d_1 d_2 d_3}) \del_{c_2} (H_{d_4 d_5 d_6}) 
\del_{e_1} (H_{f_1 f_2 f_3}) \del_{e_2} (H_{f_4 f_5 f_6}) 
\nn \\
\ &\LS 
- {\cal E}^{a_1 \cdots f_6}
{\cal E}^{b_1 \cdots e_6}
\int \d^{2n} x_0 \sqrt{g(x_0)} 
R_{a_1 a_2 b_1 b_1} 
(H_{d_1 d_2 d_3}) \Big\{ \del_{c_1} \del_{c_2} (H_{d_4 d_5 d_6}) \Big\}
\del_{e_1} (H_{f_1 f_2 f_3}) \del_{e_2} (H_{f_4 f_5 f_6}) 
\nn \\
\ &\LS 
- {\cal E}^{a_1 \cdots f_6}
{\cal E}^{b_1 \cdots e_6}
\int \d^{2n} x_0 \sqrt{g(x_0)} 
R_{a_1 a_2 b_1 b_1} 
(H_{d_1 d_2 d_3}) \del_{c_2} (H_{d_4 d_5 d_6}) 
\del_{c_1} \Big\{ \del_{e_1} (H_{f_1 f_2 f_3}) \del_{e_2} (H_{f_4 f_5 f_6})  \Big\}
\; . \label{Euler-SKT-2}
\end{align}
The first term in (\ref{Euler-SKT-2}) vanishes if there are no
boundaries on the manifold.
The second term also vanishes via the second Bianchi identity (\ref{second-B}).
The third term is zero because the derivatives are symmetric, while
the indices are anti-symmetric under the existence of 
${\cal E}^{a_1 \cdots f_6}$.
The fourth term also vanishes because the closed condition $\d H = 0$
appears as ${\cal E}^{a_1 \cdots f_6} \del_{c_1} (H_{f_1 f_2 f_3}) = 0$.
Other derivatives also yield the same result.
Thus we find that the second and the third exponents in (\ref{Euler-SKT-1})
should not contribute to the Euler characteristics and we can set
$\ell = 0$.
We conclude that the Euler characteristics on the torsional manifold
without boundary is equal to the ones on the Riemannian manifold
(\ref{Euler-ch-R}):
\begin{align}
\chi ({\cal M})
\ &= \ 
\frac{1}{(8\pi)^n n!} 
\, {\cal E}^{a_1 \cdots a_{2n}} {\cal E}^{b_1 \cdots b_{2n}}
\int_{\cal M} \!\! \d^{2n} x_0 \sqrt{g(x_0)} 
\Big( R_{a_1 a_2 b_1 b_1} \cdots R_{a_{2n-1} a_{2n} b_{2n-1} b_{2n}} \Big)
\nn \\
\ &= \ 
\frac{1}{(4\pi)^n n!} {\cal E}_{a_1 \cdots a_{2n}}
\int_{\cal M} \!\! R^{a_1 a_2} (\omega) \w \cdots \w R^{a_{2n-1} a_{2n}} (\omega)
\; . \label{Euler-SKT-3}
\end{align}


\section{Witten index in ${\cal N}=2$ quantum mechanics II}
\label{QM-Hir}

Finally we will discuss the derivation of the Hirzebruch signature on a
torsional manifold in the path integral formalism. 
We also use the ${\cal N}=2$ supersymmetric quantum mechanical path
integral, while we only insert $\Gamma_{(5)}$ into the transition
element instead of the insertion $\Gamma_{(5)} \wt{\Gamma}_{(5)}$ in
the case of the Euler characteristics.
We review the derivation of the signature on the Riemannian manifold.
Next we discuss the analysis of the signature on a torsional manifold
in the same strategy.

\subsection{Formulation}

As mentioned in the introduction, 
the Hirzebruch signature is a topological invariant which gives 
the difference between the number of self-dual forms 
and the number of anti-self-dual forms on the manifold.
Since we analyze the difference of the forms,
we analyze another Witten index defined in the ${\cal N}=2$
supersymmetric quantum mechanics in the following form
(see section {\it 14.3} in \cite{GSW}):
\begin{align}
\sigma 
\ &\equiv \ 
\lim_{\beta \to 0} 
\Tr \big\{ \Gamma_{(5)} \e^{- \beta \Scr{R}} \big\}
\ = \ 
\lim_{\beta \to 0} (-i)^{D/2} \Tr \prod_{a=1}^D 
\big( \wh{\varphi}{}^a + \wh{\ol{\varphi}}{}^a \big) 
\, \e^{- \frac{\beta}{\hbar} \wh{\Scr{H}}}
\; .
\end{align}
Here we did not insert $2^{-D/2}$ because in this system $\psi_2^a$
is also dynamical.
The chirality operators $\Gamma_{(5)}$ 
is again given in terms of the operators $\wh{\psi}_{1}^a$:
\begin{align}
\Gamma_{(5)} \ &\equiv \ 
(-i)^{D/2} \Gamma^1 \cdots \Gamma^D
\ = \ 
(-i)^{D/2} 2^{D/2} \wh{\psi}_{1}^1 \cdots \wh{\psi}_{1}^D
\ = \ 
(-i)^{D/2} \prod_{a=1}^D 
\big( \wh{\varphi}{}^a + \wh{\ol{\varphi}}{}^a \big)
\; .
\end{align}
Notice that since the non-trivial values are given when $D$ is even number, 
we find $(-i)^{2D} = 1$.
In addition, we prepare the trace formula and the complete set of the fermion
coherent states (\ref{complete-sets-fermion}).
We obtain the explicit expression of the topological invariants with respect
to the ${\cal N}=2$ quantum mechanical path integral in the same way
as (\ref{Euler}): 
\begin{align}
\sigma \ &= \ 
\lim_{\beta \to 0} 
\frac{(-i)^{D/2}}{(2 \pi \beta \hbar)^{D/2}} 
\int \d^D x_0 \sqrt{g (x_0)} \prod_{a=1}^D 
\Big( \d \ol{\eta}{}_a \d \eta^a \, \d \zeta^a \d \ol{\zeta}{}_a \Big) 
\nn \\
\ &\LS \LS \LS \times
\e^{\ol{\zeta} \zeta + \ol{\zeta} \eta - \ol{\eta} \eta + \ol{\eta} \zeta} \,
\prod_b \big( \eta^b + \ol{\zeta}{}^b \big)
\VEV{\exp \Big(- \frac{1}{\hbar} S_H^{({\rm int})} \Big)}
\; , \label{Hir-sign} 
\end{align}
where $S^{({\rm int})}$ in (\ref{Hir-sign}) is also given by (\ref{S-int-1})
which appeared in the previous subsection.
Now let us consider the fermionic measure in this path integral form.
In the same way as the Dirac index, we obtain
\begin{align}
&\int \prod_a \d \ol{\eta}{}_a \d \eta^a \d \zeta^a \d \ol{\zeta}{}_a
\, \e^{\ol{\zeta} \zeta + \ol{\zeta} \eta - \ol{\eta} \eta + \ol{\eta} \zeta}
\prod_b (\eta^b + \ol{\zeta}{}^b)
\nn \\
\ &\LS= \
(-2)^D \int \prod_a \d \ol{\eta}{}_a \d \zeta^a
\, \d (\ol{\zeta} + \eta)_a \d (\eta - \ol{\zeta})^a
\, \e^{- \half (\eta - \ol{\zeta})(\zeta - \ol{\eta})}
\prod_b (\eta^b + \ol{\zeta}{}^b)
\nn \\
\ &\LS= \
\int \prod_a \d \ol{\eta}{}_a \d \zeta^a 
\prod_b (\zeta^b - \ol{\eta}{}^b)
\; . \label{D-fer-delta}
\end{align}
This measure gives the fermionic delta function which indicates 
the coincidence of the background fermions $\zeta^a = \ol{\eta}^a$:
\begin{align}
\int \prod_a \d \ol{\eta}{}_a 
\prod_b (\zeta^b - \ol{\eta}{}^b) \, f(\ol{\eta})
\ &= \ 
f (\zeta)
\; .
\end{align}

To remove the $\beta$ dependence in the path integral measure, we
rescale the fermion
\bsubeq
\begin{gather}
\frac{1}{(\beta \hbar)^{D/2}}
\int \prod_a \d \ol{\eta}{}_a \d \zeta^a 
\prod_b (\zeta^b - \ol{\eta}{}^b)
\ \equiv \ 
\int \prod_a \d \ol{\eta}{}'_a \d \zeta'{}^a 
\prod_b (\zeta'{}^b - \ol{\eta}{}'{}^b)
\; , \\
\ol{\eta}{}^a \ \equiv \ 
\Big( \frac{1}{\beta \hbar} \Big)^{1/2} \ol{\eta}{}'{}^a
\; , \ls
\zeta^a \ \equiv \ 
\Big( \frac{1}{\beta \hbar} \Big)^{1/2} \zeta'{}^a
\; .
\end{gather}
\esubeq
Then the rescaled $S^{({\rm int})}$ (\ref{S-origin-N2}) in the path integral 
is given by (where we omit the prime symbol)
\bsubeq
\begin{align}
\sigma
\ &= \ 
\lim_{\beta \to 0} \frac{(-i)^{D/2}}{(2 \pi)^{D/2}}
\int \d^D x_0 \sqrt{g(x_0)} \prod_{a=1}^D \d \ol{\eta}{}_a \d \zeta^a
\prod_{b=1}^D (\zeta^b - \ol{\eta}{}^b) \,
\VEV{\exp \Big( - \frac{1}{\hbar} S_H^{({\rm int})} \Big) }
\; , \\
- \frac{1}{\hbar} S_H^{({\rm int})}
\ &= \ 
- \frac{1}{\beta \hbar} \int_{-1}^0 \d \tau \, 
\half \Big[ g_{mn} (x) - g_{mn} (x_0) \Big] 
\Big( \dot{q}^m \dot{q}^n + b^m c^n + a^m a^n \Big)
\nn \\
\ & \ls
- \frac{1}{\beta \hbar} \int_{-1}^0 \d \tau \, 
\dot{q}^m \Bigg( \omega_{mab} (x) \, 
\big( \ol{\eta} + \ol{\xi}{}_{\rm qu} \big)^a 
\big( \zeta + \xi_{\rm qu} \big)^b
- \half H_{mab} (x) 
\Big\{ \big( \zeta + \xi_{\rm qu} \big)^{ab} 
+ \big( \ol{\eta} + \ol{\xi}{}_{\rm qu} \big)^{ab} 
\Big\} \Bigg)
\nn \\
\ & \ls
+ \frac{1}{2 \beta \hbar} \int_{-1}^0 \d \tau \, 
R_{cdab} (\omega (x)) \, 
(\zeta + \xi_{\rm qu})^a ( \ol{\eta} + \ol{\xi}{}_{\rm qu})^b
(\zeta + \xi_{\rm qu})^c ( \ol{\eta} + \ol{\xi}{}_{\rm qu})^d
\nn \\
\ & \ls
- \frac{1}{8 \beta \hbar} \int_{-1}^0 \d \tau \, 
H_{abe} H_{cd}{}^e (x)
\Big\{ (\zeta + \xi_{\rm qu})^{abcd} 
+ (\ol{\eta} + \ol{\xi}{}_{\rm qu})^{abcd} 
- 2 (\zeta + \xi_{\rm qu})^{ab} 
(\ol{\eta} + \ol{\xi}{}_{\rm qu})^{cd} \Big\}
\nn \\
\ & \ls
- \frac{1}{6 \beta \hbar} \int_{-1}^0 \d \tau \, 
\del_m (H_{npq}) (x)
\Big\{ (\ol{\eta} + \ol{\xi}{}_{\rm qu})^m 
(\zeta + \xi_{\rm qu})^{npq}
+ (\zeta + \xi_{\rm qu})^m (\ol{\eta} + \ol{\xi}{}_{\rm qu})^{npq}
\Big\}
\nn \\
\ & \ls
- \frac{\beta \hbar}{8} \int_{-1}^0 \d \tau \, 
{\cal G}_2 (x)
\; . \label{S-int-3}
\end{align}
\esubeq
The bosonic and fermionic propagators are of order in $\beta \hbar$.
Let us truncate this action.
In the same analogy to the Dirac index,
disconnected Feynman graphs might contribute to the amplitude.
In the same way as previous case, the fermion propagator is given by
\begin{align}
\Vev{\xi_{\rm qu}^a (\sigma) \ol{\xi}{}_{\rm qu}^b (\tau)}
\ &= \ 
\beta \hbar \, \delta^{ab} \theta (\sigma - \tau)
\; . \label{f-prop-Hir}
\end{align}


\subsection{Hirzebruch signature}

\subsubsection{Riemannian manifold}

This case is quite simple.
Since there are no background interaction terms of order in
$(\beta \hbar)^{-1}$ which contribute to the disconnected graphs,
we only consider one-loop Feynman graphs.
Then, we neglect interaction terms carrying more than three
quantum fields.
We can also neglect the last line in (\ref{S-int-3})
which yields the graphs of higher order in $\beta \hbar$. 
We also use the condition by 
Riemann normal coordinate frame $\del_p g_{mn} (x_0) = \omega_{mab} (x_0) = 0$
at the point $x_0$.
We can further neglect interaction terms which are
irrelevant in the vanishing limit $\beta \to 0$.
By using the Riemann normal coordinates on the second line in (\ref{S-int-3}), 
the fermionic delta function (\ref{D-fer-delta}) and
the first Bianchi identity (\ref{first-B}) acting on the fourth line
in (\ref{S-int-3}),
we obtain a much simpler expression of the Hirzebruch signature: 
\bsubeq \label{Hir-sig-R-PI}
\begin{align}
\sigma \ &= \ 
\lim_{\beta \to 0} \frac{(-i)^{D/2}}{(2 \pi)^{D/2}}
\int \d^D x_0 \sqrt{g(x_0)} \prod_{a=1}^D \d \zeta^a \,
\VEV{\exp \Big( - \frac{1}{\hbar} S^{({\rm int})} \Big) }
\; , \\
- \frac{1}{\hbar} S^{({\rm int})}
\ &= \ 
- \frac{1}{2 \beta \hbar} 
R_{mnab} (\omega (x_0)) \, \zeta^{ab} \int_{-1}^0 \d \tau \,
q^m \dot{q}^n 
\nn \\
\ & \ls
+ \frac{1}{2 \beta \hbar} R_{abcd} (\omega (x_0)) \, \zeta^{ab}
\Bigg(
- \half \int_{-1}^0 \d \tau \, \xi_{\rm qu}^c \xi_{\rm qu}^d
- \half \int_{-1}^0 \d \tau \, \ol{\xi}{}_{\rm qu}^c \ol{\xi}{}_{\rm qu}^d
+ \int_{-1}^0 \d \tau \, \xi_{\rm qu}^c \ol{\xi}{}_{\rm qu}^d
\Bigg)
\; . \label{S-int-5}
\end{align} 
\esubeq
We should notice that the fermionic fields in the above path integral
have {\bf anti-periodic} boundary condition.
Originally the fermionic fields are introduced as the fields with
anti-periodic boundary condition (see the discussion in section 
{\it 2.4} of \cite{BvN06}), which is changed by the insertion of operators.
Now, in the form (\ref{Hir-sig-R-PI}) there are no additional
operator insertions in the path integral measure.
Thus the fermions in (\ref{Hir-sig-R-PI}) keep the anti-periodic
boundary condition.

We can easily find that the Feynman graphs will be described 
as the trace of Riemann curvature two-form in the same way as the
Pontrjagin classes. Here let us remember a property that 
the trace of odd number of Riemann curvature two-form vanishes 
$\tr ( R^{2k-1} ) = 0$.
On the other hand, the Feynman one-loop graph which contains all of three
interaction terms in the second line 
in (\ref{S-int-5}) always has odd number of the interaction vertices.
This indicates that the third interaction term in the second line
$\xi_{\rm qu}^c \ol{\xi}{}_{\rm qu}^d$ should not be connected to
the other two interactions 
($\xi_{\rm qu}^c \xi_{\rm qu}^d$ and $\ol{\xi}{}_{\rm qu}^c \ol{\xi}{}_{\rm qu}^d$) 
in the graphs. These other two terms should be connected to each other.
Furthermore, because of the anti-periodicity of the fermions,
we also find that the closed loop graphs which contain only the third
interaction $\xi_{\rm qu}^c \ol{\xi}{}_{\rm qu}^d$ vanish
in the same reason as the vanishing closed loop graphs of
$\hat{c}$-ghost in (\ref{S1-int-HA-simplest}), 
which also has the anti-periodic boundary condition.
The term in the first line exactly gives a same Feynman graphs 
as the Pontrjagin classes (\ref{S1-int-riemann}).
Summarizing these comments,
here let us again describe the action in (\ref{Hir-sig-R-PI}):
\bsubeq 
\begin{align}
- \frac{1}{\hbar} S^{({\rm int})}
\ &= \ 
- \frac{1}{\beta \hbar} 
R_{mn} \int_{-1}^0 \d \tau \, q^m \dot{q}^n 
- \frac{1}{2 \beta \hbar} R_{cd} 
\int_{-1}^0 \d \tau \, \xi_{\rm qu}^c \xi_{\rm qu}^d 
- \frac{1}{2 \beta \hbar} R_{cd}
\int_{-1}^0 \d \tau \, 
\ol{\xi}{}_{\rm qu}^c \ol{\xi}{}_{\rm qu}^d
\nn \\
\ &\equiv \
- \frac{1}{\hbar} {\cal S}_{\rm p}
- \frac{1}{\hbar} {\cal S} - \frac{1}{\hbar} \ol{\cal S}
\; , \label{S-int-6} \\
R_{cd} 
\ &\equiv \ 
\half R_{cdab} (\omega (x_0)) \, \zeta^{ab}
\ = \ 
\half R_{abcd} (\omega (x_0)) \, \zeta^{ab}
\; .
\end{align} 
\esubeq

Let us rewrite the exponent $\vev{\exp (- \frac{1}{\hbar} S^{({\rm int})})}$
in terms of the effective action $W$ in such a way as
\begin{align}
- \frac{1}{\hbar} W 
\ &= \ 
\log \VEV{\exp \Big( - \frac{1}{\hbar} S^{({\rm int})} \Big)}
\ = \ 
\sum_{k=1}^{\infty} \frac{1}{k!} 
\cVEV{\Big( - \frac{1}{\hbar} S^{({\rm int})} \Big)^k}
\nn \\
\ &\sim \ 
\sum_{k=1}^{\infty} \frac{1}{k!}
\cVEV{\Big( - \frac{1}{\hbar} {\cal S}_{\rm p} \Big)^k}
+ \sum_{k=0}^{\infty} \frac{1}{k!} 
\frac{k!}{(k/2)! (k/2)!}
\cVEV{\Big( - \frac{1}{\hbar} {\cal S} \Big)^{k/2}
\Big(- \frac{1}{\hbar} \ol{\cal S} \Big)^{k/2}}
\nn \\
\ &= \
\sum_{k=1}^{\infty} \frac{1}{k!}
\Big( - \frac{1}{\beta \hbar} \Big)^k 
R_{m_1 n_1} \cdots R_{m_k n_k} 
\int_{-1}^0 \d \tau_1 \cdots \d \tau_k
\cVEV{(q^{m_1} \dot{q}^{n_1}) (\tau_1) \cdots 
(q^{m_k} \dot{q}^{n_k}) (\tau_k)}
\nn \\
\ & \ls
+ \sum_{\ell=1}^{\infty}
\frac{1}{\ell! \ell!} \Big( - \frac{1}{2 \beta \hbar} \Big)^{2\ell}
R_{a_1 b_1} \cdots R_{a_{\ell} b_{\ell}} 
R_{c_1 d_1} \cdots R_{c_{\ell} d_{\ell}} 
\int_{-1}^0 \d \tau_1 \cdots \d \tau_{\ell}
\int_{-1}^0 \d \sigma_1 \cdots \d \sigma_{\ell}
\nn \\
\ & \LS 
\times \cVEV{
\big( \xi_{\rm qu}^{a_1} \xi_{\rm qu}^{b_1} \big) (\tau_1)
\cdots \big( \xi_{\rm qu}^{a_{\ell}} \xi_{\rm qu}^{b_{\ell}} \big) (\tau_{\ell})
\big( \ol{\xi}{}_{\rm qu}^{c_1} \ol{\xi}{}_{\rm qu}^{d_1} \big) (\sigma_1)
\cdots \big( \ol{\xi}{}_{\rm qu}^{c_{\ell}} 
\ol{\xi}{}_{\rm qu}^{d_{\ell}} \big) (\sigma_{\ell}) 
}
\; , \label{eff-act-Hir-R}
\end{align}
where we extracted terms which contribute to the Feynman graphs in the
vanishing limit $\beta \to 0$.
The bracket $\cvev{\cdots}$ gives connected Feynman graphs.
The number of the vertices ${\cal S}$ should be equal to the number of
the vertices $\ol{\cal S}$.
Because of this, we find that $k$ should be even: $k = 2 \ell$.

Since we have already analyzed the first connected graphs 
in the Pontrjagin classes (\ref{eff-Pont-Riemann}), it is easy to
analyze the first term in (\ref{eff-act-Hir-R}):
\bsubeq
\begin{align}
\text{(1st term)}
\ &= \ 
\sum_{k=1}^{\infty} \frac{1}{k!} \Big( - \frac{1}{\beta \hbar} \Big)^k
(k-1)! \, 2^{k-1} \big( - \beta \hbar \big)^k \cdot
\, R_{m_1 n_1} R_{m_2 n_2} \cdots R_{m_k n_k} 
\, g^{n_1 m_2} g^{n_2 m_3} \cdots g^{n_k m_1}
\nn \\
\ &\LS \times
\int_{-1}^0 \d \tau_1 \cdots \d \tau_k \,
\del_{\tau_1} \Delta (\tau_1, \tau_2)
\del_{\tau_2} \Delta (\tau_2, \tau_3)
\cdots 
\del_{\tau_{k-1}} \Delta (\tau_{k-1}, \tau_{k})
\del_{\tau_k} \Delta (\tau_k, \tau_1)
\nn \\
\ &\equiv \ 
\half \sum_{k=2}^{\infty} \frac{1}{k} 
\tr \big\{ (2R)^k \big\} I_k
\ = \
\half \tr \log \Big( \frac{R}{\sinh R} \Big)
\; . \label{1st-Hir-R}
\end{align}
\esubeq
Next let us here evaluate the second connected graphs in (\ref{eff-act-Hir-R}).
In order to make one-loop graphs, $\ell$ vertices ${\cal S}$ and 
$\ell$ vertices $\ol{\cal S}$ should be alternatively located on the one-loop graph
in $(\ell - 1)! \ell !$ ways.
Furthermore, there are $2^{2\ell-1}$ ways 
to contract these vertices in
terms of fermion propagator (\ref{f-prop-Hir}) 
to yield the trace of curvature two-forms 
$\tr (R^{2\ell})$ with sign $(-1)^{\ell + 1}$, which comes
from permutation of indices.
Then, the effective action (\ref{eff-act-Hir-R}) is evaluated in the
following way:
\begin{align}
\text{(2nd term)}
\ &= \ 
\sum_{\ell=1}^{\infty}
\frac{1}{\ell! \ell!} 
\Big( - \frac{1}{2 \beta \hbar} \Big)^{2 \ell} 
(\ell-1)! \ell! \, 2^{2\ell-1} \, (-1)^{\ell +1} \,
\big( \beta \hbar \big)^{2 \ell} \cdot
\tr \big( R^{2 \ell} \big) 
\nn \\
\ &\ls \ \ \times 
\int_{-1}^0 \prod_{i=1}^{\ell} \d \tau_i \, \d \sigma_i \,
\theta (\tau_1 - \sigma_1) \theta (\tau_1 - \sigma_{\ell}) \,
\theta (\tau_2 - \sigma_2) \theta (\tau_2 - \sigma_1) 
\cdots 
\theta (\tau_{\ell} - \sigma_{\ell}) \theta (\tau_{\ell} - \sigma_{\ell-1})
\nn \\
\ &\equiv \ 
\half \sum_{\ell=1}^{\infty} \frac{(-1)^{\ell + 1}}{\ell} \, 
\tr \big( R^{2 \ell} \big) J_{2 \ell}
\ = \ 
\half \tr \log \Big( \cosh R \Big)
\; . \label{2nd-Hir-R}
\end{align}
The term $\ell = 0$ does not contribute to connected graphs
because this term does not carry any background fermions.
The function $J_{2 \ell}$ is defined in such a way as
\begin{align}
J_{2 \ell}
\ &\equiv \ 
\int_{-1}^0 \prod_{i=1}^{\ell} \d \tau_i \, \d \sigma_i \,
\theta (\tau_1 - \sigma_1) \theta (\tau_1 - \sigma_{\ell}) \,
\theta (\tau_2 - \sigma_2) \theta (\tau_2 - \sigma_1) 
\cdots 
\theta (\tau_{\ell} - \sigma_{\ell}) \theta (\tau_{\ell} - \sigma_{\ell-1})
\; . \label{max-loop}
\end{align}
Thus, substituting (\ref{1st-Hir-R}) and (\ref{2nd-Hir-R}) into 
(\ref{eff-act-Hir-R}), we obtain
\begin{align}
- \frac{1}{\hbar} W 
\ &= \ 
\half \tr \log \Big( \frac{R}{\sinh R} \Big)
+ \half \tr \log \Big( \cosh R \Big)
\ = \
\half \tr \log \Big( \frac{R}{\tanh R} \Big)
\; .
\end{align}
Rescaling $\zeta^a \to \sqrt{\frac{-i}{2 \pi}} \zeta^a$,
we finally obtain the Hirzebruch signature on the Riemannian manifold
\begin{gather}
\sigma
\ = \ 
\int \d^D x_0 \sqrt{g(x_0)} \prod_{a=1}^D \d \zeta^a \,
\exp \Bigg[
\half \tr \log \Bigg( \frac{-iR/2 \pi}{\tanh (-iR/2\pi)} 
\Bigg) \Bigg]
\; , \label{Hir-R-1}
\end{gather}
or, if we integrate out the fermionic fields and using the following
formula (in the same way as (\ref{fer-contract})),
we simplify (\ref{Hir-R-1}) and obtain
\begin{gather}
\sigma
\ = \ 
\int_{\cal M} \exp \Bigg[
\half \tr \log \Bigg( \frac{iR/2 \pi}{\tanh (iR/2\pi)} 
\Bigg) \Bigg]
\; , \ls
R_{mn} \ = \ 
\half R_{mnab} (\omega) \, e^a \w e^b
\; . \label{index-Hir-R}
\end{gather}


\subsubsection{Torsional manifold}

Now let us analyze the signature on the torsional geometry.
It seems that the action (\ref{S-int-3}) carries the 
background interaction terms of order $(\beta \hbar)^{-1}$
in (\ref{S-int-3}), which
cause the divergence of the amplitude in the vanishing limit $\beta
\to 0$. 
Fortunately, however, the fermionic delta function (\ref{D-fer-delta})
removes this difficulty:
\bsubeq
\begin{align}
R_{cdab} (\omega (x_0)) \Big\{
\zeta^a \ol{\eta}{}^b \zeta^c \ol{\eta}{}^d \Big\}
\Big|_{\text{(\ref{D-fer-delta})}}
\ &= \ 
R_{[abcd]} (\omega (x_0)) \, \zeta^{abcd}
\ = \ 
0
\; , \\
H^e{}_{ab} H_{cde} (x_0)
\Big\{ \zeta^{abcd} + \ol{\eta}{}^{abcd} 
- 2 \zeta^{ab} \ol{\eta}{}^{cd} \Big\} \Big|_{\text{(\ref{D-fer-delta})}}
\ &= \ 
2 H^e{}_{ab} H_{cde} (x_0)
\Big\{ \zeta^{abcd} - \zeta^{abcd} \Big\}
\ = \ 
0
\; , \\
\del_m (H_{npq}) (x_0) \Big\{
\ol{\eta}{}^m \zeta^{npq}
+ \zeta^m \ol{\eta}{}^{npq}
\Big\} \Big|_{\text{(\ref{D-fer-delta})}}
\ &= \ 
\frac{1}{2} 
(\d H)_{mnpq} (x_0) \, \zeta^{mnpq}
\ = \ 
0 \; ,
\end{align}
\esubeq
where we used the first Bianchi identity (\ref{first-B}) and the
closed condition $\d H = 0$.
Since we find that there are no background interaction terms with
$(\beta \hbar)^{-1}$, it is sufficient to investigate the interaction
terms equipped with two quantum fields in order to generate 
the closed one-loop Feynman graphs.
Here let us study the truncation of the action in (\ref{S-int-3}) 
with the fermionic delta function (\ref{D-fer-delta}).
The first and the last lines in (\ref{S-int-3}) disappear.
The second and the third lines in (\ref{S-int-3}) are truncated to 
\bsubeq \label{expand-Hir-RH-1}
\begin{align}
&- \frac{1}{\beta \hbar} \int_{-1}^0 \d \tau \, 
\dot{q}^m \Bigg( \omega_{mab} (x) \, 
\big( \ol{\eta} + \ol{\xi}{}_{\rm qu} \big)^a 
\big( \zeta + \xi_{\rm qu} \big)^b
- \half H_{mab} (x) 
\Big\{ \big( \zeta + \xi_{\rm qu} \big)^{ab} 
+ \big( \ol{\eta} + \ol{\xi}{}_{\rm qu} \big)^{ab} 
\Big\} \Bigg) \Bigg|_{\text{(\ref{D-fer-delta})}}
\nn \\
\ &\LS= \ 
- \frac{1}{2 \beta \hbar} R_{abmn} (\omega_- (x_0)) \, \zeta^{ab} 
\int_{-1}^0 \d \tau \, q^m \dot{q}{}^n
\nn \\
\ &\LS= \ 
- \frac{1}{2 \beta \hbar} R_{mnab} (\omega_+ (x_0)) \, \zeta^{ab} 
\int_{-1}^0 \d \tau \, q^m \dot{q}{}^n
\; , \label{expand-Hir-RH-11} \\
&\frac{1}{2 \beta \hbar} \int_{-1}^0 \d \tau \, 
R_{cdab} (\omega (x)) \, 
(\zeta + \xi_{\rm qu})^a ( \ol{\eta} + \ol{\xi}{}_{\rm qu})^b
(\zeta + \xi_{\rm qu})^c ( \ol{\eta} + \ol{\xi}{}_{\rm qu})^d
\Bigg|_{\text{(\ref{D-fer-delta})}}
\nn \\
\ &\LS= \ 
\frac{1}{2 \beta \hbar} R_{abcd} (\omega (x_0)) \, \zeta^{ab}
\Bigg(
- \half \int_{-1}^0 \d \tau \, \xi_{\rm qu}^{cd}
- \half \int_{-1}^0 \d \tau \, \ol{\xi}{}_{\rm qu}^{cd}
+ \int_{-1}^0 \d \tau \, \xi_{\rm qu}^c \ol{\xi}{}_{\rm qu}^d
\Bigg)
\; . \label{expand-Hir-RH-12}
\end{align}
\esubeq
The fourth line in (\ref{S-int-3}) is also truncated to
\begin{align}
\text{(fouth line)} \Big|_{\text{(\ref{D-fer-delta})}}
\ &\sim \ 
- \frac{1}{16 \beta \hbar} \del_m \del_n (H_{abe} H_{cd}{}^e (x_0)) 
\int_{-1}^0 \d \tau \, q^m q^n
\Big\{ \zeta^{abcd} + \zeta^{abcd} - 2 \zeta^{ab} \zeta^{cd} \Big\}
\ = \ 
0 \; . \label{expand-Hir-RH-2}
\end{align}
where we used $H_{mnp} (x_0) = 0$ and $\del_m H_{abc} (x_0) \neq 0$.
The fifth line in (\ref{S-int-3}) is more complicated:
\begin{align}
\text{(fifth line)} \Big|_{\text{(\ref{D-fer-delta})}}
\ &\sim \ 
- \frac{1}{6 \beta \hbar} \int_{-1}^0 \d \tau \,
\Bigg(
q^r \del_r \del_a (H_{bcd} (x_0)) \Big\{
\ol{\xi}{}_{\rm qu}^a \zeta^{bcd}
+ 3 \zeta^{abc} \xi_{\rm qu}^d
+ \xi_{\rm qu}^a \zeta^{bcd}
+ 3 \zeta^{abc} \ol{\xi}{}_{\rm qu}^d
\Big\}
\nn \\
\ &\LS \LS \LS \ \ 
+ 3 \del_a (H_{bcd} (x_0)) \Big\{
\ol{\xi}{}_{\rm qu}^a \xi_{\rm qu}^b \, \zeta^{cd}
+ \zeta^{ab} \xi_{\rm qu}^{cd}
+ \xi_{\rm qu}^a \ol{\xi}{}_{\rm qu}^b \, \zeta^{cd}
+ \zeta^{ab} \ol{\xi}{}_{\rm qu}^{cd} \Big\}
\Bigg)
\nn \\
\ &\sim \ 
- \frac{1}{2 \beta \hbar} \del_a (H_{bcd} (x_0)) \, \zeta^{cd} 
\int_{-1}^0 \d \tau \,
\Big( \ol{\xi}{}_{\rm qu}^a \xi_{\rm qu}^b 
+ \xi_{\rm qu}^a \ol{\xi}{}_{\rm qu}^b 
\Big)
\nn \\
\ &\ls
- \frac{1}{2 \beta \hbar} \del_a (H_{bcd} (x_0)) \, \zeta^{ab} 
\int_{-1}^0 \d \tau \,
\Big( \xi_{\rm qu}^{cd} + \ol{\xi}{}_{\rm qu}^{cd} \Big)
\; . \label{expand-Hir-RH-3}
\end{align}
Notice that the first line in (\ref{expand-Hir-RH-3}) does not
contribute to the amplitudes: each term is contracted to the
terms in the same line, which yields zero amplitudes because the
background fermionic fields are anti-symmetric in the amplitudes in
such a way as
$\zeta^{abc} \zeta^{def} = - \zeta^{def} \zeta^{abc}$.
Now, we combine (\ref{expand-Hir-RH-12}) and (\ref{expand-Hir-RH-3})
to yield
\begin{align}
\text{(\ref{expand-Hir-RH-12})} 
+ \text{(\ref{expand-Hir-RH-3})} 
\ &= \ 
- \frac{1}{4 \beta \hbar} R_{cdab} (\omega_+ (x_0)) \, 
\zeta^{ab} \int_{-1}^0 \d \tau \, \xi_{\rm qu}^{cd}
- \frac{1}{4 \beta \hbar} R_{cdab} (\omega_+ (x_0)) \,
\zeta^{ab} \int_{-1}^0 \d \tau \, \ol{\xi}{}_{\rm qu}^{cd}
\nn \\
\ & \ls
+ \frac{1}{2 \beta \hbar} \Big( R_{cdab} (\omega (x_0)) 
- \del_a (H_{bcd} (x_0)) 
+ \del_b (H_{acd} (x_0)) 
\Big)
\zeta^{cd} \int_{-1}^0 \d \tau \, \xi_{\rm qu}^a \ol{\xi}{}_{\rm qu}^b
\nn \\
\ &= \ 
- \frac{1}{4 \beta \hbar} R_{cdab} (\omega_+ (x_0)) \, 
\zeta^{ab} \int_{-1}^0 \d \tau 
\Big\{ \xi_{\rm qu}^{cd} + \ol{\xi}{}_{\rm qu}^{cd} \Big\}
+ \frac{1}{2 \beta \hbar} R_{cdab} (\omega_- (x_0)) \,
\zeta^{cd} \int_{-1}^0 \d \tau \, \xi_{\rm qu}^a \ol{\xi}{}_{\rm qu}^b
\nn \\
\ &= \ 
\frac{1}{2 \beta \hbar} R_{abcd} (\omega_+ (x_0)) \, \zeta^{ab}
\Bigg(
- \half \int_{-1}^0 \d \tau \, \xi_{\rm qu}^{cd}
- \half \int_{-1}^0 \d \tau \, \ol{\xi}{}_{\rm qu}^{cd}
+ \int_{-1}^0 \d \tau \, \xi_{\rm qu}^c \ol{\xi}{}_{\rm qu}^d
\Bigg)
\; . \label{expand-Hir-RH-4}
\end{align}
Then, we rewrite the action (\ref{S-int-3}) by
summarizing (\ref{expand-Hir-RH-11}), (\ref{expand-Hir-RH-3})
 and (\ref{expand-Hir-RH-4}) in the following form:
\begin{align}
- \frac{1}{\hbar} S_H^{({\rm int})}
\ &= \ 
- \frac{1}{2 \beta \hbar} 
R_{mnab} (\omega_+ (x_0)) \, \zeta^{ab} \int_{-1}^0 \d \tau \,
q^m \dot{q}^n 
\nn \\
\ & \ls
+ \frac{1}{2 \beta \hbar} R_{abcd} (\omega_+ (x_0)) \, \zeta^{ab}
\Bigg(
- \half \int_{-1}^0 \d \tau \, \xi_{\rm qu}^c \xi_{\rm qu}^d
- \half \int_{-1}^0 \d \tau \, \ol{\xi}{}_{\rm qu}^c \ol{\xi}{}_{\rm qu}^d
+ \int_{-1}^0 \d \tau \, \xi_{\rm qu}^c \ol{\xi}{}_{\rm qu}^d
\Bigg)
\; . \label{S-Hir-RH}
\end{align}
This form is exactly same as (\ref{S-int-5}).
Then the path integral of (\ref{S-Hir-RH}) is also given in the same
as (\ref{Hir-R-1}):
\bsubeq \label{Hir-RH-1}
\begin{gather}
\sigma
\ = \ 
\int \d^D x_0 \sqrt{g(x_0)} \prod_{a=1}^D \d \zeta^a \,
\exp \Bigg[
\half \tr \log \Bigg( \frac{-iR^{(+)}/2 \pi}{\tanh (-iR^{(+)}/2\pi)} 
\Bigg) \Bigg]
\; , \\
R^{(+)}_{mn} \ = \ 
\half R_{mnab} (\omega_+ (x_0)) \, \zeta^a \zeta^b 
\; ,
\end{gather}
\esubeq
or, if we integrate out the fermionic fields and using the following
formula (in the same way as (\ref{fer-contract}))
\begin{align}
\int \prod_{a=1}^D \d \zeta^a \, \zeta^{a_1 \cdots a_D}
\ &= \ 
\int \d \zeta^1 \d \zeta^2 \cdots \d \zeta^D \,
\zeta^1 \zeta^2 \cdots \zeta^D \cdot {\cal E}^{a_1 a_2 \cdots a_D}
\ = \
(-1)^{D/2} {\cal E}^{a_1 a_2 \cdots a_D}
\; , 
\end{align}
we simplify (\ref{Hir-RH-1}) and obtain
\bsubeq \label{index-Hir-RH}
\begin{gather}
\sigma
\ = \ 
\int \exp \Bigg[
\half \tr \log \Bigg( \frac{iR^{(+)}/2 \pi}{\tanh (iR^{(+)}/2\pi)} 
\Bigg) \Bigg]
\; , \\
\tr (R_{(+)}^k) 
\ = \ 
R^{(+)}_{m_1 n_1} R^{(+)}_{m_2 n_2} \cdots R^{(+)}_{m_k n_k}
\, g^{n_1 m_2} g^{n_2 m_3} \cdots g^{n_k m_1} 
\; , \\
R^{(+)}_{mn} \ = \ 
\half R_{mnab} (\omega_+ (x_0)) \, e^a \w e^b
\; .
\end{gather}
\esubeq


\section{Summary and discussions} \label{summary}

In this paper we have studied various topological invariants on the 
torsional manifold in the framework of supersymmetric quantum
mechanical path integral formalism. 
First we constructed the ${\cal N}=1$ supersymmetric quantum
mechanics (\ref{SUSY-real-A}) 
whose target space corresponds to the torsional manifold.
We extended this to the ${\cal N}=2$ quantum mechanics 
(\ref{SUSY-complex-H}) with 
introducing a closed condition of the torsion.
Next we described the transition elements which appear in the
calculation of the Witten index. 
Following the work \cite{BvN06},
we rewrote the transition elements from the Hamiltonian formalism to
the Lagrangian formalism (\ref{transition-N2}) in the ${\cal N}=2$ case,
and (\ref{transition-N1}) in the ${\cal N}=1$ case.
Since we have already known these topological invariants on the Riemannian
manifold in the framework of the quantum mechanical path integral,
we applied the same formalism to the analyses of the Witten indices.
Then we realized the formulation of the Dirac index on the
torsional manifold (\ref{D-index-SKT})
which have already been investigated by Mavromatos
\cite{Mav88}, Yajima \cite{Yajima89}, Peeters and Waldron \cite{PW99},
and so forth.
The point is that we should carefully use the Riemann normal
coordinate frame on the spin connection (and the affine connection)
equipped with torsion.
We also analyzed the Euler characteristic (\ref{Euler-SKT-3})
and the Hirzebruch signature (\ref{index-Hir-RH}) on the torsional manifold.
These modified values should also be topological invariants 
because we started from the well-defined supersymmetric algebras
(\ref{SUSY-algebra-N1}) in the ${\cal N}=1$ case 
and (\ref{SUSY-algebra-N2}) in the ${\cal N}= 2$ case, respectively.
In these systems we can define the bosonic and fermionic states whose
energy levels are degenerated. We should also find the zero energy
eigenstates, which gives the Witten index as the topological value.
We evaluated these Witten indices in various supersymmetric systems.

The most significant result in this paper is that the Euler
characteristic (\ref{Euler-SKT-3}) is not modified even in the
presence of torsion, while the Dirac index (\ref{D-index-SKT}) and the
Hirzebruch signature (\ref{index-Hir-RH}) are.
Then we conclude that 
if the compactified manifold has the Bismut torsion (\ref{const-0-3})
with the constraint (\ref{const-5-4}) in string theory compactification scenarios,
the numbers of generation in the four-dimensional effective theory
is not changed from the numbers of generation derived from the 
corresponding Calabi-Yau manifold without the torsion.

In this paper we imposed the closed condition $\d H = 0$ on the totally
anti-symmetric torsion.
Peeters and Waldron \cite{PW99} have already investigated 
the Dirac index on a four-dimensional geometry 
with boundary in the presence of a totally anti-symmetric torsion $H$, 
and have discussed the role of $\d H$ in the Feynman graphs.
The four-form $\d H$ can be described as the Nieh-Yan four-form 
${\cal N} (e, H) = \d ( e^A \w H_A )$, which appears in \cite{NY82} and 
is applied to the analysis of the chiral anomaly \cite{CZ97}, and the
Dirac index \cite{PW99}.
To complete the analysis of the index theorems on a torsional geometry 
in the presence of non-vanishing $\d H$ is of particular importance 
when we study the string theory compactified on a $G$-structure
manifold \cite{IP0008, GMW0302, KY0605}.

This four-form $\d H$ also appears and plays a 
crucial role in the anomaly cancellation mechanism in heterotic string
theory (see \cite{AW83, GSW, BvN06} as instructive references).
In the usual anomaly cancellation in heterotic string, 
the Bianchi identity of the NS-NS three-form $H$ is given 
in terms of the Riemann curvature two-form 
and the field strength of the gauge field \cite{GS84}:
$\d H = - \alpha' [ \tr \{ R (\omega) \w R (\omega) \} - \tr ( F \w F ) ]$.
In the presence of non-vanishing $H$-flux,
the spin connection $\omega$ in the Bianchi identity 
is modified to $\omega_{+MAB} = \omega_{MAB} + H_{MAB}$
and the Bianchi identity is rewritten such as
\begin{align}
\d H \ &= \ 
- \alpha' \Big[
\tr \big\{ R (\omega_+) \w R (\omega_+) \big\}
- \tr \big( F \w F \big) \Big]
\; . \label{dH-Bianchi-H}
\end{align}
The modification of the Bianchi identity (\ref{dH-Bianchi-H}) was, for instance, 
investigated by Hull \cite{Hull86} in the framework of the worldsheet sigma model.
Bergshoeff and de Roo applied (\ref{dH-Bianchi-H}) to the supergravity
Lagrangian with higher-order $\alpha'$ corrections \cite{BdR89}.
Recent papers follow this modification and analyze the structures 
in the effective theories from the heterotic string 
(see, for instance, \cite{CCDLMZ0211, BBDG0301, CCDL0306, BT0509,
  BBFTY0604, KY0605, BTY0612} and references therein).
Since, even in the presence of the condition $\d H = 0$,
 we have completed the derivation of topological invariants which
will contribute to the anomaly in string theory,
we will be able to derive the above modified Bianchi identity in the
flux compactification scenarios in an explicit way. 


\section*{Acknowledgements}

The author would like to thank Piljin Yi for the discussions
at some stages of this project.
He would also like to thank 
Yuji Tachikawa, Seiji Terashima and Kenske Tsuda for valuable discussions.
He also thanks Marco Serone for correspondence.
This work is supported by the Grant-in-Aid for the 21st Century COE
``{\sc Center for Diversity and Universality in Physics}'' from the Ministry
of Education, Culture, Sports, Science and Technology (MEXT) of
Japan.


\begin{appendix}

\section*{Appendix}

\section{Convention} \label{app-convention}

We introduce vielbeins $e_M{}^A$ and their inverses $E_A{}^M$, 
which come from the spacetime metric $g_{MN}$ and the metric $\eta_{AB}$ on 
orthogonal frame via
$g_{MN} = \eta_{AB} \, e_M{}^A \, e_N{}^B$ and
$\eta_{AB} = g_{MN} \, E_A{}^M \, E_B{}^N$.
By using these geometrical variables, let us define the covariant
derivatives $D_M (\omega,\Gamma)$ in such a way as
\bsubeq \label{cov-affine}
\begin{align}
D_M (\Gamma) A_N 
\ &= \ \del_M A_N - \Gamma^P{}_{NM} A_P
\; , \\
D_M ({\Gamma}) A^N 
\ &= \ \del_M A^N + \Gamma^N{}_{PM} A^P
\; , \\
D_M (\Gamma) g_{NP}
\ &\equiv \ 
0 \ = \ 
\del_M g_{NP} - \Gamma^Q{}_{NM} g_{QP}
- \Gamma^Q{}_{PM} g_{NQ}
\; , \\
D_M (\Gamma) g^{NP}
\ &\equiv \ 
0 \ = \ 
\del_M g^{NP} + \Gamma^N{}_{QM} g^{QP}
+ \Gamma^P{}_{QM} g^{NQ}
\; , \\
D_M (\omega, \Gamma) e_N{}^A \ &\equiv \ 0 
\ = \ 
\del_M e_N{}^A + \omega_M{}^A{}_B \, e_N{}^B
- \Gamma^P{}_{NM} \, e_P{}^A 
\; , \\
D_M (\omega, \Gamma) E_A{}^N \ &\equiv \ 0 
\ = \ 
\del_M E_A{}^N - E_B{}^N \, \omega_M{}^B{}_A 
+ \Gamma^N{}_{PM} \, E_A{}^P
\; , \\
[ D_M (\Gamma), D_N (\Gamma) ] A_Q \ &= \ 
- R^P{}_{QMN} (\Gamma) 
A_P + 2 T^P{}_{MN} \, D_Q (\Gamma) A_P
\; , \\
R^P{}_{QMN} (\Gamma) \ &= \ 
\del_M^{\phantom{P}} \Gamma^P{}_{QN} 
- \del_N^{\phantom{P}} \Gamma^P{}_{QM}
+ \Gamma^P{}_{RM} \Gamma^R{}_{QN} 
- \Gamma^P{}_{RN} \Gamma^R{}_{QM}
\; .
\end{align}
\esubeq
Note that $A_M$ in the above equations are vector.
$\Gamma^P{}_{MN}$ is the affine connection whose two lower indices are
not symmetric in general case. The anti-symmetric part of the affine
connection $\Gamma^P{}_{[MN]}$ 
is defined as a torsion $T^P{}_{MN}$, while the symmetric
part $\Gamma^P{}_{(MN)}$ is given in terms of the Levi-Civita connection
$\Gamma^P_{0MN}$ and torsion terms in the following way:
\bsubeq
\begin{align}
\Gamma^P{}_{MN}
\ &= \ 
\Gamma^P{}_{(MN)}
+ \Gamma^P{}_{[MN]}
\; , \\
\Gamma^P{}_{[NM]} \ &= \ T^P{}_{NM}
\; , \ls
\Gamma^P{}_{(MN)}
\ = \ 
\Gamma^P_{0MN}
- T_M{}^P{}_N - T_N{}^P{}_M
\; , \\
\Gamma^P_{0MN} \ &= \ 
\half g^{PQ} \big( \del_M g_{QN} + \del_N g_{MQ} - \del_Q g_{MN} \big)
\; .
\end{align}
\esubeq
Then the affine connection is also given in terms of the 
Levi-Civita connection and the other:
\begin{align}
\Gamma^P{}_{MN}
\ &= \ 
\Gamma^P_{0MN}
+ K^P{}_{MN}
\; , \ls
K^P{}_{MN}
\ \equiv \ 
T^P{}_{MN} - T_M{}^P{}_N - T_N{}^P{}_M
\; .
\end{align}
The tensor $K^P{}_{MN}$ is called the contorsion.

We also introduce the covariant derivative induced 
by the local Lorentz transformation acting on a generic field $\phi^i$ as
\begin{align}
D_M (\omega) \phi^i \ = \ 
\Big\{ \delta^i_j \, \del_M 
- \frac{i}{2} \omega_M{}^{AB} 
(\Sigma_{AB})^i{}_j \Big\} \phi^j
\; , \label{cov-Lorentz}
\end{align}
where $\Sigma_{AB}$ is the Lorentz generator whose explicit form
depends on the representation of the field $\phi^i$.
The curvature tensor associated with this covariant derivative is
given in terms of the spin connection
\bsubeq 
\begin{gather}
[ D_M (\omega), D_N (\omega) ] \phi \ = \ 
- \frac{i}{2} R^{AB}{}_{MN} (\omega) \, \Sigma_{A B} \phi
\; , \\
R^{A B}{}_{MN} (\omega)
\ = \ 
\del_M \omega_N{}^{A B} - \del_N \omega_M{}^{A B}
+ \omega_M{}^A{}_C \, \omega_N{}^{C B}
- \omega_N{}^A{}_C \, \omega_M{}^{C B}
\; .
\end{gather}
\esubeq
We also describe the first and second Bianchi identity on Riemann
tensor:
\bsubeq \label{Bianchi-Riemann}
\begin{align}
\text{1st:} \ls
0 \ &= \ 
R^M{}_{NPQ} (\Gamma_0)
+ R^M{}_{PQN} (\Gamma_0) + R^M{}_{QNP} (\Gamma_0)
\; , \label{first-B} \\
\text{2nd:} \ls 
0 \ &= \ 
\nabla_M R^N{}_{PQR} (\Gamma_0)
+ \nabla_Q R^N{}_{PRM} (\Gamma_0)
+ \nabla_R R^N{}_{PMQ} (\Gamma_0)
\; . \label{second-B}
\end{align}
\esubeq


\section{Formulae} \label{app-formula}


In the formulation of discretized and continuum path integral in
quantum mechanics, we define a number of functions without
ambiguities \cite{BvN06}. Here let us summarize functions which appear
in propagators and their derivatives in the quantum mechanics.
\bsubeq \label{formula-delta}
\begin{gather}
\Delta (\sigma , \tau)
\ = \ 
\sigma (\tau + 1) \theta (\sigma - \tau)
+ \tau (\sigma + 1 ) \theta (\tau - \sigma)
\ = \ 
\Delta (\tau , \sigma)
\; , \\
\theta (\sigma - \tau) \big|_{\tau = \sigma}
\ = \ 
\half
\; , \ls
\theta (\tau - \sigma)
\ = \ 
- \theta (\sigma - \tau) + 1
\; , \\
\del_{\sigma} \theta (\sigma - \tau)
\ = \ 
\delta (\sigma - \tau)
\; , \ls
\del_{\sigma}^2 \Delta (\sigma , \tau)
\ = \ 
\delta (\sigma - \tau)
\; , \\
\int_{-1}^0 \d \sigma \int_{-1}^0 \d \tau \,
\Delta (\sigma, \tau)
\ = \ - \frac{1}{12}
\; , \ls
\int_{-1}^0 \d \sigma \int_{-1}^0 \d \tau \, 
\delta (\sigma - \tau) \, \theta (\sigma - \tau) \, \theta (\tau - \sigma)
\ = \ 
\frac{1}{4}
\; .
\end{gather}
\esubeq
Notice that $\delta (\sigma - \tau)$ should be regarded as the
``Kronecker delta'' instead of the delta function because this
function appears in the discretized form of the path integral 
and we should take the continuum limit carefully.


By using the above basic functions, we should compute various kinds of
integral when we analyze loop diagrams in the path integral formalism.
In this paper we mainly use a set of useful formulae which appear in
the derivation of invariant polynomials such as the Dirac genus, the
Chern characters, the Hirzebruch signature, and so forth.
Here we only list the formula for these invariant polynomials.
When we derive the Dirac genus, we use the integral $I_k$ defined as
\bsubeq
\begin{gather}
I_k \ \equiv \ 
\int_{-1}^0 \d \tau_1 \cdots \int_{-1}^0 \d \tau_k \,
\del_{\tau_1} \Delta (\tau_1 , \tau_{2})
\del_{\tau_2} \Delta (\tau_2 , \tau_{3})
\cdots
\del_{\tau_{k-1}} \Delta (\tau_{k-1} , \tau_{k})
\del_{\tau_k} \Delta (\tau_k , \tau_{1})
\; , \\
\del_{\tau_i} \Delta (\tau_i , \tau_{i+1})
\ = \ 
\tau_i + \theta (\tau_i - \tau_{i+1})
\; , \ls
\sum_{k=2}^{\infty} \frac{y^k}{k} I_k
\ = \ 
\log \frac{y/2}{\sinh (y/2)}
\; .
\end{gather}
\esubeq
The following two integrals play important roles in the derivations of
the Chern classes and the Hirzebruch signature:
\bsubeq
\begin{gather}
\int_{-1}^0 \d \sigma_1 \int_{-1}^0 \d \sigma_2 \cdots \int_{-1}^0 \d \sigma_k
\, \theta (\sigma_1 - \sigma_2) \theta (\sigma_2 - \sigma_3) \cdots
\theta (\sigma_{k-1} - \sigma_k) \theta (\sigma_k - \sigma_1)
\ = \ 0
\; , \\
\int_{-1}^0 \d \sigma_1 \int_{-1}^0 \d \sigma_2 \cdots \int_{-1}^0 \d \sigma_k
\, \theta (\sigma_1 - \sigma_2) \theta (\sigma_2 - \sigma_3) \cdots
\theta (\sigma_{k-1} - \sigma_k) 
\ = \ \frac{1}{k!}
\; ,
\end{gather}
\esubeq
for $k \geq 2$.
We also use the following integral when we derive the
Hirzebruch signature:
\bsubeq \label{Hir-theta}
\begin{gather}
J_{2 \ell}
\ = \ 
\int_{-1}^0 \prod_{i=1}^{\ell} \d \tau_i \, \d \sigma_i \,
\theta (\tau_1 - \sigma_{\ell}) \theta (\tau_1 - \sigma_1) \, 
\theta (\tau_2 - \sigma_1) \theta (\tau_2 - \sigma_2) 
\cdots 
\theta (\tau_{\ell} - \sigma_{\ell-1}) \theta (\tau_{\ell} - \sigma_{\ell}) 
\; , \\
\sum_{\ell=1}^{\infty} \frac{(-1)^{\ell + 1}}{\ell} y^{2 \ell} J_{2 \ell}
\ = \ 
\log \big( \cosh y \,\big)
\; .
\end{gather}
\esubeq

\end{appendix}


}

\begin{thebibliography}{99}

\bibitem{G0509}
M.~Gra\~{n}a,
``{\sl Flux compactifications in string theory: a comprehensive review},''
Phys. Rept. {\bf 423} (2006) 91
[arXiv:hep-th/0509003].

\bibitem{DK0610}
M.R.~Douglas and S.~Kachru,
  ``{\sl Flux compactification},''
  arXiv:hep-th/0610102.

\bibitem{BKLS0610}
R.~Blumenhagen, B.~K\"{o}rs, D.~L\"{u}st and S.~Stieberger,
  ``{\sl Four-dimensional string compactifications with D-branes, orientifolds and
  fluxes},''
  arXiv:hep-th/0610327.

\bibitem{KKLT0301}
  S.~Kachru, R.~Kallosh, A.~Linde and S.P.~Trivedi,
  ``{\sl De Sitter vacua in string theory},''
  Phys. Rev.  D {\bf 68} (2003) 046005
  [arXiv:hep-th/0301240].

\bibitem{Str86}
  A.~Strominger,
  ``{\sl Superstrings with torsion},''
Nucl. Phys.  B {\bf 274} (1986) 253.

\bibitem{CHSW85}
P.~Candelas, G.T.~Horowitz, A.~Strominger and E.~Witten,
  ``{\sl Vacuum configurations for superstrings},''
  Nucl. Phys.  B {\bf 258} (1985) 46.

\bibitem{Sala89}
S.~Salamon,
``{\sl Riemannian geometry and holonomy groups},''
Pitman Research Notes Math. {\bf 201} (Longman, 1989).

M.~Falcitelli, A.~Farinola and S.~Salamon,
``{\sl Almost-Hermitian geometry},''
Diff. Geom. Appl. {\bf 4} (1994) 259.

S.~Salamon,
``{\sl Almost parallel structures},''
in Proceedings, Congress in memory of Alfred Gray, to appear in AMS
Contemporary Mathematics; Contemp. Math. {\bf 288} (2001), pp.162-181 
[arXiv:math.DG/0107146].


S.~Chiossi and S.~Salamon,
``{\sl The intrinsic torsions of $SU(3)$ and $G_2$ structures},''
in Proc. conf. Differential Geometry Valencia 2001; Differential
Geometry, Valencia 2001, World Sci. Publishing, 2002, pp.115-133
[arXiv:math.DG/0202282].

\bibitem{GMW0302}
  J.P.~Gauntlett, D.~Martelli and D.~Waldram,
  ``{\sl Superstrings with intrinsic torsion},''
  Phys. Rev. D {\bf 69} (2004) 086002
  [arXiv:hep-th/0302158].

\bibitem{Bismut89}
J.-M.~Bismut,
``{\sl A local index theorem for non K\"{a}hler manifolds},''
    Math. Ann. {\bf 284} (1989) 681.

\bibitem{IP0010}
S.~Ivanov and G.~Papadopoulos,
  ``{\sl Vanishing theorems and string backgrounds},''
  Class. Quant. Grav.  {\bf 18} (2001) 1089
  [arXiv:math.DG/0010038].

\bibitem{Pap0201}
G.~Papadopoulos,
  ``{\sl KT and HKT geometries in strings and in black hole moduli spaces},''
in the proceedings of Workshop on Special Geometric Structures in
String Theory, Bonn, Germany, pp.8-11 Sep 2001
  [arXiv:hep-th/0201111].

\bibitem{BT0509}
  K.~Becker and L.-S.~Tseng,
  ``{\sl Heterotic flux compactifications and their moduli},''
Nucl. Phys. B {\bf 741} (2006) 162
  [arXiv:hep-th/0509131].


\bibitem{BTY0612}
M.~Becker, L.-S.~Tseng and S.-T.~Yau,
  ``{\sl Moduli space of torsional manifolds},''
  arXiv:hep-th/0612290.

\bibitem{Alvarez-gaume83a}
  L.~Alvarez-Gaum\'{e},
  ``{\sl Supersymmetry and the Atiyah-Singer index theorem},''
  Commun. Math. Phys.  {\bf 90} (1983) 161.

  L.~Alvarez-Gaum\'{e},
  ``{\sl A note on the Atiyah-Singer index theorem},''
  J. Phys. A: Math. Gen. {\bf 16} (1983) 4177.

\bibitem{ASZ84}
  O.~Alvarez, I.M.~Singer and B.~Zumino,
  ``{\sl Gravitational anomalies and the family's index theorem},''
  Commun. Math. Phys. {\bf 96} (1984) 409.

\bibitem{FOY86}
K.~Fujikawa, S.~Ojima and S.~Yajima,
  ``{\sl On the simple evaluation of chiral anomalies in the path integral
  approach},''
  Phys. Rev. D {\bf 34} (1986) 3223.

\bibitem{EGH}
T.~Eguchi, P.B.~Gilkey and A.J.~Hanson,
``{\sl Gravitation, gauge theories and differential geometry},''
  Phys. Rept. {\bf 66} (1980) 213.

\bibitem{Nakahara}
M.~Nakahara, 
``{\sl Geometry, topology and physics},''
Institute of Physics Publishing (1990), Bristol.

\bibitem{GSW}
  M.B.~Green, J.H.~Schwarz and E.~Witten,
  ``{\sl Superstring theory},'' 
Cambridge University Press (1987).

\bibitem{Mav88}
  N.E.~Mavromatos,
  ``{\sl A note on the Atiyah-Singer index theorem for manifolds with totally
  antisymmetric $H$ torsion},''
 J. Phys. A: Math. Gen. {\bf 21} (1988) 2279.

\bibitem{Yajima89}
S.~Yajima,
  ``{\sl Gravitational anomalies with curl vanishing torsion},''
Hiroshima University Preprint HUPD-8901 (1989).

\bibitem{PW99}
  K.~Peeters and A.~Waldron,
  ``{\sl Spinors on manifolds with boundary: APS index theorems with torsion},''
  JHEP {\bf 9902} (1999) 024
  [arXiv:hep-th/9901016].

\bibitem{SS0010}
  C.A.~Scrucca and M.~Serone,
  ``{\sl A note on the torsion dependence of D-brane RR couplings},''
  Phys. Lett.  B {\bf 504} (2001) 47
  [arXiv:hep-th/0010022].

\bibitem{KY0605}
T.~Kimura and P.~Yi,
  ``{\sl Comments on heterotic flux compactifications},''
  JHEP {\bf 0607} (2006) 030
  [arXiv:hep-th/0605247].


\bibitem{dBPSN9509}
  J.~de Boer, B.~Peeters, K.~Skenderis and P.~van Nieuwenhuizen,
   ``{\sl Loop calculations in quantum-mechanical nonlinear sigma models sigma
  models with fermions and applications to anomalies},''
  Nucl. Phys. B {\bf 459} (1996) 631
  [arXiv:hep-th/9509158].

\bibitem{BvN06}
F.~Bastianelli and P.~van Nieuwenhuizen,
``{\sl Path integrals and anomalies in curved space},''
Cambridge University Press (2006).

\bibitem{IP0008}
  S.~Ivanov and G.~Papadopoulos,
  ``{\sl A no-go theorem for string warped compactifications},''
  Phys. Lett.  B {\bf 497} (2001) 309
  [arXiv:hep-th/0008232].

\bibitem{Braden85}
H.W.~Braden,
  ``{\sl Supersymmetry with torsion},''
  Phys. Lett.  {\bf 163B} (1985) 171,
  [Erratum-ibid.\ B {\bf 167} (1986) 485].

H.W.~Braden,
  ``{\sl Sigma models with torsion},''
  Annals Phys.  {\bf 171} (1986) 433.

\bibitem{RW86}
R.~Rohm and E.~Witten,
  ``{\sl The antisymmetric tensor field in superstring theory},''
  Annals Phys.  {\bf 170} (1986) 454.

\bibitem{Gua0401}
M.~Gualtieri,
``{\sl Generalized complex geometry},''
arXiv:math.DG/0401221.

\bibitem{KL0407}
A.~Kapustin and Y.~Li,
  ``{\sl Topological sigma-models with $H$-flux and twisted generalized complex
  manifolds},''
  arXiv:hep-th/0407249.

\bibitem{NY82}
H.T.~Nieh and M.L.~Yan,
  ``{\sl An identity in Riemann-Cartan geometry},''
  J. Math. Phys.  {\bf 23} (1982) 373.

\bibitem{CZ97}
  O.~Chandia and J.~Zanelli,
  ``{\sl Topological invariants, instantons and chiral anomaly on spaces with
  torsion},''
  Phys. Rev. D {\bf 55} (1997) 7580
  [arXiv:hep-th/9702025].

O.~Chandia and J.~Zanelli,
  ``{\sl Supersymmetric particle in a spacetime with torsion and the index
  theorem},''
  Phys. Rev. D {\bf 58} (1998) 045014
  [arXiv:hep-th/9803034].

\bibitem{AW83}
  L.~Alvarez-Gaum\'{e} and E.~Witten,
  ``{\sl Gravitational anomalies},''
Nucl. Phys.  B {\bf 234} (1984) 269.

  L.~Alvarez-Gaum\'{e} and P. Ginsparg,
``{\sl The structure of gauge and gravitational anomalies},''
Annals Phys.  {\bf 161} (1985) 423
  [Erratum-ibid. {\bf 171} (1986) 233].

\bibitem{GS84}
M.B.~Green and J.H.~Schwarz,
  ``{\sl Anomaly cancellation in supersymmetric $D=10$ gauge theory and superstring
  theory},''
  Phys. Lett.  B {\bf 149} (1984) 117.

\bibitem{Hull86}
C.M.~Hull,
  ``{\sl Anomalies, ambiguities and superstrings},''
  Phys. Lett. B {\bf 167} (1986) 51.

C.M.~Hull,
``{\sl Compactifications of the heterotic superstring},''
Phys. Lett. B {\bf 178} (1986) 357.

\bibitem{BdR89}
E.A.~Bergshoeff and M.~de Roo,
``{\sl The quartic effective action of the heterotic string and supersymmetry},''
  Nucl. Phys. B {\bf 328} (1989) 439.



\bibitem{CCDLMZ0211}
G.L.~Cardoso, G.~Curio, G.~Dall'Agata, D.~L\"{u}st, P.~Manousselis and G.~Zoupanos,
  ``{\sl Non-K\"{a}hler string backgrounds and their five torsion classes},''
  Nucl. Phys. B {\bf 652} (2003) 5,
  [arXiv:hep-th/0211118].

\bibitem{BBDG0301}
  K.~Becker, M.~Becker, K.~Dasgupta and P.S.~Green,
  ``{\sl Compactifications of heterotic theory on non-K\"{a}hler complex manifolds.
  I},''
  JHEP {\bf 0304} (2003) 007
  [arXiv:hep-th/0301161].

\bibitem{CCDL0306}
G.L.~Cardoso, G.~Curio, G.~Dall'Agata and D.~L\"{u}st,
  ``{\sl BPS action and superpotential for heterotic string compactifications  with
  fluxes},''
  JHEP {\bf 0310} (2003) 004
  [arXiv:hep-th/0306088].

\bibitem{BBFTY0604}
  J.-X.~Fu and S.-T.~Yau,
  ``{\sl The theory of superstring with flux on non-K\"{a}hler manifolds and the
  complex Monge-Amp\`{e}re equation},''
 arXiv:hep-th/0604063.

K.~Becker, M.~Becker, J.-X.~Fu, L.-S.~Tseng and S.-T.~Yau,
  ``{\sl Anomaly cancellation and smooth non-K\"{a}hler solutions in
  heterotic string theory},''
  Nucl. Phys. B {\bf 751} (2006) 108
  [arXiv:hep-th/0604137].


\end{thebibliography}
\end{document}